%% file: main.tex
\documentclass[a4paper,12pt,parskip=half,english]{scrartcl}

\include{header}

\begin{document}
\input{chapters/summery}
\input{chapters/introduction}
\input{chapters/basics}
\input{chapters/spatial}

\input{chapters/time}

\input{chapters/example}
\input{chapters/conclusion}

\bibliographystyle{plain}
\bibliography{literature}
\end{document}

%% file: header.tex
\usepackage{babel}
\usepackage[T1]{fontenc}
\usepackage{lmodern,textcomp,latexsym}
\usepackage{amsthm,amssymb,amsmath,mathrsfs}
\usepackage{graphicx,psfrag,color,rotate}
\usepackage{paralist,threeparttable}
\usepackage[footnotesize]{caption}
\usepackage{alphalph}
\usepackage{longtable,booktabs}
\usepackage{tabularx,nth}
\usepackage{tikz}
\usepackage{pifont}
\usepackage{hyperref}

\DeclareCaptionStyle{table_new}{font={footnotesize},justification=raggedright}
\setdefaultenum{1)}{\theenumi.1)}{}{}

\renewcommand{\vec}{\boldsymbol}
\renewcommand{\d}[1][]{\ensuremath{\,\mathrm{d}#1}}

\newcommand{\indi}[1]{\ensuremath{\operatorname{#1}}}
\newcommand{\op}[1]{\ensuremath{\operatorname{#1}}}

\makeatletter
\renewcommand*{\env@matrix}[1][*\c@MaxMatrixCols c]{%
  \hskip -\arraycolsep
  \let\@ifnextchar\new@ifnextchar
  \array{#1}}
\makeatother

\newcommand{\s}{{\mathfrak{s}}}
\newcommand{\cross}{\vec{\times}}

\graphicspath{{./pictures/}}

\DeclareOldFontCommand{\it}{\normalfont\itshape}{\mathit}

%% file: chapters/summery.tex
\begin{center}
\large{\textbf{A strain-gradient formulation for fiber reinforced polymers: Hybrid phase-field model for porous-ductile fracture }}

{\large M.~Dittmann\footnote{Corresponding author. E-mail address: maik.dittmann@uni-siegen.de, christian.hesch@uni-siegen.de}, J.~Schulte, F.~Schmidt and C.~Hesch\(^{*}\)}

{\small
Chair of Computational Mechanics, University of Siegen, Siegen, Germany\\}

\end{center}

\vspace*{-0.1cm}\textbf{Abstract} 

A novel numerical approach to analyze the mechanical behavior within composite materials including the inelastic regime up to final failure is presented. Therefore, a second-gradient theory is combined with phase-field methods to fracture. In particular, we assume that the polymeric matrix material undergoes ductile fracture, whereas continuously embedded fibers undergo brittle fracture as it is typical e.g.\ for roving glass reinforced thermoplastics. A hybrid phase-field approach is developed and applied along with a modified Gurson-Tvergaard-Needelman GTN-type plasticity model accounting for a temperature-dependent growth of voids on microscale. The mechanical response of the arising microstructure of the woven fabric gives rise to additional higher-order terms, representing homogenized bending contributions of the fibers. Eventually, a series of tests is conducted for this physically comprehensive multifield formulation to investigate different kinds and sequences of failure within long fiber reinforced polymers.


%% file: chapters/introduction.tex
\section{Introduction}\label{sec:introduction}

In the past decade, lightweight designs of composite materials have gained increasing attention in research including the field of computational engineering. This is primarily due to the wide range of industrial applications of composites. In the last years, novel machines for additive manufacturing have been designed and introduced to the market, which are able to construct tailor made composites with controllable mechanical properties by adding fibers during the manufacturing process into the produced parts. New challenges arise for the implementation of this class of materials in a suitable simulation environment to obtain realistic predictions of the material behavior. In particular, numerical investigations of complex structures made of composite materials, which undergo large deformations with regard to thermomechanical effects are demanding and of high interest at the same time. In the present work, we focus on long fiber reinforced thermoplastics which are complex materials with pronounced deformations and temperature dependent material properties. In addition, various damage and fracture mechanisms have to be investigated to formulate realistic simulation models. 

Fiber reinforced materials can be understood as materials with a detailed microstructure since the length scales of the embedded fibers are small compared to the surrounding continuum. A most general framework to incorporate the effects of microstructures within a continuum formulation has been presented in the pioneering and most fundamental work on higher-gradient theories of Mindlin \cite{mindlin1964,mindlin1965}, see also the work of Germain \cite{germain1973b}, Toupin \cite{toupin1962,toupin1964} as well as Eringen \cite{eringen1999}. A general non-linear framework on higher-order theories has been proposed by Javili et al.\ \cite{steinmann2013}. Based on these most general formulations, specific mechanical problems like elastic nets and woven fabrics have been addressed in Steigmann et al.\ \cite{steigmann1991,steigmann2012,steigmann2015,steigmann2018}, see also dell'Isola et al.\ \cite{hesch2019a} for the application on panthographic structures. In Schulte et al.\ \cite{hesch2019} \textcolor{blue}{and \cite{duong2020} }, a model of a woven fabric as presented in Steigmann \cite{steigmann2018} has been embedded into the Kirchhoff-Love shell theory. Therein, the in-plane flexural resistance of the fibers is taken into account in addition to first gradient anisotropic effects. Within this previous work, we were able to demonstrate on the basis of experimental results that a classical Cauchy continuum theory without higher-gradient contributions can only be adapted to a specific fiber orientation and load case, but never independently on the orientation. In contrast, the proposed formulation as generalized continuum with higher-gradient contributions allows for an independent modeling without recalibration of the material for the specific fiber direction. Further discussions on higher-order contributions for the constitutive modeling of composites have been presented in Asmanoglo and Menzel \cite{menzel2017a,menzel2017c}, using the framework as provided in the preliminary work of Spencer and Soldatos \cite{spencer2007} and Soldatos \cite{soldatos2010}. Note that we will show here, that the formulation as proposed by dell'Isola et al.\ \cite{hesch2019a} can be recast into the formulation of Asmanoglo and Menzel \cite{menzel2017a,menzel2017c}.

Fiber reinforced polymers are exposed to various damage mechanisms. A large number of phenomenological and micromechanical approaches exists in literature for the modeling of damage within polymers used here as matrix material. To describe such phenomena, the material behavior at the microscale must be incorporated within the continuum formulation. In particular, the growth of microvoids prior to final rupture at the macroscale needs to be considered as rooted in the pioneering works of Gurson \cite{gurson1975,gurson1977}. Therein, a macroscopic yield surface has been developed by homogenization of a porous representative volume element with assumed rigid plastic flow, which degrades with increasing void fraction. Later, this model was modified by Tvergaard \cite{tvergaard1982}, Tvergaard and Needleman \cite{tvergaard1984}, Needleman and Tvergaard \cite{needleman1984}, Leblond et al.\ \cite{leblond1995}, Nahshon and Hutchinson \cite{nahshon2008}, Xue et al.\ \cite{xue2010}, Li et al.\ \cite{li2011} and Huespe et al.\ \cite{huespe2012} to account for damage growth, where the yield criterion function has been extended by introducing new material parameters to account for nucleation and coalescence effects. In H\"utter et al.\ \cite{huetter2013} and Reusch et al.\ \cite{reusch2003a,reusch2003b}, non-local Gurson-models to overcome the nonphysical mesh sensitivity in the softening materials have been presented. Actual work on the application of this model towards polyamide, as often used for fiber reinforced thermoplastics, can be found e.g.\ in Selles et al.\ \cite{selles2016} and Cayzac et al.\ \cite{cayzac2013}.

A model for initiation and propagation of ductile fracture using the damage plasticity theory has been proposed in Bai and Wierzbicki \cite{bai2008}. Moreover, a large number of purely phenomenological approaches exist in literature describing ductile fracture in the context of continuum damage mechanics, see Lemaitre \cite{lemaitre1985,lemaitre1992}, Lemaitre and Chaboche \cite{lemaitre1990}, Steinmann et al.\ \cite{steinmann1994b}, de Borst et al.\ \cite{deborst1999}, Besson \cite{besson2010}, Enakoutsa et al.\ \cite{enakoutsa2006}, Larsson et al.\ \cite{larsson2015}, Seabra et al.\ \cite{seabra2011} and Br\"unig et al.\ \cite{bruenig2011,bruenig2015}. To calculate fracture with complex crack topologies within an efficient computational environment, phase-field methodologies have been developed and applied to multiphysical environments, see, among many other, Miehe et al.\ \cite{miehe2010d}, Hesch et al.\ \cite{hesch2014b,hesch2016b}, Borden et al.\ \cite{borden2014}, Kuhn et al.\ \cite{kuhn2015}, Verhoosel and De Borst \cite{verhoosel2013}, Paggi and Reinoso \cite{paggi2017b}, Teichtmeister et al.\ \cite{teichtmeister2017}, Zhang et al.\ \cite{zhang2017}, Dittmann et al.\ \cite{dittmann2017a,dittmann2018b}, Heider and Markert \cite{heider2017}, Bryant and Sun \cite{bryant2018} and Aldakheel et al.\ \cite{aldakheel2018} for brittle fracture.

Phase-field methodologies have been successfully extended towards ductile fracture by coupling the gradient damage mechanism with models of elastoplasticity. The extension to finite strains is considered in \cite{aldakheel2014,aldakheel2016,miehe2017,dittmann2017b,dittmann2018b,krueger2020} based on the variational principle. In Alessi et al.\ \cite{alessi2018} a comparative study between different phase-field models of fracture coupled with plasticity is outlined. As already stated in the context of damage for polymers, see Gurson \cite{gurson1977}, pressure effects should be included in the modeling of failure in ductile materials to account for complex phenomena at the microscale, such as nucleation, growth and coalescence of microvoids. This has been observed experimentally in Gurland and Plateau \cite{gurland1963}. To this end, Aldakheel et al.\ \cite{aldakheel2017b} extend the phase-field modeling of fracture towards porous finite plasticity to account for this complex phenomena at the microscale as well as for the final rupture at the macroscale. Thermomechanical extensions of the Gurson type model with brittle crack propagation in thermoelastic solids have been shown in Dittmann et al.\ \cite{dittmann2018b} and Miehe et al.\ \cite{miehe2015a}. An extension towards finite strain thermo-porous-plasticity based on the phase-field approach has been recently developed in Dittmann et al.\ \cite{dittmann2019b}.

The paper is structured as follows: The governing equations for the coupled problem are outlined in Section \ref{sec:basics}, whereas algorithmic issues are addressed in Section \ref{sec:spatial} and \ref{sec:time}. In Section \ref{sec:example}, a variety of representative numerical examples is presented including a verification of the implementation. Finally, conclusions are drawn in Section \ref{sec:conclusions}. 

%% file: chapters/basics.tex
\section{Governing equations}\label{sec:basics}
In this section we give a brief summary of the fundamental equations for the modeling of thermomechanical damage in fiber reinforced composites. In order to provide a clear representation of mathematical operations used therein we define the gradient with respect to the reference and current configuration \(\nabla(\bullet)\) and \(\nabla_x(\bullet)\) of a vector field \(\vec{a}\) as
\begin{equation}
[\nabla\vec{a}]_{iJ}=\frac{\partial[\vec{a}]_i}{\partial[\vec{X}]_J}\quad\text{and}\quad[\nabla_x\vec{a}]_{ij}=\frac{\partial[\vec{a}]_i}{\partial[\vec{x}]_j}
\end{equation}
and of a second-order tensor field \(\vec{A}\) as
\begin{equation}
[\nabla\vec{A}]_{iJK}=\frac{\partial[\vec{A}]_{iJ}}{\partial[\vec{X}]_K}\quad\text{and}\quad[\nabla_x\vec{A}]_{iJk}=\frac{\partial[\vec{A}]_{iJ}}{\partial[\vec{x}]_k}.
\end{equation}
The divergence operator with respect to the reference configuration \(\nabla\cdot(\bullet)\) of a second-order tensor field \(\vec{A}\) and third-order tensor field \(\vec{\mathfrak{A}}\) is defined as 
\begin{equation}
[\nabla\cdot\vec{A}]_{i}=\frac{\partial[\vec{A}]_{iJ}}{\partial[\vec{X}]_J}\quad\text{and}\quad[\nabla\cdot\vec{\mathfrak{A}}]_{iJ}=\frac{\partial[\vec{\mathfrak{A}}]_{iJK}}{\partial[\vec{X}]_K},
\end{equation}
respectively. Moreover, the double contractions of a second-order and third-order tensor, i.e.\ \(\vec{a}=\vec{A}:\vec{\mathfrak{A}}\) and \(\vec{b}=\vec{\mathfrak{A}}:\vec{A}\), are defined as \([\vec{a}]_k=[\vec{A}]_{ij}:[\vec{\mathfrak{A}}]_{ijk}\) and \([\vec{b}]_i=[\vec{\mathfrak{A}}]_{ijk}:[\vec{A}]_{jk}\), respectively.

\subsection{Primary fields and state variables}
We consider a fiber reinforced composite as a three-dimensional continuum body which occupies the domain \(\mathcal{B}_0\subset\mathbb{R}^3\) referred to as reference configuration. Assuming that the deformation of the fibers coincides with deformation of the matrix material, we introduce
\begin{equation}
\vec{\varphi}(\vec{X},t):\mathcal{B}_0\times\mathcal{T}\, \rightarrow\, \mathbb{R}^3\quad\text{with}\quad\vec{x}=\vec{\varphi}(\vec{X},t)
\end{equation}
as a common field mapping at time \(t\in\mathcal{T}=[0,T]\) points \(\vec{X}\in\mathcal{B}_0\) onto points \(\vec{x}\in\mathcal{B}\) of the current configuration. The material deformation gradient is defined by \(\vec{F}=\nabla\vec{\varphi}(\vec{X},t)\) with its determinant \(J=\det(\vec{F})>0\). Moreover, the absolute temperature 
\begin{equation}
\theta(\vec{X},t):\mathcal{B}_0\times\mathcal{T}\, \rightarrow\, \mathbb{R}
\end{equation}
is introduced as a further common field representing the thermal state of the matrix as well as the fiber material. 

Regarding the different damage behavior, the above common variables are supplemented by variables describing porous plasticity and ductile fracture of the matrix material and brittle fracture of the fiber material. For the matrix material we introduce the equivalent plastic strain and its dual, the dissipative resistance force
\begin{equation}
\alpha(\vec{X},t):\mathcal{B}_0\times\mathcal{T}\, \rightarrow\, \mathbb{R}\quad\text{and}\quad r^{\mathrm{p}}(\vec{X},t):\mathcal{B}_0\times\mathcal{T}
\end{equation}
along with the plastic deformation map
\begin{equation}
\vec{F}^{\mathrm{p}}(\vec{X},t):\mathcal{B}_0\times\mathcal{T}\, \rightarrow\, \mathbb{R}^{3\times3}\quad\text{with}\quad J^{\mathrm{p}}=\det(\vec{F}^{\mathrm{p}})\geq1.
\end{equation}
and the void volume fraction 
\begin{equation}
f(\vec{X},t):\mathcal{B}_0\times\mathcal{T}\, \rightarrow\, \mathbb{R}\quad\text{with}\quad f=\frac{\text{volume of voids}}{\text{total matrix volume}}\geq f_0,
\end{equation}
where \(f_0\) denotes the initial porosity. In addition, the crack phase-field is described by an order parameter 
\begin{equation}
\s(\vec{X},t):\mathcal{B}_0\times\mathcal{T}\, \rightarrow\, \mathbb{R}\quad\text{with}\quad \s\in[0,1]\quad\text{and}\quad\dot{\s}\geq0,
\end{equation}
where the value \(\s=0\) refers to the undamaged and \(\s=1\) to the fully ruptured state of the matrix material.

Assuming that the fiber reinforcement exhibits a woven structure, we introduce a dual crack phase-field for the fiber material. To be specific, the order parameters
\begin{equation}
\s_{\mathrm{L}}(\vec{X},t):\mathcal{B}_0\times\mathcal{T}\, \rightarrow\, \mathbb{R}\quad\text{with}\quad \s_{\mathrm{L}}\in[0,1]\quad\text{and}\quad\dot{\s_{\mathrm{L}}}\geq0
\end{equation}
and
\begin{equation}
\s_{\mathrm{M}}(\vec{X},t):\mathcal{B}_0\times\mathcal{T}\, \rightarrow\, \mathbb{R}\quad\text{with}\quad \s_{\mathrm{M}}\in[0,1]\quad\text{and}\quad\dot{\s_{\mathrm{M}}}\geq0
\end{equation}
describe the crack phase-field of the fiber aligned in \(\vec{L}\)-direction and \(\vec{M}\)-direction, respectively, where \(\vec{L}\) and \(\vec{M}\) are constant,  orthogonal unit vector fields \textcolor{black}{within the body} in the reference configuration.

The above introduced variables characterize a multifield setting for the formulation of temperature-dependent micro- and macromechanical damage in fiber reinforced composites based on seven independent fields
\begin{equation}
\mathfrak{U}=[\vec{\varphi},\theta,\alpha,r^{\mathrm{p}},\s,\s_{\mathrm{L}},\s_{\mathrm{M}}],
\end{equation}
the finite deformation map \(\vec{\varphi}\), the absolute temperature field \(\theta\), the equivalent plastic strain field \(\alpha\), the dissipative plastic resistance force \(r^{\mathrm{p}}\), the crack phase-field \(\s\) of the matrix material and the dual crack phase-field \([\s_{\mathrm{L}},\s_{\mathrm{M}}]\) of the fiber material. Moreover, the Lagrangian plastic deformation map \(\vec{F}^{\mathrm{p}}\) and the void volume fraction \(f\) will be condensed within the balance equations.

\subsection{Kinematics and deformation measures}
In a first step we derive the required deformation measures related to the matrix and fiber material. To this end, we apply a multiplicative split of the deformation gradient and its determinant as usual in non-linear elastoplasticity and obtain the elastic parts as
\begin{equation}
\vec{F}^{\mathrm{e}} = \vec{F}(\vec{F}^{\mathrm{p}})^{-1}\quad\text{and}\quad J^{\mathrm{e}} = J(J^{\mathrm{p}})^{-1}
\end{equation}
which can also be defined in terms of the elastic parts of the principal stretches \(\lambda_a^{\mathrm{e}}\) with \(a=\{1,2,3\}\) and the principal directions of the left and right stretch tensors \(\vec{n}_a\) and \(\vec{N}_a\) as
\begin{equation}
\vec{F}^{\mathrm{e}}=\sum\limits_a\lambda_a^{\mathrm{e}}\, \vec{n}_a\otimes\vec{N}_a\quad\text{and}\quad J^{\mathrm{e}}=\prod\limits_{a}\lambda_a^{\mathrm{e}}.
\end{equation}
Since the elastoplastic response of the matrix material relies on different mechanisms for the deviatoric and volumetric contributions, it is convenient to introduce the isochoric elastic parts of the principal stretches
\begin{equation}
\bar{\lambda}_a^{\mathrm{e}}=(J^{\mathrm{e}})^{-1/3}\lambda_a^{\mathrm{e}}=\prod\limits_{b}(\lambda_b^{\mathrm{e}})^{-1/3}\lambda_a^{\mathrm{e}}.
\end{equation}
Following the ansatz proposed in Hesch \& Weinberg \cite{hesch2014a}, fracture insensitive parts of the elastic principal stretches are given as
\begin{equation}\label{lambdaMat}
\tilde{\lambda}_a^{\mathrm{e}}=(\lambda^{\mathrm{e}}_a)^{g(\s)}\quad\text{and}\quad\tilde{\bar{\lambda}}_a^{\mathrm{e}}=(\bar{\lambda}^{\mathrm{e}}_a)^{g(\s)}, 
\end{equation}
where \(g=a_{\mathrm{g}}((1-\s)^3-(1-\s)^2)-2(1-\s)^3+3(1-\s)^2\) is an adjustable degradation function via the modeling parameter \(a_{\mathrm{g}}\). Assuming that fracture requires a local state of tensile/shear deformation as considered, e.g.\ in Amor et al.\ \cite{amor2009} and Dittmann et al.\ \cite{dittmann2018,dittmann2019b}, we define the elastic fracture insensitive part of the isochoric deformation gradient and the Jacobian determinant as
\begin{equation}
\tilde{\bar{\vec{F}}}^{\mathrm{e}} = \sum\limits_{a}\tilde{\bar{\lambda}}^{\mathrm{e}}_a\, \vec{n}_a\otimes\vec{N}_a\quad\text{and}\quad 
\tilde{J}^{\mathrm{e}}=\begin{cases}\prod\limits_{a}\tilde{\lambda}^{\mathrm{e}}_{a} & \text{if} \quad \prod\limits_{a}\lambda^{\mathrm{e}}_{a} > 1\\ \prod\limits_{a}\lambda^{\mathrm{e}}_{a} & \text{else}\end{cases}\, .
\end{equation}

Concerning the fiber material, we introduce
\begin{equation}
\lambda_{\mathrm{L}}=\|\vec{l}\|=\|\vec{F}\vec{L}\|\quad\text{and}\quad\lambda_{\mathrm{M}}=\|\vec{m}\|=\|\vec{F}\vec{M}\|
\end{equation}
as stretch of the respective fiber and
\begin{equation}
\begin{aligned}
\varphi&=\mathrm{acos}(\tilde{\vec{l}}\cdot\tilde{\vec{m}})-\frac{\pi}{2}\\
&=\mathrm{acos}\left(\frac{(\vec{F}\vec{L})\cdot(\vec{F}\vec{M})}{\|\vec{F}\vec{L}\|\|\vec{F}\vec{M}\|}\right)-\frac{\pi}{2}
\end{aligned}
\end{equation}
as change of the angle between both fibers. Here \(\vec{l}=\lambda_{\mathrm{L}}\tilde{\vec{l}}\) and \(\vec{m}=\lambda_{\mathrm{M}}\tilde{\vec{m}}\) are deformed fiber configurations decomposed into fiber stretches and normalized fiber directions. To describe fiber bending, the gradients of the deformed fiber vectors, i.e.\ \(\nabla\vec{l}=\nabla\vec{F}\vec{L}\) and \(\nabla\vec{m}=\nabla\vec{F}\vec{M}\), have to be taken into account. In particular, we consider 
\begin{equation}
\nabla\vec{l}\vec{L}=\lambda_{\mathrm{L}}\nabla\tilde{\vec{l}}\vec{L}+(\nabla\lambda_{\mathrm{L}}\cdot\vec{L})\tilde{\vec{l}}
\quad\text{and}\quad\nabla\vec{m}\vec{M}=\lambda_{\mathrm{M}}\nabla\tilde{\vec{m}}\vec{M}+(\nabla\lambda_{\mathrm{M}}\cdot\vec{M})\tilde{\vec{m}}
\end{equation}
which are projections of the fiber configuration gradients onto the initial fiber direction, cf.\ Asmanoglo and Menzel \cite{menzel2017a}. These expressions include terms related to stretch gradients of the fibers as well as fiber curvatures. We introduce the curvature measure for the fiber initially aligned in \(\vec{L}\)-direction as
\begin{equation}
\begin{aligned}
\vec{\kappa}_{\mathrm{L}}&=\frac{1}{\lambda_{\mathrm{L}}}\nabla\tilde{\vec{l}}\vec{L}\\
&=\frac{1}{\lambda^2_{\mathrm{L}}}(\nabla\vec{l}-\tilde{\vec{l}}\otimes\nabla\lambda_{\mathrm{L}})\vec{L}\\
&=\frac{1}{\|\vec{F}\vec{L}\|^2}\left(\nabla\vec{F}\vec{L}-\frac{\vec{F}\vec{L}}{\|\vec{F}\vec{L}\|}\otimes\left(\frac{\vec{F}\vec{L}}{\|\vec{F}\vec{L}\|}\otimes\vec{L}\right):\nabla\vec{F}\right)\vec{L}
\end{aligned}
\end{equation}
and for the fiber initially aligned in \(\vec{M}\)-direction as
\begin{equation}
\begin{aligned}
\vec{\kappa}_{\mathrm{M}}&=\frac{1}{\lambda_{\mathrm{M}}}\nabla\tilde{\vec{m}}\vec{M}\\
&=\frac{1}{\lambda^2_{\mathrm{M}}}(\nabla\vec{m}-\tilde{\vec{m}}\otimes\nabla\lambda_{\mathrm{M}})\vec{M}\\
&=\frac{1}{\|\vec{F}\vec{M}\|^2}\left(\nabla\vec{F}\vec{M}-\frac{\vec{F}\vec{M}}{\|\vec{F}\vec{M}\|}\otimes\left(\frac{\vec{F}\vec{M}}{\|\vec{F}\vec{M}\|}\otimes\vec{M}\right):\nabla\vec{F}\right)\vec{M}.
\end{aligned}
\end{equation}
Assuming that the fiber material is brittle compared to the matrix material and that fiber rupture requires a local tensile state, we formulate fracture insensitive parts of the stretches as
\begin{equation}
\tilde{\lambda}_{\mathrm{L}}=\begin{cases}(\lambda_{\mathrm{L}})^{g_{\mathrm{L}}(\s_{\mathrm{L}})} & \text{if} \quad \lambda_{\mathrm{L}} > 1 \\ \lambda_{\mathrm{L}} & \text{else}\end{cases}\quad\text{and}\quad
\tilde{\lambda}_{\mathrm{M}}=\begin{cases}(\lambda_{\mathrm{M}})^{g_{\mathrm{M}}(\s_{\mathrm{M}})} & \text{if} \quad \lambda_{\mathrm{M}} > 1 \\ \lambda_{\mathrm{M}} & \text{else}\end{cases}\, ,
\end{equation}
where \(g_{\mathrm{L}}=a_{\mathrm{g_L}}((1-\s_{\mathrm{L}})^3-(1-\s_{\mathrm{L}})^2)-2(1-\s_{\mathrm{L}})^3+3(1-\s_{\mathrm{L}})^2\) and \(g_{\mathrm{M}}=a_{\mathrm{g_M}}((1-\s_{\mathrm{M}})^3-(1-\s_{\mathrm{M}})^2)-2(1-\s_{\mathrm{M}})^3+3(1-\s_{\mathrm{M}})^2\) are degradation functions with modeling parameters \(a_{\mathrm{g_L}}\) and \(a_{\mathrm{g_M}}\), cf.\ (\ref{lambdaMat}). Next, we formulate a corresponding measure related to the shear deformation of the fiber material as
\begin{equation}
\tilde{\phi}=g_{\mathrm{L}}(\s_{\mathrm{L}})g_{\mathrm{M}}(\s_{\mathrm{M}})\tan(\varphi).
\end{equation}
Note that this deformation measure is completely degraded in case of single fiber rupture even if the remaining fiber is undamaged. Eventually, fracture insensitive measures of the fiber curvatures read
\begin{equation}
\tilde{\vec{\kappa}}_{\mathrm{L}}=g_{\mathrm{L}}(\s_{\mathrm{L}})\vec{\kappa}_{\mathrm{L}}\quad\text{and}\quad\tilde{\vec{\kappa}}_{\mathrm{M}}=g_{\mathrm{M}}(\s_{\mathrm{M}})\vec{\kappa}_{\mathrm{M}}.
\end{equation}

\subsection{Variational formulation}
Next, we propose the constitutive framework for thermomechanical damage in fiber reinforced composites. To be specific, we introduce constitutive energetic and dissipative response functions based on the above definitions and derive the required relations and evolution laws to formulate the multifield variational problem.

\subsubsection{Energetic response}
The stored thermoelastic energy density of the composite material is defined by the functional
\begin{equation}\label{eq:storedEnergy}
\begin{aligned}
\Psi&=\zeta\Psi_{\mathrm{mat}}^{\mathrm{e},\theta}(\tilde{\bar{\vec{F}}}^{\mathrm{e}},\tilde{J}^{\mathrm{e}},\theta)
+\frac{1-\zeta}{2}\Psi_{\mathrm{fib}}^{\mathrm{e},\theta}(\tilde{\lambda}_{\mathrm{L}},\tilde{\lambda}_{\mathrm{M}},\tilde{\phi},\tilde{\vec{\kappa}}_{\mathrm{L}},\tilde{\vec{\kappa}}_{\mathrm{M}},\theta)\\
&=\zeta\left(\Psi_{\mathrm{mat}}^{\mathrm{e}}(\tilde{\bar{\vec{F}}}^{\mathrm{e}},\tilde{J}^{\mathrm{e}},\theta)+\Psi_{\mathrm{mat}}^{\theta}(\theta)\right)
+\frac{1-\zeta}{2}\left(\Psi_{\mathrm{fib}}^{\mathrm{e}}(\tilde{\lambda}_{\mathrm{L}},\tilde{\lambda}_{\mathrm{M}},\tilde{\phi},\tilde{\vec{\kappa}}_{\mathrm{L}},\tilde{\vec{\kappa}}_{\mathrm{M}},\theta)+\Psi_{\mathrm{fib}}^{\theta}(\theta)\right),
\end{aligned}
\end{equation}
where \(\zeta\in[0,1]\) is the volume fraction of the matrix material. The elastic contribution to the stored energy function of the matrix material is decomposed into volumetric and deviatoric parts
\begin{equation}
\Psi_{\mathrm{mat}}^{\mathrm{e}}=
\Psi_{\mathrm{mat}}^{\mathrm{e,iso}}\left(\tilde{\bar{\vec{F}}}^{\mathrm{e}}(\lambda_1^{\mathrm{e}},\lambda_2^{\mathrm{e}},\lambda_3^{\mathrm{e}},\s),\theta\right)+\Psi_{\mathrm{mat}}^{\mathrm{e,vol}}\left(\tilde{J}^{\mathrm{e}}(\lambda_1^{\mathrm{e}},\lambda_2^{\mathrm{e}},\lambda_3^{\mathrm{e}},\s),\theta\right).
\end{equation}
As a representative non-linear constitutive law, a modified Ogden material model with the associated strain energy density function
\begin{equation}\label{eq:ogden}
\Psi_{\mathrm{mat}}^{\mathrm{e,iso}}=\sum\limits_a\sum\limits_b\frac{\mu_b}{\alpha_b}\left((\tilde{\bar{\lambda}}_a^{\mathrm{e}})^{\alpha_b}-1\right)
\end{equation}
and
\begin{equation}
\Psi_{\mathrm{mat}}^{\mathrm{e,vol}}=\frac{\kappa}{\beta^2}\left(\beta\ln(\tilde{J}^{\mathrm{e}})+(\tilde{J}^{\mathrm{e}})^{-\beta}-1\right)-3\frac{\epsilon\kappa}{\gamma}(\theta-\theta_0)\left((\tilde{J}^{\mathrm{e}})^{\gamma}-1\right)
\end{equation}
is used for the numerical examples. The parameters \(\mu_b\) and \(\alpha_b\) with \(b=\{1,\hdots,N\}\) are related to the shear modulus and the parameters \(\kappa\) and \(\beta\) are related to the bulk modulus. Moreover, \(\theta_0\) is a reference temperature and the parameters \(\epsilon\) and \(\gamma\) are related to the thermal expansion coefficient. Assuming that the fiber portion in both directions is identical, the corresponding elastic contribution of the fiber material is defined by
\begin{equation}
\begin{aligned}
\Psi_{\mathrm{fib}}^{\mathrm{e}}&=\frac{1}{2}a\left((\tilde{\lambda}_{\mathrm{L}}-1)^2+(\tilde{\lambda}_{\mathrm{M}}-1)^2\right)+b\,\tilde{\phi}^2\\
&\quad+\frac{1}{2}\left(\tilde{\vec{\kappa}}_{\mathrm{L}}\cdot\vec{c}\,\tilde{\vec{\kappa}}_{\mathrm{L}}+\tilde{\vec{\kappa}}_{\mathrm{M}}\cdot\vec{c}\, \tilde{\vec{\kappa}}_{\mathrm{M}}\right)
+a\upsilon(\theta-\theta_0)\left((\tilde{\lambda}_{\mathrm{L}}-1)+(\tilde{\lambda}_{\mathrm{M}}-1)\right)
\end{aligned}
\end{equation} 
where \(a\) and \(b\) are stiffness parameters related to stretch and shear of the fiber material and \(\upsilon\) denotes the thermal expansion coefficient. Moreover, the stiffness tensor related to fiber curvature is given as
\begin{equation}
\vec{c}=c_{\#}(\tilde{\vec{l}}\otimes\tilde{\vec{l}}+\tilde{\vec{m}}\otimes\tilde{\vec{m}})+c_{\perp}\tilde{\vec{n}}\otimes\tilde{\vec{n}}\quad\text{with}\quad\tilde{\vec{n}}=\tilde{\vec{l}}\times\tilde{\vec{m}}
\end{equation}
taking into account a geometric dependency via the stiffness parameters \(c_{\#}\) and \(c_{\perp}\), \textcolor{black}{which can be interpreted as the in-plane and out-of-plane bending stiffness,} see Section \ref{sec:inplane}, \textcolor{black}{\cite{hesch2019}, \cite{menzel2017a}, and \cite{delIsola2015}} for details.  Next, the purely thermal contributions to the stored energy of the matrix and the fiber material 
are defined by 
\begin{equation}
\Psi_{\mathrm{mat}}^{\theta}=c_{\mathrm{mat}}\left(\theta-\theta_0-\theta\ln\left(\frac{\theta}{\theta_0}\right)\right)
\end{equation}
and
\begin{equation}\label{eq:thermalEnergyFib}
\Psi_{\mathrm{fib}}^{\theta}=2c_{\mathrm{fib}}\left(\theta-\theta_0-\theta\ln\left(\frac{\theta}{\theta_0}\right)\right),
\end{equation}
respectively. Therein, \(c_{\mathrm{mat}}\) and \(c_{\mathrm{fib}}\) are constant parameters representing specific heat capacities of the respective material. 

The evolution of the stored thermoelastic energy is given by
\begin{equation}
\begin{aligned}
\frac{\d}{\d t}\Psi&=\zeta\left(\sum\limits_a\frac{\partial\Psi_{\mathrm{mat}}^{\mathrm{e}}}{\partial\lambda_a^{\mathrm{e}}}\dot{\lambda}_a^{\mathrm{e}}
+\frac{\partial\Psi_{\mathrm{mat}}^{\mathrm{e}}}{\partial\s}\dot{\s}+\frac{\partial(\Psi_{\mathrm{mat}}^{\mathrm{e}}+\Psi_{\mathrm{mat}}^{\theta})}{\partial\theta}\dot{\theta}\right)\\
&\quad+\frac{1-\zeta}{2}\left(\frac{\partial\Psi_{\mathrm{fib}}^{\mathrm{e}}}{\partial\vec{F}}\dot{\vec{F}}+\frac{\partial\Psi_{\mathrm{fib}}^{\mathrm{e}}}{\partial\nabla\vec{F}}\nabla\dot{\vec{F}}
+\frac{\partial\Psi_{\mathrm{fib}}^{\mathrm{e}}}{\partial\s_{\mathrm{L}}}\dot{\s}_{\mathrm{L}}+\frac{\partial\Psi_{\mathrm{fib}}^{\mathrm{e}}}{\partial\s_{\mathrm{M}}}\dot{\s}_{\mathrm{M}}
+\frac{\partial(\Psi_{\mathrm{fib}}^{\mathrm{e}}+\Psi_{\mathrm{fib}}^{\theta})}{\partial\theta}\dot{\theta}\right).
\end{aligned}
\end{equation}
Regarding the partial derivatives therein, we introduce first relations related to the Kirchhoff stress \(\vec{\tau}=\vec{\tau}_{\mathrm{mat}}+\vec{\tau}_{\mathrm{fib}}\) as 
\begin{equation}
\begin{aligned}
\vec{\tau}_{\mathrm{mat}}&=\vec{\tau}^{\mathrm{dev}}_{\mathrm{mat}}+\vec{\tau}^{\mathrm{vol}}_{\mathrm{mat}}\\
&=\sum\limits_a\left(\tau_{\mathrm{mat},a}^{\mathrm{dev}}+\tau_{\mathrm{mat},a}^{\mathrm{vol}}\right)\vec{n}_{a}\otimes\vec{n}_{a}\\
&=\zeta\sum\limits_a\lambda_a^{\mathrm{e}}\left(\frac{\partial\Psi_{\mathrm{mat}}^{\mathrm{e,iso}}}{\partial\lambda_a^{\mathrm{e}}}+\frac{\partial\Psi_{\mathrm{mat}}^{\mathrm{e,vol}}}{\partial\lambda_a^{\mathrm{e}}}\right)\vec{n}_{a}\otimes\vec{n}_{a}
\end{aligned}
\end{equation}
and
\begin{equation}
\begin{aligned}
\vec{\tau}_{\mathrm{fib}}=\frac{1-\zeta}{2}\left(\frac{\partial\Psi_{\mathrm{fib}}^{\mathrm{e}}}{\partial\tilde{\lambda}_{\mathrm{L}}}\frac{\partial\tilde{\lambda}_{\mathrm{L}}}{\partial\vec{F}}
+\frac{\partial\Psi_{\mathrm{fib}}^{\mathrm{e}}}{\partial\tilde{\lambda}_{\mathrm{M}}}\frac{\partial\tilde{\lambda}_{\mathrm{M}}}{\partial\vec{F}}
+\frac{\partial\Psi_{\mathrm{fib}}^{\mathrm{e}}}{\partial\tilde{\phi}}\frac{\partial\tilde{\phi}}{\partial\vec{F}}
+\frac{\partial\Psi_{\mathrm{fib}}^{\mathrm{e}}}{\partial\tilde{\vec{\kappa}}_{\mathrm{L}}}\frac{\partial\tilde{\vec{\kappa}}_{\mathrm{L}}}{\partial\vec{F}}
+\frac{\partial\Psi_{\mathrm{fib}}^{\mathrm{e}}}{\partial\tilde{\vec{\kappa}}_{\mathrm{M}}}\frac{\partial\tilde{\vec{\kappa}}_{\mathrm{M}}}{\partial\vec{F}}\right)\vec{F}^{\mathrm{T}},
\end{aligned}
\end{equation}
the higher-order stress of the fiber material as
\begin{equation}
\vec{\mathfrak{P}}_{\mathrm{fib}}=\frac{1-\zeta}{2}\left(\frac{\partial\Psi_{\mathrm{fib}}^{\mathrm{e}}}{\partial\tilde{\vec{\kappa}}_{\mathrm{L}}}\frac{\partial\tilde{\vec{\kappa}}_{\mathrm{L}}}{\partial\nabla\vec{F}}
+\frac{\partial\Psi_{\mathrm{fib}}^{\mathrm{e}}}{\partial\tilde{\vec{\kappa}}_{\mathrm{M}}}\frac{\partial\tilde{\vec{\kappa}}_{\mathrm{M}}}{\partial\nabla\vec{F}}\right),
\end{equation}
the driving force of the respective crack phase-field as
\begin{equation}
\mathcal{H}=-\zeta\frac{\partial\Psi^{\mathrm{e}}_{\mathrm{mat}}}{\partial\s},\quad
\mathcal{H}_{\mathrm{L}}=-\frac{1-\zeta}{2}\frac{\partial\Psi^{\mathrm{e}}_{\mathrm{fib}}}{\partial\s_{\mathrm{L}}},\quad
\mathcal{H}_{\mathrm{M}}=-\frac{1-\zeta}{2}\frac{\partial\Psi^{\mathrm{e}}_{\mathrm{fib}}}{\partial\s_{\mathrm{M}}}
\end{equation}
and the specific entropy as
\begin{equation}\label{eq:entropy}
\begin{aligned}
\eta&=\eta_{\mathrm{mat}}+\eta_{\mathrm{fib}}\\
&=-\zeta\frac{\partial(\Psi_{\mathrm{mat}}^{\mathrm{e}}+\Psi_{\mathrm{mat}}^{\theta})}{\partial\theta}-\frac{1-\zeta}{2}\frac{\partial(\Psi_{\mathrm{fib}}^{\mathrm{e}}+\Psi_{\mathrm{fib}}^{\theta})}{\partial\theta}.
\end{aligned}
\end{equation}
Moreover, we introduce a dissipation function 
\begin{equation}\label{eq:dissipation}
\mathcal{D}_{\mathrm{int}}=\nu_{\mathrm{p_{mat}}}\vec{\tau}_{\mathrm{mat}}:\vec{d}^{\mathrm{p}}+\nu_{\mathrm{f_{mat}}}\mathcal{H}\dot{\s}+\nu_{\mathrm{f_{fib}}}(\mathcal{H}_{\mathrm{L}}\dot{\s}_{\mathrm{L}}+\mathcal{H}_{\mathrm{M}}\dot{\s}_{\mathrm{M}}),
\end{equation}
to account for a transfer of dissipated energy due to plastification and fracture into the thermal field, where \(\vec{d}^{\mathrm{p}}\) denotes the Eulerian plastic rate of deformation tensor. The above relations are derived in a thermodynamically consistent manner by assuming that the dissipated energy is completely transfered into the thermal field, i.e.\ by setting \(\nu_{\mathrm{p_{mat}}}=\nu_{\mathrm{f_{mat}}}=\nu_{\mathrm{f_{fib}}}=1\). Note, however, that the plastic dissipation factor \(\nu_{\mathrm{p_{mat}}}\) is typically chosen in the range of \(85\%\) to \(95\%\) in the context of thermoplasticity, see e.g.\ \cite{simo1992f,zdebel1987,lehmann1985}. In addition, based on  experimental observations it is reasonable to set fracture dissipation factors to \(\nu_{\mathrm{f_{mat}}}<1\) and \(\nu_{\mathrm{f_{fib}}}<1\), see the discussion related to an energy transfer into the thermal field in \cite{dittmann2018b,schulte2018} and the references therein.

To model the plastic and fracture mechanical response, we introduce an auxiliary functional as
\begin{equation}\label{eq:auxEnergy}
\widehat{\Psi}=\zeta\left(\widehat{\Psi}_{\mathrm{mat}}^{\mathrm{p}}(\alpha,\nabla\alpha,\theta)+\widehat{\Psi}_{\mathrm{mat}}^{\mathrm{f}}(\s,\nabla\s,\alpha)\right)+\frac{1-\zeta}{2}\widehat{\Psi}_{\mathrm{fib}}^{\mathrm{f}}(\s_{\mathrm{L}},\nabla\s_{\mathrm{L}},\s_{\mathrm{M}},\nabla\s_{\mathrm{M}}).
\end{equation}
The plastic contribution \(\widehat{\Psi}_{\mathrm{mat}}^{\mathrm{p}}\) describes the response of isotropic strain-gradient hardening related to the matrix material. To be specific, we focus on the equivalent plastic strain \(\alpha\) and its gradient
\(\nabla\alpha\) with the particular form
\begin{equation}
\widehat{\Psi}_{\mathrm{mat}}^{\mathrm{p}}(\alpha,\nabla\alpha,\theta)=\int\limits_{0}^{\alpha}y(\bar{\alpha},\theta)\d \bar{\alpha}+y_0(\theta)\frac{l^2_{\mathrm{p}}}{2}\|\nabla\alpha\|^2.
\end{equation}
Here, \(l_{\mathrm{p}}\) is a plastic length scale related to a strain-gradient hardening effect and accounts for size effects to overcome the nonphysical mesh sensitivity of the localized plastic deformation in softening materials, as outlined in \cite{aldakheel2016}. Moreover, \(y(\alpha,\theta)\) is an isotropic local hardening function taken from \cite{cayzac2013,selles2016} and adapted to thermoplasticity following \cite{simo1992f,reis2016,dacosta2019}. In particular, we use the saturation-type function
\begin{equation}\label{eq:hardeningSatFunction}
y(\alpha,\theta)=y_{0}(\theta) + y_{1}(\theta)\mathrm{exp}[\omega_{\mathrm{p1}}\alpha] + y_{2}(\theta)(1-\mathrm{exp}[-\omega_{\mathrm{p2}}\alpha])  , 
\end{equation}
with the three temperature-dependent material parameters \(y_0>0\), \(y_1\ge0\) and \(y_{2}\ge0\) defined as
\begin{equation}
\begin{aligned}
y_{0}(\theta)&=y_{0}(\theta_{\mathrm{ref}})(1-\omega_{\mathrm{t0}}(\theta-\theta_{\mathrm{ref}})),\\
y_{1}(\theta)&=y_{1}(\theta_{\mathrm{ref}})(1-\omega_{\mathrm{t1}}(\theta-\theta_{\mathrm{ref}})),\\
y_{2}(\theta)&=y_{2}(\theta_{\mathrm{ref}})(1-\omega_{\mathrm{t2}}(\theta-\theta_{\mathrm{ref}})).
\end{aligned}
\end{equation}
Note that this formulation is typically applied for polyamide which is often used as matrix material of composite structures. 
Therein, the initial yield stress \(y_0+y_1\) determines the threshold of the effective elastic response, \(y_{2}(\theta)(1-\mathrm{exp}[-\omega_{\mathrm{p2}}\alpha])\) describes an initial hardening stage and \(y_{1}(\theta)\mathrm{exp}[\omega_{\mathrm{p1}}\alpha]\) allows for the simulation of large stretches of fibrils which leads to an abrupt increase of stress. This phenomenon is often called rheo-hardening. Moreover, \(\omega_{\mathrm{p1}}\) and \(\omega_{\mathrm{p2}}\) are saturation parameters and \(\omega_{\mathrm{t}0}\), \(\omega_{\mathrm{t}1}\) and \(\omega_{\mathrm{t}2}\) are thermal hardening/softening parameters. Note that since we are only interested in the mean mechanical effects of the semi-crystalline matrix material, we consider a unified elastoplastic model with averaged material parameters taken from the multimechanism model in \cite{cayzac2013,selles2016}. Next, we formulate fracture contributions for the matrix as well as the fiber material. Therefore, we approximate a sharp crack surface \(\Gamma_{\bullet}\) by a regularized functional\footnote{The \(\bullet\) indicates the matrix material or the respective fiber direction.}
\begin{equation}\label{reg}
\widehat{\Gamma}_{\bullet}(\s_{\bullet},\nabla\s_{\bullet})=\int\limits_{\mathcal{B}_0}\widehat{\gamma}_{\bullet}(\s_{\bullet},\nabla\s_{\bullet})\d V\quad\text{with}\quad\widehat{\gamma}_{\bullet}(\s_{\bullet},\nabla\s_{\bullet})=\frac{1}{2l_{\mathrm{f}_{\bullet}}}(\s_{\bullet}^2+l_{\mathrm{f}_{\bullet}}^2\|\nabla\s_{\bullet}\|^2),
\end{equation}
based on a specific crack regularization profile \(\widehat{\gamma}_{\bullet}\) defined per unit volume of the reference configuration and the fracture length scale \(l_{\mathrm{f}_{\bullet}}\) which controls the regularization. Concerning ductile fracture of the matrix material, we require that \(l_{\mathrm{p}}\geq l_{\mathrm{f}}\) such that the regularized crack zone lies inside of the plastic zone. Using the regularization given in \eqref{reg}, the approximated fracture energy of the composite material reads 
\begin{equation}
W^{\mathrm{f}}\approx\int\limits_{\mathcal{B}_0}\zeta g_{\mathrm{c}}(\alpha)\widehat{\gamma}(\s,\nabla\s)+\frac{1-\zeta}{2}\left(g_{\mathrm{c_L}}\widehat{\gamma}_{\mathrm{L}}(\s_{\mathrm{L}},\nabla\s_{\mathrm{L}})+g_{\mathrm{c_M}}\widehat{\gamma}_{\mathrm{M}}(\s_{\mathrm{M}},\nabla\s_{\mathrm{M}})\right)\d V.
\end{equation}
Here, \(g_{\mathrm{c}_{\bullet}}\) denotes the Griffith-type critical energy density required to create fracture within the respective material. For the matrix material, the critical energy density is decomposed additively into elastic and plastic contributions as
\begin{equation}\label{eq:critFractureEnergy}
g_{\mathrm{c}}(\alpha)=g_{\mathrm{c,p}}+g_{\mathrm{c,e}}\exp(-\omega_{\mathrm{f}}\alpha),
\end{equation}
where \(\omega_{\mathrm{f}}\) is a modeling parameter. Summarized, the phase-field fracture contributions are given in terms of crack density functions as
\begin{equation}
\begin{aligned}
\widehat{\Psi}^{\mathrm{f}}_{\mathrm{mat}}&=g_{\mathrm{c}}(\alpha)\widehat\gamma(\s,\nabla\s)\\
&=\frac{g_{\mathrm{c}}(\alpha)}{2l_{\mathrm{f}}}(\s^2+l^2_{\mathrm{f}}\|\nabla\s\|^2)
\end{aligned}
\end{equation}
and
\begin{equation}
\begin{aligned}
\widehat{\Psi}^{\mathrm{f}}_{\mathrm{fib}}&=g_{\mathrm{c_L}}\widehat{\gamma}_{\mathrm{L}}(\s_{\mathrm{L}},\nabla\s_{\mathrm{L}})+g_{\mathrm{c_M}}\widehat{\gamma}_{\mathrm{M}}(\s_{\mathrm{M}},\nabla\s_{\mathrm{M}})\\
&=\frac{g_{\mathrm{c_L}}}{2l_{\mathrm{f_L}}}(\s_{\mathrm{L}}^2+l^2_{\mathrm{f_L}}\|\nabla\s_{\mathrm{L}}\|^2) +\frac{g_{\mathrm{c_M}}}{2l_{\mathrm{f_M}}}(\s_{\mathrm{M}}^2+l^2_{\mathrm{f_M}}\|\nabla\s_{\mathrm{M}}\|^2).
\end{aligned}
\end{equation}
Eventually, we obtain dissipative resistance forces of the plastic field and the respective crack phase-field via the variational derivatives of \(\widehat{\Psi}\) with respect to \(\alpha\) and \(\s_{\bullet}\) as
\begin{equation}
r^{\mathrm{p}}=\zeta\delta_{\alpha}\widehat{\Psi}^{\mathrm{p}}_{\mathrm{mat}}=\zeta(\partial_{\alpha}\widehat{\Psi}^{\mathrm{p}}_{\mathrm{mat}}-\nabla\cdot(\partial_{\nabla\alpha}\widehat{\Psi}^{\mathrm{p}}_{\mathrm{mat}}))
\end{equation}
and
\begin{equation}
r^{\mathrm{f}}_{\bullet}=\delta_{\s_{\bullet}}\widehat{\Psi} = \partial_{\s_{\bullet}}\widehat{\Psi}-\nabla\cdot(\partial_{\nabla\s_{\bullet}}\widehat{\Psi}).
\end{equation}

\subsubsection{Dissipative response}
Regarding the porous elastoplastic material behavior, we consider a GTN type function \cite{gurson1977,tvergaard1982,needleman1984} which implicitly defines the effective scalar stress \(\bar{\sigma}:=\bar{\sigma}(\vec{\sigma}_{\mathrm{mat}},f)\) in terms of the Cauchy stress tensor \(\vec{\sigma}_{\mathrm{mat}}=\vec{\tau}_{\mathrm{mat}}/J\) and the void volume fraction \(f\)
\begin{equation}\label{eq:gursonyield}
\Upsilon^\mathrm{G}(\bar{\sigma},\vec{\sigma}_{\mathrm{mat}},f)=\frac{\sigma^2_{\mathrm{eq}}}{{\bar{\sigma}}^2}+2 q_1 f\mathrm{cosh}\left[\frac{3}{2} q_2 \frac{p}{\bar{\sigma}}\right]-\left(1+( q_1 f)^2\right)=0.
\end{equation}
Here, \(\sigma_{\mathrm{eq}}=\sqrt{3/2}\, \| \vec{\tau}^{\mathrm{dev}}_{\mathrm{mat}} / J \|\) denotes the von Mises equivalent stress, \(p=\frac{1}{3}\mathrm{tr}[\vec{\tau}_{\mathrm{mat}}/J] \) the pressure and \(q_{1/2}\) are fitting parameters. Note that for \(q_1=0\) the influence of the pressure and the void volume fraction vanishes, i.e.\ \( \bar{\sigma}=\sigma_\mathrm{eq}\). With the effective stress \(\bar{\sigma}\) and the dissipative resistance force $r^{\mathrm{p}}$ we define the plastic yield function as
\begin{equation}\label{eq:yield}
\Phi^{\mathrm{p}}\left(\bar{\sigma}(\vec{\sigma}_{\mathrm{mat}},f),r^{\mathrm{p}}\right)=\bar{\sigma}-r^{\mathrm{p}}  .
\end{equation}
Focusing on void growth and thereby neglecting other influences such as void nucleation or void softening due to shear, the evolution form of the void growth reads \(\dot{f}=(1-f)\mathrm{tr}[\vec{d}^{\mathrm{p}}]\). Following \cite{miehe2016b}, the current void volume fraction is given in terms of the plastic deformation as
\begin{equation}
f=1-\frac{1-f_{0}}{J^{\mathrm{p}}}.
\end{equation} 
A plastic Lagrange multiplier \(\lambda^{\mathrm{p}}\) is introduced to enforce the Karush-Kuhn-Tucker conditions 
\begin{equation}\label{eq:kktp}
\lambda^{\mathrm{p}}\geq 0,\quad\Phi^{\mathrm{p}}\leq 0,\quad\lambda^{\mathrm{p}}\Phi^{\mathrm{p}} = 0.
\end{equation}
For the incorporation of the fracture mechanical behavior, we define crack threshold functions as 
\begin{equation}
\Phi^{\mathrm{f}}_{\bullet}(\mathcal{H}_{\bullet}-r^{\mathrm{f}}_{\bullet})=\mathcal{H}_{\bullet}-r^{\mathrm{f}}_{\bullet}
\end{equation}
where the energetic driving forces \(\mathcal{H}_{\bullet}\) are bounded by crack resistance forces \(r^{\mathrm{f}}_{\bullet}\) dual to the crack phase-field variables \(\s_{\bullet}\). Similar to plasticity, we introduce fracture Lagrange multipliers \(\lambda^{\mathrm{f}}_{\bullet}\) to enforce the Karush-Kuhn-Tucker conditions of the respective crack phase-field
\begin{equation}\label{eq:kktf}
\lambda^{\mathrm{f}}_{\bullet}\geq 0,\quad\Phi^{\mathrm{f}}_{\bullet}\leq 0,\quad\lambda^{\mathrm{f}}_{\bullet}\Phi^{\mathrm{f}}_{\bullet} = 0.
\end{equation}
Based on the concept of maximum dissipation and the set \(\mathfrak{C}=[\vec{\sigma}_{\mathrm{mat}},r^{\mathrm{p}},\mathcal{H}-r^{\mathrm{f}},\mathcal{H}_{\mathrm{L}}-r^{\mathrm{f}}_{\mathrm{L}},\mathcal{H}_{\mathrm{M}}-r^{\mathrm{f}}_{\mathrm{M}},\lambda^{\mathrm{p}},\lambda^{\mathrm{f}},\lambda^{\mathrm{f}}_{\mathrm{L}},\lambda^{\mathrm{f}}_{\mathrm{M}}]\), we define an extended dissipation potential and obtain a constrained optimization problem as
\begin{equation}\label{disPot}
\begin{aligned}
V=\underbrace{\op{sup}}_{\mathfrak{C}}\, \big[&\vec{\sigma}_{\mathrm{mat}}:\vec{d}^{\mathrm{p}}-(1-f)r^{\mathrm{p}}\dot{\alpha}+(\mathcal{H}-r^{\mathrm{f}})\dot{\s}+(\mathcal{H}_{\mathrm{L}}-r^{\mathrm{f}}_{\mathrm{L}})\dot{\s}_{\mathrm{L}}
+(\mathcal{H}_{\mathrm{M}}-r^{\mathrm{f}}_{\mathrm{M}})\dot{\s}_{\mathrm{M}} \\
&-\lambda^{\mathrm{p}}\Phi^{\mathrm{p}}(\vec{\sigma}_{\mathrm{mat}},r^{\mathrm{p}})-\lambda^{\mathrm{f}}\Phi^{\mathrm{f}}(\mathcal{H}-r^{\mathrm{f}})
-\lambda^{\mathrm{f}}_{\mathrm{L}}\Phi^{\mathrm{f}}_{\mathrm{L}}(\mathcal{H}_{\mathrm{L}}-r^{\mathrm{f}}_{\mathrm{L}})-\lambda^{\mathrm{f}}_{\mathrm{M}}\Phi^{\mathrm{f}}_{\mathrm{M}}(\mathcal{H}_{\mathrm{M}}-r^{\mathrm{f}}_{\mathrm{M}})\big] 
\end{aligned}
\end{equation}
where the Lagrange multipliers \(\lambda^{\mathrm{p}}\) and \(\lambda^{\mathrm{f}}_{\bullet}\) control the non-smooth evolution of plasticity and fracture, respectively. Then, the associated plastic evolution equations follows as
\begin{equation}\label{eq:plasticRate}
\vec{d}^{\mathrm{p}}=\lambda^{\mathrm{p}}\frac{\partial\Phi^{\mathrm{p}}}{\partial\vec{\sigma}_{\mathrm{mat}}}\quad\text{and}\quad\dot{\alpha}=-\frac{\lambda^{\mathrm{p}}}{1-f}\frac{\partial{\Phi}^{\mathrm{p}}}{\partial r^{\mathrm{p}}}
\end{equation}
and the evolution equation of the respective crack phase-field as
\begin{equation}\label{eq:fractureRate}
\dot{\s}_{\bullet}=\lambda^{\mathrm{f}}_{\bullet}\frac{\partial\Phi^{\mathrm{f}}_{\bullet}}{\partial(\mathcal{H}_{\bullet}-r^{\mathrm{f}}_{\bullet})}.
\end{equation}
A penalty regularization of the Lagrange multipliers can be utilized as follows\footnote{The Macaulay brackets are defined by \(\langle x\rangle=(x+|x|)/2\).}
\begin{equation}
\lambda^{\mathrm{p}}=\frac{1}{\eta_{\mathrm{p}}} {\langle\Phi^{\mathrm{p}}  \rangle}^{\mathrm{n}_\mathrm{p}}\geq0,\quad\lambda^{\mathrm{f}}=\frac{1}{\eta_{\mathrm{f}}}\langle\Phi^{\mathrm{f}}\rangle\geq0
,\quad\lambda^{\mathrm{f}}_{\mathrm{L}}=\frac{1}{\eta_{\mathrm{f_L}}}\langle\Phi^{\mathrm{f}}_{\mathrm{L}}\rangle\geq0\quad\text{and}\quad\lambda^{\mathrm{f}}_{\mathrm{M}}=\frac{1}{\eta_{\mathrm{f_M}}}\langle\Phi^{\mathrm{f}}_{\mathrm{M}}\rangle\geq0,
\end{equation}
where \(\eta_{\mathrm{p}}, \mathrm{n}_\mathrm{p} \) and \(\eta_{\mathrm{f}_{\bullet}}\) are material parameters which characterize the viscosity of plastification and crack propagation. Note that in the sense of continuum setting as defined in (\ref{eq:storedEnergy}) and (\ref{eq:auxEnergy}), the rates obtained in (\ref{eq:plasticRate}) and (\ref{eq:fractureRate}) are weighted by the respective volume fraction \(\zeta\) and \((1-\zeta)/2\), respectively. 

\subsubsection{Heat conduction}
Regarding the heat transfer within the composite material, we introduce a relation for the Piola-Kirchhoff heat flux vector as
\begin{equation}\label{eq:heatFlux}
\vec{Q}(\vec{F},\theta,\s)=-\vec{K}(\vec{F},\s)\nabla\theta
\end{equation}
which is known as Duhamel's law of heat conduction. The thermal conductivity tensor is defined as
\begin{equation}\label{eq:condTensor}
\vec{K}=\left(K(1-\s)+K^{\mathrm{conv}}\s\right)\vec{F}^{-1}\vec{F}^{-\mathrm{T}}.
\end{equation}
In case of fracture, the conduction degenerates locally such that we achieve a pure convection problem and the heat transfer depends mainly on the crack opening width of the matrix material. Here, we formulate the conductivity tensor \(\vec{K}\) in terms of the phase-field parameter \(\s\). Moreover, \(K=\zeta K_{\mathrm{mat}}+(1-\zeta)K_{\mathrm{fib}}\) is a average conductivity parameter related to the composite material and \(K^{\mathrm{conv}}\) is a convection parameter.

\subsubsection{Localization}
To identify and collect the internal contributions to the boundaries of the mechanical field, i.e.\ the resulting bending moments and normal stress contributions, we derive the internal virtual work as
\begin{equation}
\delta W^{\mathrm{e,int}}=\int\limits_{\mathcal{B}_0}\delta\Psi^{\mathrm{e}}_{\mathrm{mat}}+\delta\Psi^{\mathrm{e}}_{\mathrm{fib}}\d V=\int\limits_{\mathcal{B}_0}\vec{\tau}:\nabla_x\delta\vec{\varphi}+\vec{\mathfrak{P}}:\hspace{-0.7mm}\cdot\,\nabla\nabla\delta\vec{\varphi}\d V.
\end{equation}
The usage of the transformation 
\begin{equation}
\vec{\tau}:\nabla_x\delta\vec{\varphi}=\vec{\tau}\vec{F}^{\mathrm{-T}}:\nabla\delta\vec{\varphi}
\end{equation}
along with integration by parts yields\footnote{In index notation the internal virtual work yields \begin{equation}
\delta W^{\mathrm{e,int}}=\int\limits_{\mathcal{B}_0} -(\tau_{ij}F^{\mathrm{-T}}_{jK})_{,K}\delta\varphi_{i}   + (\tau_{ij}F^{\mathrm{-T}}_{jK}\delta\varphi_{i})_{,K}   - \mathfrak{P}_{iJK,K}\delta\varphi_{i,J} +  (\mathfrak{P}_{iJK}\delta\varphi_{i,J})_{,K} \d V
\end{equation} }
\color{black}
\begin{equation}
\delta W^{\mathrm{e,int}}=\int\limits_{\mathcal{B}_0}-\nabla\cdot(\vec{\tau}\vec{F}^{\mathrm{-T}})\cdot\delta\vec{\varphi}+\nabla\cdot(\delta\vec{\varphi}\cdot\vec{\tau}\vec{F}^{\mathrm{-T}})-\nabla\cdot\vec{\mathfrak{P}}:\nabla\delta\varphi+\nabla\cdot(\nabla\delta\vec{\varphi}:\vec{\mathfrak{P}})\d V
\end{equation}
\color{black}
and a second integration by parts related to the third term yields
\color{black}
\begin{equation}
\delta W^{\mathrm{e,int}}=\int\limits_{\mathcal{B}_0}\nabla\cdot(\nabla\cdot\vec{\mathfrak{P}}-\vec{\tau}\vec{F}^{\mathrm{-T}})\cdot\delta\vec{\varphi}+\nabla\cdot(\delta\vec{\varphi}\cdot(\vec{\tau}\vec{F}^{\mathrm{-T}}-\nabla\cdot\vec{\mathfrak{P}}))
+\nabla\cdot(\nabla\delta\vec{\varphi}:\vec{\mathfrak{P}})\d V.
\end{equation}
\color{black}
In a last step, we apply the divergence theorem for the second and third term such that we obtain
\color{black}
\begin{equation}
\delta W^{\mathrm{e,int}}=\int\limits_{\mathcal{B}_0}\nabla\cdot(\nabla\cdot\vec{\mathfrak{P}}-\vec{\tau}\vec{F}^{\mathrm{-T}})\cdot\delta\vec{\varphi}\d V
+\int\limits_{\partial\mathcal{B}_0}((\vec{\tau}\vec{F}^{\mathrm{-T}}-\nabla\cdot\vec{\mathfrak{P}})\vec{N})\cdot\delta\vec{\varphi}+\vec{\mathfrak{P}}\vec{N}:\nabla\delta\vec{\varphi}\d A.
\end{equation}
\color{black}
Note that also contributions in tangential direction at the boundaries can be considered such that further integrations by parts incorporates the boundaries \(\partial^2\mathcal{B}_0\) and \(\partial^3\mathcal{B}_0\) which represent curves and points, see e.g.\ Schulte et al.\ \cite{hesch2019} and Javili et al.\ \cite{steinmann2013}. Assuming that the principle of virtual work \(\delta W^{\mathrm{e,int}}-\delta W^{\mathrm{e,ext}}=0\) is valid with respect to the corresponding functional spaces of admissible solution and test functions, the external contribution can be formulated as
\begin{equation}
\delta W^{\mathrm{e,ext}}=\int\limits_{\mathcal{B}_0}\vec{B}\cdot\delta\vec{\varphi}+\int\limits_{\Gamma_0^{T}}\bar{\vec{T}}\cdot\delta\vec{\varphi}\d A+\int\limits_{\Gamma_0^{M}}\bar{\vec{M}}:\nabla\delta\vec{\varphi}\d A,
\end{equation}
where \(\vec{B}\) is a given body force per unit volume of the reference configuration. Eventually, we obtain the local form of the mechanical problem as
\begin{equation}
\nabla\cdot(\vec{\tau}\vec{F}^{\mathrm{-T}}-\nabla\cdot\vec{\mathfrak{P}})+\vec{B}=\vec{0}
\end{equation}
supplemented by boundary conditions
\begin{equation}\label{eq:mechDir}
\begin{aligned}
\vec{\varphi}&=\bar{\vec{\varphi}}\quad \text{on}\quad\Gamma_0^{\varphi}\\
\nabla\vec{\varphi}\vec{N}&=\nabla\bar{\vec{\varphi}}\vec{N}\quad \text{on}\quad\Gamma_0^{\nabla\varphi}\\
(\vec{\tau}\vec{F}^{\mathrm{-T}}-\nabla\cdot\vec{\mathfrak{P}})\vec{N}&=\bar{\vec{T}}\quad \text{on}\quad\Gamma_0^{T}\\
\vec{\mathfrak{P}}\vec{N}&=\bar{\vec{M}}\quad \text{on}\quad\Gamma_0^{M}
\end{aligned}
\end{equation}
with prescribed fields \(\bar{\vec{\varphi}}\) and \(\nabla\bar{\vec{\varphi}}\vec{N}\) at the mechanical Dirichlet boundaries \(\Gamma_0^{\varphi}\) and \(\Gamma_0^{\nabla\varphi}\). As usual for fourth-order boundary value problems, the entire boundary is decomposed twice, i.e.\ \(\Gamma_0=\Gamma_0^{\varphi}\cup\Gamma_0^{T}\) with \(\Gamma_0^{\varphi}\cap\Gamma_0^{T}=\emptyset\) and \(\Gamma_0=\Gamma_0^{\nabla\varphi}\cup\Gamma_0^{M}\) with \(\Gamma_0^{\nabla\varphi}\cap\Gamma_0^{M}=\emptyset\). For details related to the enforcement of the gradient condition given in \eqref{eq:mechDir}\(_2\) see e.g.\ Schu{\ss} et al.\ \cite{schuss2018}.

\subsubsection{Coupled problem}
Based on the derivations concerning the mechanical field within the previous section, the set of admissible test functions related to \(\mathfrak{U}\) is given as 
\begin{equation}
\delta\mathfrak{U}=[\delta\vec{\varphi},\delta\theta,\delta\alpha,\delta r^{\mathrm{p}},\delta\s,\delta\s_{\mathrm{L}},\delta\s_{\mathrm{M}}],
\end{equation}
i.e.\ variations of the deformation, the absolute temperature, the equivalent plastic strain, the dual plastic resistance force, the crack phase-field of the matrix material and the variables of the dual crack phase-field of the fiber material, where their spaces are defined as 
\begin{equation}
\begin{aligned}
\mathcal{V}^{\varphi}&=\{\delta\vec{\varphi}\in H^2(\mathcal{B}_0)\,|\,\delta\vec{\varphi}=\vec{0}\,\text{on}\,\Gamma_0^{\varphi},\,\nabla\delta\vec{\varphi}\vec{N}=\vec{0}\,\text{on}\,\Gamma_0^{\nabla\varphi}\},\\
\mathcal{V}^{\theta} &= \{\delta\theta\;\in H^1(\mathcal{B}_0)\,|\,\delta\theta\;=0\;\text{on}\,\Gamma_0^{\theta}\},\\
\mathcal{V}^{\alpha} &= \{\delta\alpha\;\in H^1(\mathcal{B}_0)\,|\,\delta\alpha\;=0\;\text{on}\,\Gamma_0^{\alpha}\},\\
\mathcal{V}^{r^{\mathrm{p}}} &= \{\delta r^{\mathrm{p}}\;\in \mathcal{L}^2(\mathcal{B}_0)\},\\
\mathcal{V}^{\s} &= \{\delta\s\;\in H^1(\mathcal{B}_0)\,|\,\delta\s\;=0\;\text{on}\,\bar{\Gamma}\},\\
\mathcal{V}^{\s_{\mathrm{L}}} &= \{\delta\s_{\mathrm{L}}\;\in H^1(\mathcal{B}_0)\,|\,\delta\s_{\mathrm{L}}\;=0\;\text{on}\,\bar{\Gamma}_{\mathrm{L}}\},\\
\mathcal{V}^{\s_{\mathrm{M}}} &= \{\delta\s_{\mathrm{M}}\;\in H^1(\mathcal{B}_0)\,|\,\delta\s_{\mathrm{M}}\;=0\;\text{on}\,\bar{\Gamma}_{\mathrm{M}}\}
\end{aligned}
\end{equation}
included within the Sobolev functional space of square integrable functions and derivatives \(H^k\) with \(k\geq0\). Then, the weak form of the coupled multifield problem reads
\begin{equation}\label{eq:weak}
\begin{aligned}
&\int\limits_{\mathcal{B}_0}\nabla_{x}\delta\vec{\varphi}:\vec{\tau}+\nabla\nabla\delta\vec{\varphi}:\hspace{-0.7mm}\cdot\,\vec{\mathfrak{P}}_{\mathrm{fib}}-\delta\vec{\varphi}\cdot\vec{B}\d V -\int\limits_{\Gamma_0^{T}}\delta\vec{\varphi}\cdot\bar{\vec{T}}\d A-\int\limits_{\Gamma_0^{M}}\nabla\delta\vec{\varphi}:\bar{\vec{M}}\d A=0,\\
&\int\limits_{\mathcal{B}_0}\delta\theta(\theta\dot{\eta}-\mathcal{D}_{\mathrm{int}}-\mathcal{R})-\nabla\delta\theta\cdot\vec{Q}\d V-\int\limits_{\Gamma_0^{Q}}\delta\theta \bar{Q}\d A=0,\\
&\int\limits_{\mathcal{B}_0}\left(\delta\alpha(\zeta y-r^{\mathrm{p}})+\zeta y_0l_{\mathrm{p}}^2\nabla\delta\alpha\cdot\nabla\alpha\right)\d V=0,\\
&\int\limits_{\mathcal{B}_0}\delta r^{\mathrm{p}}\left(\eta_{\mathrm{p}}\dot{\alpha}-\frac{\chi_{\mathrm{p}}(\Phi^{\mathrm{p}})^{\mathrm{n}_\mathrm{p}}}{1-f}\right)\d V=0,\\
&\int\limits_{\mathcal{B}_0}\delta\s\eta_{\mathrm{f}}\dot{\s}-\delta\s\chi_{\mathrm{f}}\left(\mathcal{H}-\frac{\zeta g_{\mathrm{c}}}{l_{\mathrm{f}}}\s\right)+\chi_{\mathrm{f}}\zeta g_{\mathrm{c}}\nabla\delta\s\cdot\nabla\s\d V=0,\\
&\int\limits_{\mathcal{B}_0}\delta\s_{\mathrm{L}}\eta_{\mathrm{f_L}}\dot{\s}_{\mathrm{L}}-\delta\s_{\mathrm{L}}\chi_{\mathrm{f_L}}\left(\mathcal{H}_{\mathrm{L}}-\frac{(1-\zeta)g_{\mathrm{c_L}}}{2l_{\mathrm{f_L}}}\s_{\mathrm{L}}\right)+\chi_{\mathrm{f_L}}\frac{1-\zeta}{2}g_{\mathrm{c_L}}\nabla\delta\s_{\mathrm{L}}\cdot\nabla\s_{\mathrm{L}}\d V=0,\\
&\int\limits_{\mathcal{B}_0}\delta\s_{\mathrm{M}}\eta_{\mathrm{f_M}}\dot{\s}_{\mathrm{M}}-\delta\s_{\mathrm{M}}\chi_{\mathrm{f_M}}\left(\mathcal{H}_{\mathrm{M}}-\frac{(1-\zeta)g_{\mathrm{c_M}}}{2l_{\mathrm{f_M}}}\s_{\mathrm{M}}\right)+\chi_{\mathrm{f_M}}\frac{1-\zeta}{2}g_{\mathrm{c_M}}\nabla\delta\s_{\mathrm{M}}\cdot\nabla\s_{\mathrm{M}}\d V=0,
\end{aligned}
\end{equation}
where \(\mathcal{R}\) is a given heat supply per unit volume of the reference configuration and \(\bar{Q}\) is a heat supply across the thermal Neumann boundary \(\Gamma_0^{Q}\). For each other field, homogeneous Neumann boundary conditions are applied and appropriate thermal Dirichlet boundary conditions are formulated in terms of prescribed temperature \(\bar{\theta}\), see Table \ref{tab:strong}. Note that we neglect inertia terms within the mechanical balance equation, i.e.\ we consider only quasi static problems. Additionally, internal conditions for the crack phase-field equations are given by
\begin{equation}
\s_{\bullet}=1\quad\text{on}\quad\bar{\Gamma}_{\bullet}\subset\widehat{\Gamma}_{\bullet},
\end{equation}
ensuring that a fully broken state remains broken. The Karush-Kuhn Tucker conditions in (\ref{eq:kktp}) and (\ref{eq:kktf}), are evaluated by inserting local variables defined as
\begin{equation}
\chi_{\mathrm{p}}= :
\left\{
\begin{array}{ll}
1\quad\text{for}\quad\Phi^{\mathrm{p}}>0\\
0\quad\text{otherwise}
\end{array}
\right.\quad\text{and}\quad
\chi_{\mathrm{f_{\bullet}}}= :
\left\{
\begin{array}{ll}
1\quad\text{for}\quad\Phi_{\bullet}^{\mathrm{f}}>0\\
0\quad\text{otherwise}
\end{array}
\right. .
\end{equation}
Using local variables \(\chi_{\mathrm{f_{\bullet}}}\) in (\ref{eq:weak}), we demand \(\dot{\s}_{\bullet}\geq0\) for thermodynamical consistency, avoiding a transfer of dissipated energy back into the mechanical field. This prevents healing effects, which may be taken into account as well. We can also set \(\chi_{\mathrm{f}_{\bullet}}\equiv1\) and restrict only the fully broken state, i.e.\ we allow for healing until the respective crack phase-field reaches the value one.
\begin{table}
\footnotesize
\begin{tabular}{@{}p{1\textwidth}@{}}
\hline
\begin{compactenum}
\item Stress equilibrium
\vspace{-1mm}
\begin{equation}
\nabla\cdot(\vec{\tau}\vec{F}^{\mathrm{-T}}-\nabla\cdot\vec{\mathfrak{P}})+\vec{B}=\vec{0}
\end{equation}
\item Kirchhoff stress
\vspace{-3mm}
\begin{equation}
\vec{\tau}=\vec{\tau}_{\mathrm{mat}}+\vec{\tau}_{\mathrm{fib}},\quad
\vec{\tau}_{\mathrm{mat}}=\zeta\sum\limits_a\lambda_a^{\mathrm{e}}\frac{\partial\Psi^{\mathrm{e}}_{\mathrm{mat}}}{\partial\lambda^{\mathrm{e}}_a}\vec{n}_a\otimes\vec{n}_a,\quad
\vec{\tau}_{\mathrm{fib}}=\frac{1-\zeta}{2}\frac{\partial\Psi^{\mathrm{e}}_{\mathrm{fib}}}{\partial\vec{F}}\vec{F}^{\mathrm{T}}
\end{equation}
\item Higher-order stress
\vspace{-2mm}
\begin{equation}
\vec{\mathfrak{P}}=\frac{1-\zeta}{2}\frac{\partial\Psi^{\mathrm{e}}_{\mathrm{fib}}}{\partial\nabla\vec{F}} 
\end{equation}
\item Energy balance  
\vspace{-1mm}
\begin{equation}
\theta\dot{\eta}+\nabla\cdot\vec{Q}-\mathcal{D}_{\mathrm{int}}-\mathcal{R}=0
\end{equation}
\item Entropy
\vspace{-3mm}
\begin{equation}
\eta=\eta_{\mathrm{mat}}+\eta_{\mathrm{fib}},\quad
\eta_{\mathrm{mat}}=-\zeta\frac{\partial\left(\Psi_{\mathrm{mat}}^{\mathrm{e}}+\Psi_{\mathrm{mat}}^{\theta}\right)}{\partial\theta},\quad
\eta_{\mathrm{fib}}=-\frac{1-\zeta}{2}\frac{\partial\left(\Psi_{\mathrm{fib}}^{\mathrm{e}}+\Psi_{\mathrm{fib}}^{\theta}\right)}{\partial\theta}\notag
\end{equation}
\item Dissipation
\vspace{-1mm}
\begin{equation}
\mathcal{D}_{\mathrm{int}}=\nu_{\mathrm{p_{mat}}}\vec{\tau}_{\mathrm{mat}}:\vec{d}^{\mathrm{p}}+\nu_{\mathrm{f_{mat}}}\mathcal{H}\dot{\s}+\nu_{\mathrm{f_{fib}}}(\mathcal{H}_{\mathrm{L}}\dot{\s}_{\mathrm{L}}+\mathcal{H}_{\mathrm{M}}\dot{\s}_{\mathrm{M}})
\end{equation}
\item Piola-Kirchhoff heat flux
\vspace{-2mm}
\begin{equation}
\vec{Q}=-\vec{K}\nabla\theta,\quad\vec{K}=\left(K(1-\s)+K^{\mathrm{conv}}\s\right)\vec{F}^{-1}\vec{F}^{-\mathrm{T}}
\end{equation}
\item Plastic strain
\vspace{-2mm}
\begin{equation}
\vec{d}^{\mathrm{p}}-\lambda^{\mathrm{p}}\frac{\partial\Phi^{\mathrm{p}}}{\partial\vec{\sigma}_{\mathrm{mat}}}=\vec{0},\quad\lambda^{\mathrm{p}}=\frac{1}{\eta_{\mathrm{p}}}{\langle\Phi^{\mathrm{p}}\rangle}^{\mathrm{n}_\mathrm{p}},\quad\vec{\sigma}_{\mathrm{mat}}=\vec{\tau}_{\mathrm{mat}}/J
\end{equation}
\item Equivalent plastic strain
\vspace{-2mm}
\begin{equation}
-\dot{\alpha}-\frac{\lambda^{\mathrm{p}}}{1-f}\frac{\partial\Phi^{\mathrm{p}}}{\partial r^{\mathrm{p}}}=0
\end{equation}
\item Plastic resistance force
\vspace{-1mm}
\begin{equation}
r^{\mathrm{p}}=\zeta\delta_{\alpha}\widehat{\Psi}^{\mathrm{p}}_{\mathrm{mat}}
\end{equation}
\item Crack phase-field equations
\vspace{-3mm}
\begin{equation}
\dot{\s}_{\bullet}-\lambda_{\bullet}^{\mathrm{f}}\frac{\partial\Phi_{\bullet}^{\mathrm{f}}}{\partial(\mathcal{H}_{\bullet}-r^{\mathrm{f}}_{\bullet})}=0,\quad\lambda_{\bullet}^{\mathrm{f}}=\frac{1}{\eta_{\mathrm{f_{\bullet}}}}\langle\Phi^{\mathrm{f}}_{\bullet}\rangle
\end{equation}
\item Crack phase-field driving forces
\vspace{-2mm}
\begin{equation}
\mathcal{H}_{\bullet}=-\frac{\partial\Psi^{\mathrm{e}}}{\partial\s_{\bullet}}
\end{equation}
\item Fracture resistance forces
\vspace{-1mm}
\begin{equation}
r^{\mathrm{f}}_{\bullet}=\delta_{\s_{\bullet}}\widehat{\Psi}
\end{equation}
\item Dirichlet and Neumann conditions
\vspace{-2mm}
\textcolor{black}{
\begin{equation}
\begin{aligned}
\vec{\varphi} &= \bar{\vec{\varphi}}(\vec{X},t)\;\text{on}\ \Gamma_{0}^{\varphi},\qquad &&(\vec{\tau}\vec{F}^{\mathrm{-T}}-\nabla\cdot\vec{\mathfrak{P}})\vec{N} = \bar{\vec{T}}(\vec{X},t)\ \textrm{on}\ \Gamma^{T}_0\\
\nabla\vec{\varphi}\vec{N}&=\nabla\bar{\vec{\varphi}}(\vec{X},t)\vec{N}\; \text{on}\ \Gamma_0^{\nabla\varphi} ,\qquad && \vec{\mathfrak{P}}\vec{N}=\bar{\vec{M}}(\vec{X},t)\ \textrm{on}\ \Gamma^{M}_0\\
\theta&=\bar{\theta}(\vec{X},t)\;\text{on}\ \Gamma_{0}^{\theta},\qquad &&-\vec{Q}\cdot\vec{N}=\bar{Q}\ \textrm{on}\ \Gamma^{Q}_0\\
\alpha&=0\;\text{on}\ \Gamma^{\alpha}_0,\qquad &&\nabla\alpha\cdot\vec{N}=0\ \textrm{on}\ \Gamma_{0}^{\nabla\alpha}\\
\s_{\bullet}&=1\;\text{on}\;\bar{\Gamma}_{\bullet},\qquad &&\nabla\s_{\bullet}\cdot\vec{N}=0\;\text{on}\;\Gamma_{0}
\end{aligned}
\end{equation}}
\vspace{-2mm}
\item Initial conditions
\vspace{-2mm}
\begin{equation}
\vec{\varphi}(\vec{X},0)=\vec{\varphi}_0,\,\,\, \dot{\vec{\varphi}}(\vec{X},0)=\vec{v}_0,\,\,\, \theta(\vec{X},0)=\theta_0,\,\,\, \alpha(\vec{X},0)=0,\,\,\, r^p(\vec{X},0)=0,\,\,\, \s_{\bullet}(\vec{X},0)=0
\end{equation}
\end{compactenum}\\
\hline
\end{tabular}
\vspace{-3mm}
\caption{Strong formulation of the coupled problem} 
\label{tab:strong}
\end{table}

%% file: chapters/spatial.tex
\section{Isogeometric discretization}\label{sec:spatial}
Concerning the spatial discretization, the domain \(\mathcal{B}_0\) is subdivided into a finite set of non-overlapping elements \(e\in\mathbb{N}\) such that 
\begin{equation}
\mathcal{B}_0 \approx \mathcal{B}_0^{\mathrm{h}} = \bigcup\limits_{e\in\mathbb{N}}\mathcal{B}_{e}. 
\end{equation}
Due to the incorporation of curvature contributions into the fiber material, the mechanical part of the variational problem requires approximation functions which are globally at least \(C^1\)-continuous to satisfy \(\vec{\varphi}^{\mathrm{h}},\delta\vec{\varphi}^{\mathrm{h}}\in\mathcal{H}^2(\mathcal{B}_0^{\mathrm{h}})\). To meet this continuity requirement an isogeometric analysis approach which employs Non-Uniform Rational B-splines (NURBS) of order \(p_{\alpha}\geq 2\) can be applied. Accordingly, rational approximations of the deformed geometry \(\vec{\varphi}\) and its variation \(\delta\vec{\varphi}\) are defined as 
\begin{equation}\label{eq:approxGeo}
\vec{\varphi}^{\mathrm{h}}=\sum\limits_{A\in\mathcal{I}}R^{A}\vec{q}_{A} \quad\text{and}\quad\delta \vec{\varphi}^{\mathrm{h}} =\sum\limits_{A\in\mathcal{I}}R^{A}\delta\vec{q}_{A},  
\end{equation}
respectively, where \(\vec{q}_{A}\in\mathbb{R}^3\) and \(\delta\vec{q}_{A}\in\mathbb{R}^3\). Moreover, the approximations of the crack phase-fields \(\s_{\bullet}\) and the temperature field \(\theta\) as well as
their variations \(\delta \s_{\bullet}\) and \(\delta\theta\) read
\begin{equation}\label{eq:approxPF}
\s_{\bullet}^{\mathrm{h}}=\sum\limits_{A\in\mathcal{I}}R^{A}\s_{\bullet,A} \quad\text{and}\quad\delta\s_{\bullet}^{\mathrm{h}} =\sum\limits_{A\in\mathcal{I}}R^{A}\delta\s_{\bullet,A}
\end{equation}
and
\begin{equation}\label{eq:approxTemp}
\theta^{\mathrm{h}}=\sum\limits_{A\in\mathcal{I}}R^{A}\theta_{A} \quad\text{and}\quad\delta \theta^{\mathrm{h}} =\sum\limits_{A\in\mathcal{I}}R^{A}\delta\theta_{A},
\end{equation}
where \(\s_{\bullet,A},\delta\s_{\bullet,A},\theta_{A},\delta\theta_{A}\in\mathbb{R}\). Introducing global shape functions \(R^A:\mathcal{B}_0^{\mathrm{h}}\rightarrow\mathbb{R}\) associated with control points \(A\in\mathcal{I}=\{1,\,\hdots,\,\mathfrak{N}\}\), NURBS based shape functions read
\begin{equation}\label{eq:nurbs}
R^{A} := R^{\vec{i}}(\vec{\xi}) = \frac{\prod\limits_{\alpha=1}^{3}B^{i_{\alpha}}(\xi^{\alpha})w_{\vec{i}}}{\sum\limits_{\vec{j}}\prod\limits_{\alpha=1}^{3}B^{j_{\alpha}}(\xi^{\alpha})w_{\vec{j}}},
\end{equation}
where \(B^{i_{\alpha}}\) are univariate non-rational B-splines defined on a parametric domain which is subdivided by the knot vector \([\xi_{1}^{\alpha},\xi_{2}^{\alpha},\hdots,\xi_{\mathfrak{N}_{\alpha}+p_{\alpha}+1}^{\alpha}]\), \(\mathfrak{N}=\mathfrak{N}_1\mathfrak{N}_2\mathfrak{N}_3\). The recursive definition of a single univariate B-spline is given as follows
\begin{equation}
B^{i_{\alpha}}_{p_{\alpha}}(\xi^{\alpha})=\frac{\xi^{\alpha}-\xi^{\alpha}_{i_{\alpha}}}{\xi^{\alpha}_{i_{\alpha}+p_{\alpha}}-\xi^{\alpha}_{i_{\alpha}}}B^{i_{\alpha}}_{p_{\alpha}-1}(\xi^{\alpha})+\frac{\xi^{\alpha}_{i_{\alpha}+p_{\alpha}+1}-\xi^{\alpha}}{\xi^{\alpha}_{i_{\alpha}+p_{\alpha}+1}-\xi^{\alpha}_{i_{\alpha}+1}}B^{i_{\alpha}+1}_{p_{\alpha}-1}(\xi^{\alpha}),
\end{equation}
beginning with
\begin{equation}
B_{0}^{i_{\alpha}}(\xi^{\alpha})=\left\{\begin{array}{l}1\quad\text{if}\;\xi_{i_{\alpha}}\leq\xi^{\alpha}<\xi^{\alpha}_{i_{\alpha}+1} \\ 0\quad\text{otherwise}\end{array}\right. . 
\end{equation}
Moreover, \(w_{\vec{j}}\) are corresponding NURBS weights. For further details on the construction of NURBS based shape functions as well as the construction of local refinements related to the IGA concept, see e.g.\ Cottrell et al.\ \cite{cottrell2009}, Bornemann and Cirak \cite{bornemann2013}, Hesch et al.\ \cite{hesch2016b} and Dittmann \cite{dittmann2017b}. 

Next, the hardening variable \(\alpha\) and its variation \(\delta\alpha\) are approximated as
\begin{equation}\label{eq:approxHard}
\alpha^{\mathrm{h}}=\sum\limits_{i\in\mathcal{J}}N^{i}\alpha_{i},\quad\delta\alpha^{\mathrm{h}}=\sum\limits_{i\in\mathcal{J}}N^{i}\delta \alpha_{i} 
\end{equation}
and the dual driving force \(r^{\mathrm{p}}\) to the hardening variable and its variation \(\delta r^{\mathrm{p}}\) are approximated as
\begin{equation}\label{eq:approxrp}
r^{\mathrm{p,h}}=\sum\limits_{i\in\mathcal{J}}N^{i}r^{\mathrm{p}}_{i},\quad\delta r^{\mathrm{p,h}}= \sum\limits_{i\in\mathcal{J}}N^{i}\delta r^{\mathrm{p}}_{i} ,
\end{equation}
where we make use of linear shape functions \(N^{i}\) defined on the physical mesh representation of the NURBS geometry with nodes \(i\in\mathcal{J}=\{1,\,\hdots,\,\mathfrak{n}\}\) and the corresponding number of nodes \(\mathfrak{n}\). 

\textit{Remark}: The more natural choice using the same NURBS shape functions for the approximation of the hardening variable \(\alpha\) and the dual driving force \(r^{\mathrm{p}}\) leads to oscillations within both fields, indicating stability issues. The above described scheme using quadratic shape function \(R^{A}\) and linear shape functions \(N^{i}\) has shown to be stable and numerically robust within our numerical examples, cf.\ Dittmann et al.\ \cite{dittmann2018}.

Inserting (\ref{eq:approxGeo})-(\ref{eq:approxTemp}) along with (\ref{eq:approxHard}) and (\ref{eq:approxrp}) into (\ref{eq:weak}) yields the semi-discrete set of coupled equations
\begin{equation}\label{eq:semiDiscr}
\begin{aligned}
\delta\vec{q}_{A}\cdot &\left[\int\limits_{\mathcal{B}_0}\vec{\tau}^{\mathrm{h}}\nabla_x R^{A}+\mathfrak{P}^{\mathrm{h}}:\nabla\nabla R^{A}\,\d V-\vec{F}^{\mathrm{ext},A}\right]=0,\\
\delta\theta_{A} &\left[\int\limits_{\mathcal{B}_0}\left(\dot{\eta}^{\mathrm{h}} R^{A} R^{B}\theta_{B}- R^A \mathcal{D}_{\mathrm{int}}^{\mathrm{h}}-\nabla R^A \vec{Q}^{\mathrm{h}}\right)\d V-Q^{\mathrm{ext},A} \right]=0, \\
\delta\alpha_{i} &\left[\int\limits_{\mathcal{B}_0}N^{i}(\zeta y^{\mathrm{h}}-N^{j}r^{\mathrm{p}}_{j})\,\d V+ K^{ij}_{\alpha}\alpha_{j}\right]=0,\\
\delta r^{\mathrm{p}}_{i} &\left[M^{ij}_{r^{\mathrm{p}}}\dot{\alpha}_{j}-\int\limits_{\mathcal{B}_0} \chi_{\mathrm{p}} N^{i}\frac{(\Phi^{\mathrm{p,h}})^{n_{\mathrm{p}}}}{J^{\mathrm{h}} (1-f^{\mathrm{h}})}\,\d V\right]=0,\\
\delta\s_{\bullet,A} &\left[M^{AB}_{\s_{\bullet}}\dot{\s}_{\bullet,B}-\int\limits_{\mathcal{B}_0}R^{A}\mathcal{H}_{\bullet}^{\mathrm{h}}\,\d V+K^{AB}_{\s_{\bullet}}\s_{\bullet,B}\right]=0.
\end{aligned}
\end{equation}
Therein, \(\vec{\tau}^{\mathrm{h}}\), \(\mathfrak{P}^{\mathrm{h}}\), \(\mathcal{\eta}^{\mathrm{h}}\) and \(\mathcal{H}_{\bullet}^{\mathrm{h}}\) are semi-discrete versions of the Kirchhoff stress tensor, the higher-order stress tensor, the local entropy and the phase-field driving forces obtained via the partial derivatives of the semi-discrete stored energy density 
\begin{equation}
\Psi^{\mathrm{h}}=\Psi^{\mathrm{h}}(\tilde{\bar{\vec{F}}}^{\mathrm{e,h}},\tilde{J}^{\mathrm{e,h}},\theta^{\mathrm{h}},\tilde{\lambda}_{\mathrm{L}}^{\mathrm{h}},\tilde{\lambda}_{\mathrm{M}}^{\mathrm{h}},\tilde{\phi}^{\mathrm{h}},\tilde{\vec{\kappa}}_{\mathrm{L}}^{\mathrm{h}},\tilde{\vec{\kappa}}_{\mathrm{M}}^{\mathrm{h}}),
\end{equation}
cf.\ (\ref{eq:storedEnergy})-(\ref{eq:entropy}). \(\mathcal{D}_{\mathrm{int}}^{\mathrm{h}}\) and \(\vec{Q}^{\mathrm{h}}\) are semi-discrete definitions of the dissipation density and heat flux, cf.\ (\ref{eq:dissipation}), (\ref{eq:heatFlux}) and (\ref{eq:condTensor}). Moreover, the semi-discrete external contributions in (\ref{eq:semiDiscr})\(_1\) and (\ref{eq:semiDiscr})\(_2\) are formulated as
\begin{equation}
\vec{F}^{\mathrm{ext},A} = \int\limits_{\mathcal{B}_0}R^{A}\vec{B}\,\d V+\int\limits_{\Gamma_0^{T}}R^{A}\bar{\vec{T}}\,\d A+\int\limits_{\Gamma_0^{M}}\bar{\vec{M}}\nabla R^{A}\,\d A
\end{equation}
and
\begin{equation}
Q^{\mathrm{ext},A} = \int\limits_{\mathcal{B}_0}R^{A}\mathcal{R}\,\d V+\int\limits_{\partial\mathcal{B}_0^{\theta_{\mathrm{n}}}}R^{A}\bar{Q}\,\d A.
\end{equation}
The coefficients of the matrices in (\ref{eq:semiDiscr})\(_3\) and (\ref{eq:semiDiscr})\(_4\) take the form 
\begin{equation}\label{eq:semiDiscr4}
K_{\alpha}^{ij}=\zeta y_{0}l_{\mathrm{p}}^{2}\int\limits_{\mathcal{B}_0}\nabla N^{i}\cdot\nabla N^{j}\d V\quad\text{and}\quad
M_{r^{\mathrm{p}}}^{ij}=\eta_{\mathrm{p}}\int\limits_{\mathcal{B}_0}N^{i}N^{j}\d V,
\end{equation}
whereas the matrices in (\ref{eq:semiDiscr})\(_5\) are given by
\begin{equation}
\begin{aligned}
M_{\s_{\bullet}}^{AB}&=\eta_{\mathrm{f_{\bullet}}}\int\limits_{\mathcal{B}_0}R^{A}R^{B}\d V,\\
K_{\s}^{AB}&=\frac{\zeta}{l_{\mathrm{f}}}\int\limits_{\mathcal{B}_0}g_{\mathrm{c}}^{\mathrm{h}}\chi_{\mathrm{f}}\left(R^{A}R^{B}+l_{\mathrm{f}}^{2}\nabla R^{A}\cdot\nabla R^{B}\right)\d V,\\
K_{\s_{\mathrm{L}}}^{AB}&=\frac{1-\zeta}{2l_{\mathrm{f_{\mathrm{L}}}}}\int\limits_{\mathcal{B}_0}g_{\mathrm{c}_{\mathrm{L}}}^{\mathrm{h}}\chi_{\mathrm{f}_{\mathrm{L}}}\left(R^{A}R^{B}+l_{\mathrm{f}_{\mathrm{L}}}^{2}\nabla R^{A}\cdot\nabla R^{B}\right)\d V,\\
K_{\s_{\mathrm{M}}}^{AB}&=\frac{1-\zeta}{2l_{\mathrm{f_{\mathrm{M}}}}}\int\limits_{\mathcal{B}_0}g_{\mathrm{c}_{\mathrm{M}}}^{\mathrm{h}}\chi_{\mathrm{f}_{\mathrm{M}}}\left(R^{A}R^{B}+l_{\mathrm{f}_{\mathrm{M}}}^{2}\nabla R^{A}\cdot\nabla R^{B}\right)\d V.
\end{aligned}
\end{equation}
Eventually, the semi-discrete functions \(\widehat{y}^{\mathrm{h}}\), \(\Phi^{\mathrm{p,h}}\) and \(g_{\mathrm{c}}^{\mathrm{h}}\) denote the local hardening, the plastic yield and the critical fracture energy density, cf.\ (\ref{eq:hardeningSatFunction}), (\ref{eq:yield}) and (\ref{eq:critFractureEnergy}).

%% file: chapters/time.tex
\section{Temporal discretization}\label{sec:time}
In a final step, the semi-discrete coupled problem (\ref{eq:semiDiscr}) has to be discretized in time to obtain a set of non-linear algebraic equations to be solved via a Newton-Raphson method. Therefore, we subdivide the considered time interval \(\mathcal{T}\) into a sequence of times \(t_{0},\hdots,t_{n},t_{n+1},\hdots,T\), where \((\bullet)_{n}\) and \((\bullet)_{n+1}\) denote the value of a given physical quantity at time \(t_{n}\) and \(t_{n+1}\), respectively. Assume that the discrete set of state variables at \(t_{n}\) given by \(\{\vec{q}_{A,n},\theta_{A,n},\alpha_{i,n}, r^{\mathrm{p}}_{i,n},\s_{A,n},\s_{\mathrm{L},A,n},\s_{\mathrm{M},A,n}\}\) and the local plastic deformation variable \(\vec{F}_{n}^{\mathrm{p,h}}\) at time \(t_{n }\) are known and the time step size \(\Delta t = t_{n+1}-t_{n}\) is given. Then, the goal is to determine the corresponding fields at time \(t_{n+1}\) via the algorithmic approximation to the weak formulation (\ref{eq:semiDiscr}) defined as

\begin{equation}\label{eq:discr}
\begin{aligned}
\delta\vec{q}_{A}\cdot &\left[\int\limits_{\mathcal{B}_0}\vec{\tau}^{\mathrm{h}}_{n+1}(\nabla_x R^{A})_{n+1}+\mathfrak{P}^{\mathrm{h}}_{n+1}\nabla\nabla R^{A}\,\d V-\vec{F}^{\mathrm{ext},A}_{n+1}\right]=0,\\
\delta\theta_{A} &\left[\int\limits_{\mathcal{B}_0}\left(\frac{\eta^{\mathrm{h}}_{n+1}-\eta^{\mathrm{h}}_{n}}{\Delta t} R^{A} R^{B}\theta_{B,n+1}- R^A \mathcal{D}_{\mathrm{int},n+1}^{\mathrm{h}}-\nabla R^A \vec{Q}_{n+1}^{\mathrm{h}}\right)\d V-Q^{\mathrm{ext},A}_{n+1} \right]=0, \\
\delta\alpha_{i} &\left[\int\limits_{\mathcal{B}_0}N^{i}(\zeta y^{\mathrm{h}}_{n+1}-N^{j}r^{\mathrm{p}}_{j,n+1})\,\d V+ K^{ij}_{\alpha}\alpha_{j,n+1}\right]=0,\\
\delta r^{\mathrm{p}}_{i} &\left[M^{ij}_{r^{\mathrm{p}}}\frac{\alpha_{j,n+1}-\alpha_{j,n}}{\Delta t}-\int\limits_{\mathcal{B}_0} \chi_{\mathrm{p},n+1} N^{i}\frac{(\Phi^{\mathrm{p,h}}_{n+1})^{n_{\mathrm{p}}}}{J_{n+1}^{\mathrm{h}} (1-f_{n+1}^{\mathrm{h}})}\,\d V\right]=0,\\
\delta\s_{\bullet,A} &\left[M^{AB}_{\s_{\bullet}}\frac{\s_{\bullet,B,n+1}-\s_{\bullet,B,n}}{\Delta t}-\int\limits_{\mathcal{B}_0}R^{A}\mathcal{H}_{\bullet,n+1}^{\mathrm{h}}\,\d V+K^{AB}_{\s_{\bullet},n+1}\s_{\bullet,B,n+1}\right]=0.
\end{aligned}
\end{equation}
Therein, a full-discrete definition of the internal dissipation is given by
\begin{equation}
\begin{aligned}
\mathcal{D}^{\mathrm{h}}_{\mathrm{int},n+1}=\nu_{\mathrm{p_{mat}}}\vec{\tau}^{\mathrm{h}}_{\mathrm{mat},n+1}&:\vec{d}^{\mathrm{p,h}}_{n+1}+\nu_{\mathrm{f_{mat}}}\mathcal{H}_{n+1}\frac{\s^{\mathrm{h}}_{n+1}-\s^{\mathrm{h}}_{n}}{\Delta t}\\
&+\nu_{\mathrm{f_{fib}}}\left(\mathcal{H}_{\mathrm{L},n+1}\frac{\s^{\mathrm{h}}_{\mathrm{L},n+1}-\s^{\mathrm{h}}_{\mathrm{L},n}}{\Delta t}+\mathcal{H}_{\mathrm{M},n+1}\frac{\s^{\mathrm{h}}_{\mathrm{M},n+1}-\s^{\mathrm{h}}_{\mathrm{M},n}}{\Delta t}\right),
\end{aligned}
\end{equation}
Using small values for the plastic viscosity parameter \(\eta_{\mathrm{p}}\), we obtain 
\begin{equation}
\vec{\tau}^{\mathrm{h}}_{n+1}:\vec{d}^{\mathrm{p,h}}_{n+1}\approx J^{\mathrm{h}}_{n+1} (1-f^{\mathrm{h}}_{n+1})\, r^{\mathrm{p,h}}_{n+1}\frac{\alpha^{\mathrm{h}}_{n+1}-\alpha^{\mathrm{h}}_{n}}{\Delta t}
\end{equation}
such that the internal dissipation can be recast as
\begin{equation}
\begin{aligned}
\mathcal{D}^{\mathrm{h}}_{\mathrm{int},n+1} :=  \nu_{\mathrm{p_{mat}}}J^{\mathrm{h}}_{n+1}& (1-f^{\mathrm{h}}_{n+1})\, r^{\mathrm{p,h}}_{n+1}N^{a}\frac{\alpha_{a,n+1}-\alpha_{a,n}}{\Delta t} + \nu_{\mathrm{f_{mat}}}\, \mathcal{H}^{\mathrm{h}}_{n+1} R^A \frac{\s_{A,n+1}-\s_{A,n}}{\Delta t}\\
&+\nu_{\mathrm{f_{fib}}}\left(\mathcal{H}^{\mathrm{h}}_{\mathrm{L},n+1} R^A\frac{\s_{\mathrm{L},A,n+1}-\s_{\mathrm{L},A,n}}{\Delta t}+\mathcal{H}^{\mathrm{h}}_{\mathrm{M},n+1} R^A\frac{\s_{\mathrm{M},A,n+1}-\s_{\mathrm{M},A,n}}{\Delta t}\right)
\end{aligned}
\end{equation}
for practical reasons. 

To solve the above multifield problem, we apply a staggered solution scheme, i.e.\ the displacement field along with the plastic and hardening fields \(\{\vec{q}_{A,n+1}, \alpha_{i,n+1}, r^{\mathrm{p}}_{i,n+1}, \vec{F}_{n+1}^{\mathrm{p,h}}\}\), the crack phase-fields \(\s_{\bullet,A,n+1}\) and the temperature field \(\theta_{A,n+1}\) are solved successively. For the time integration of the plastic evolution equations, the construction of a return mapping algorithm is most crucial. Therefore, we define a trial state as\footnote{For the sake of readability, we neglect the labeling of the spatial approximation in the following.}
\begin{equation}
{\vec{F}}^{\mathrm{e}}_{\mathrm{tr}} = {\vec{F}}_{n+1} (\vec{F}^{\mathrm{p}}_n)^{-1}
\end{equation}
assuming that no further plastic deformation occurs within the time step. Based on this trial state, we evaluate the yield criteria (\ref{eq:yield}). If \(\Phi^{\mathrm{p}}_{\mathrm{tr}}\leq0\), then the process is purely elastic and the elastic trial state is the solution. If on the other hand \(\Phi^{\mathrm{p}}_{\mathrm{tr}}>0\), then the trial state is not admissible and a plastic correction is required. Therefore, we apply an exponential integration scheme regarding (\ref{eq:plasticRate}) which leads to
\begin{equation}
\lambda^{\mathrm{e}}_{a,n+1} = \lambda^{\mathrm{e}}_{a,\mathrm{tr}}  \mathrm{exp}\left[-\Delta t \lambda^{\mathrm{p}}_{n+1} n_{a,n+1}\right]\quad\text{with}\quad n_{a,n+1}=\frac{\partial\Phi^{\mathrm{p}}_{n+1}}{\partial \sigma_{\mathrm{mat},a,n+1}},
\end{equation}
where \(\sigma_{\mathrm{mat},a,n+1}=(\tau_{\mathrm{mat},a,n+1}^{\mathrm{dev}}+\tau_{\mathrm{mat},a,n+1}^{\mathrm{vol}})/J_{n+1}\). Note that in contrast to standard von Mises plasticity $\vec{n}_{n+1} \neq \vec{n}_{\mathrm{tr}}$ and \(\|\vec{n}_{n+1}\|\neq 1\), i.e.\ the plastic correction has to be performed by the Lagrange multiplier \(\lambda^{\mathrm{p}}_{n+1}\) as well as the components \(n_{a,n+1}\) which can be obtained by solving the non-linear relations
\begin{equation}
\begin{aligned}
\widehat\Phi^{\mathrm{p}}_{n+1} - \eta_{\mathrm{p}} \lambda^{\mathrm{p}}_{n+1} = 0 \quad \text{and} \quad  \frac{\partial\Phi^{\mathrm{p}}_{n+1}}{\partial\sigma_{\mathrm{mat},a,n+1}} - n_{a,n+1}= 0  
\end{aligned}
\end{equation}
via an internal Newton-Raphson iteration. In addition, the void volume fraction $f_{n+1}$ is locally calculated by
\begin{equation}
f_{n+1} = \mathrm{max} \big\{ f_{0}   ,1- (1-f_0) / J^{\mathrm{p}}_{n+1} \big\}.
\end{equation}
For further details on the return map algorithm see Dittmann et al.\ \cite{dittmann2019b}.

%% file: chapters/example.tex
\section{Numerical examples}\label{sec:example}
In this section we investigate the accuracy and performance of the proposed formulation for endless fiber reinforced polymers. We start with a verification of the higher-order contributions of the fiber material by the means of two simple bending tests. Subsequently, a series of tensile tests demonstrates the capability of the proposed hybrid phase-field model to investigate different failure mechanisms for a prototypical fiber reinforced composite depending on the fiber configuration. This study is completed by thermal investigations on the damage behavior of the model and its impact on final failure. Without loss of generality we apply a Neo-Hookian model for the matrix material within all examples, i.e.\ we set \(b=1\) and \(\alpha_1=2\) in (\ref{eq:ogden}).

\subsection{Bending Test }
\input{chapters/bending}

\subsection{Tension Test }
\input{chapters/tension}

%% file: chapters/bending.tex
This first examples is dedicated to the verification of the higher-order, bending contributions of the fiber material. In particular, we investigate the in-plane bending behavior using a benchmark from Schulte et al.\ \cite{hesch2019}, originally used for the verification of gradient shell formulations, as well as the out-of-plane bending behavior using a four point bending test. Therefore, we consider a purely elastic behavior of the material, i.e.\ we neglect thermoplastic effects and fracture. 

\subsubsection{In-plane bending test}\label{sec:inplane}
\begin{figure}[ht]
\begin{center}
\footnotesize
\psfrag{a}[c][c]{\(\vartheta\)}
\psfrag{m}[c][c]{}
\psfrag{A}[c][cl]{\(\vec{e}_2\)}
\psfrag{B}[c][c]{\(\vec{e}_1\)}
\psfrag{C}[t][c]{\(\vec{e}_3\)}
\includegraphics[width=0.85\textwidth]{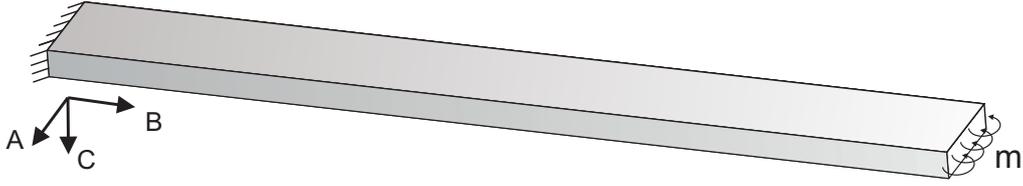}
\end{center}
\caption{\textbf{In-plane bending test.} Problem setting. The lines illustrate the fiber structure.}
\label{fig:bendingInRef}
\end{figure}

\textcolor{black}{We consider a benchmark example from \cite{hesch2019},  where the left edge of a Kirchhoff-Love shell  is clamped while the right edge is subjected to an external in-plane torque,  chosen to match a reference analytical solution.    To verify the proposed formulation in terms of in-plane bending stiffness parameterization,  we take the shell deformation result obtained in \cite{hesch2019}, extrude it to the corresponding 3D geometry and calculate the energy. The plate is of size \(L \times W \times H =10\,\mathrm{mm} \times 1\,\mathrm{mm} \times 0.5\,\mathrm{mm}\) and is discretized by \(8 \times 2\times1\) quadratic B-spline based elements,   see Figure \ref{fig:bendingInRef}}.  Furthermore, we assume that the plate consists of a single fiber bundle with a cross section of  \(A=HW=0.5\,\mathrm{mm^{2}}\) and a tensile stiffness of \(E_{\mathrm{fib}}=79000\,\mathrm{N}/\mathrm{mm}^2\). The area moments of inertia of the fiber bundle with respect to the \(\vec{e}_3\)-axis and the \(\vec{e}_2\)-axis are given by \(I_{\vec{e}_3}=HW^{3}/12=0.0417\,\mathrm{mm^{4}}\) and \(I_{\vec{e}_2}=WH^{3}/12=0.0104 \,\mathrm{mm^{4}}\), respectively. Using these quantities we calibrate the bending stiffness parameters as \(c_{\#}=E_{\mathrm{fib}}I_{\vec{e}_3}/A=6583.3333\,\mathrm{N}\) and \textcolor{black}{ \(c_{\perp}=E_{\mathrm{fib}}I_{\vec{e}_2}/A=2212\,\mathrm{N}\)}.

\begin{figure}[h]
\begin{center}
\scriptsize
\psfrag{0}[c][c]{\(0.080\)}
\psfrag{2}[c][c]{\(0.081\)}
\psfrag{1}[c][c]{[$\mathrm{J/mm^3}$]}
\includegraphics[width=0.49\textwidth]{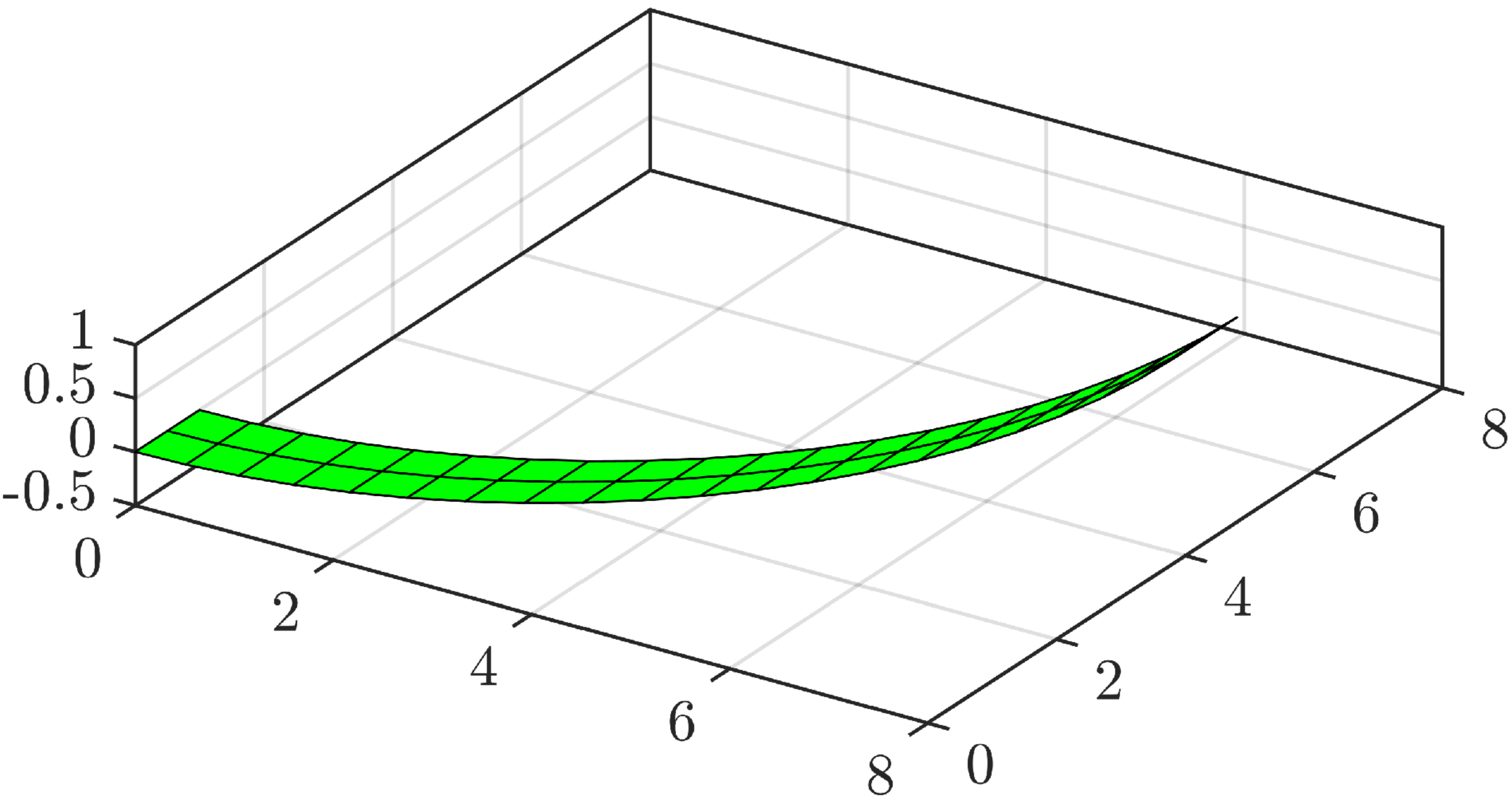}
\includegraphics[width=0.49\textwidth]{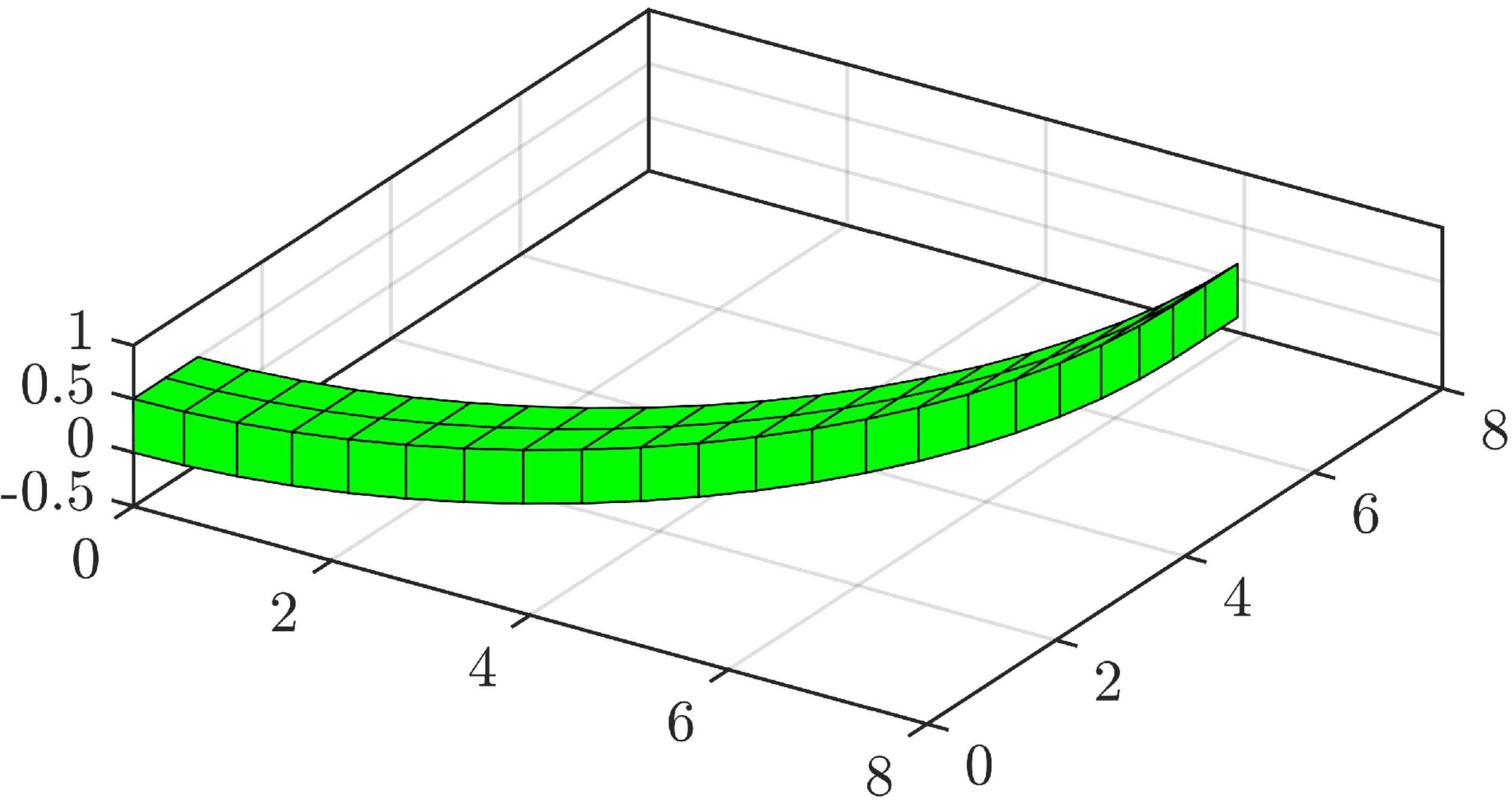}
 \\
\vspace{2mm}
\includegraphics[width=0.38\textwidth]{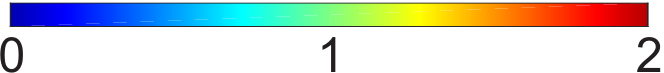}
\\
\end{center}
\caption{\textbf{In-plane bending test.} Strain energy distribution of the Kirchhoff-Love shell formulation (left) and higher-gradient continuum formulation (right).}
\label{fig:contishell}
\end{figure}

In Figure \ref{fig:contishell}, the strain energy density is depicted for both the Kirchhoff-Love shell formulation as well as the proposed higher-order continuum formulation. Therein, we can observe the same homogeneous distributions which verifies the calibration of the \textcolor{black}{in-plane} bending stiffness parameter \(c_{\#}\). Note that the parameter $c_{\perp}$ does not contribute to the simulation result, but will be investigated within the next example.

\subsubsection{Four point bending test}
\begin{figure}[ht]
\begin{center}
\footnotesize
\psfrag{a}[c][c]{\(\vartheta\)}
\includegraphics[width=0.85\textwidth]{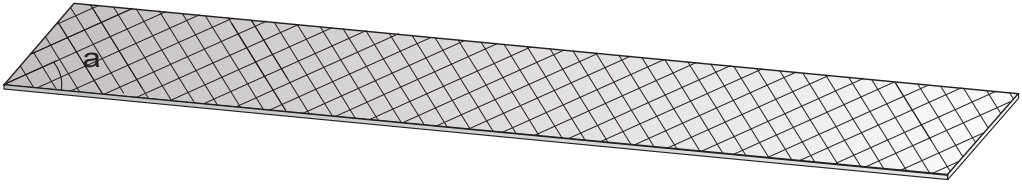}
\end{center}
\caption{\textbf{In-plane bending test.} Problem setting. The lines illustrate the fiber structure.}
\label{fig:bending4PointRef}
\end{figure}

\begin{figure}[h]
\begin{center}
\includegraphics[width=0.55\textwidth]{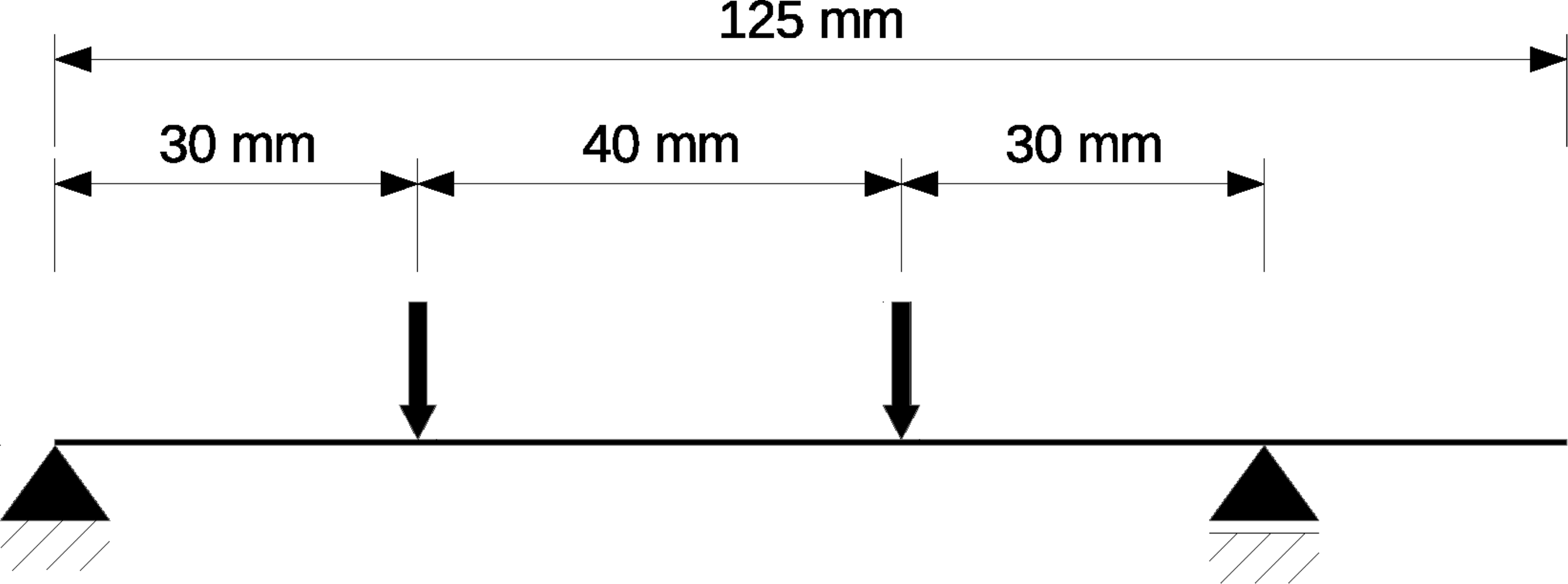}
\caption{\textbf{Four point bending test.} Boundary conditions of the four point bending test. }\label{fig:fourpoint_setup}
\end{center}
\end{figure}

Next, the out-of-plane bending behavior of the fiber material is investigated using a four point bending test. Therefore, we consider again a rectangular geometry of size \( L \times W \times H =125\,\mathrm{mm} \times 25\,\mathrm{mm} \times 0.5\,\mathrm{mm}\) discretized by \(50 \times 10\times2\) quadratic B-spline based elements. The \textcolor{black}{bidirectional} composite material has a matrix volume ratio of $\zeta=0.53$ and the fibers are aligned in the \(\vartheta=0^{\circ}\) configuration, see Figure  \ref{fig:bending4PointRef} for the details on the fiber orientation. The four point bending test as shown in Figure \ref{fig:fourpoint_setup} leads to a pure out-of-plane bending deformation of the structure. In particular, we prevent the displacement in upward direction for the outer support points and prescribe a displacement in downward direction for the inner contact points. Additionally, the left support point is horizontally fixed, whereas we allow sliding for the other contact points. The material setting of the matrix material reads \(\mu=1630.4\,\mathrm{N/mm^2} \) and \(\alpha_1=2\) for the deviatoric part and \(\kappa = 6250\,\mathrm{N/mm^2}\) and \(\beta=-2\) for the volumetric part, which corresponds to a Young's modulus of \(E_{\indi{mat}}=4500 \, \mathrm{N/mm^2}\) and a Poisson's ratio of \(\nu = 0.38\).

Two different settings of the fiber material properties are applied assuming a single layer of fibers over thickness direction. Firstly, we set the tensile stiffness of the fibers to \(a = E_{\indi{fib}}=79000\,\mathrm{N/mm^2}\) and the bending stiffness to \(c_{\perp}=0\,\mathrm{N}\). Secondly, we set the tensile stiffness of the fibers to zero and adjust the bending stiffness as \(c_{\perp}=E_{\indi{fib}}H^2/12=1645.83\,\mathrm{N}\).  

\begin{figure}[htb]
\begin{center}
\includegraphics[width=0.75\textwidth]{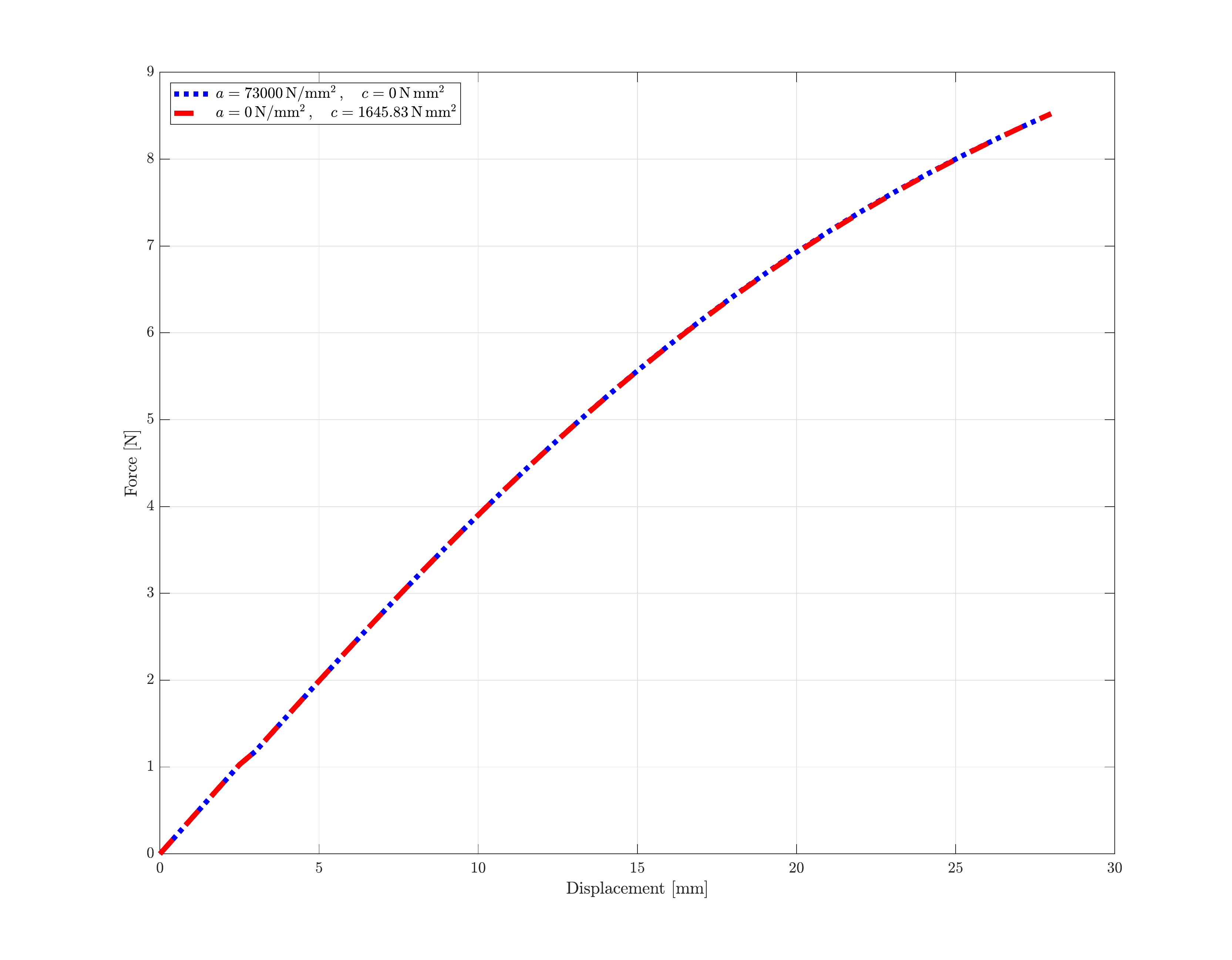}
\caption{\textbf{Four point bending test.} Force-displacement curves for bending tests. }\label{fig:fourpoint_results}
\end{center}
\end{figure}

The applied bending stiffness of the continuum fiber model correlates to the out-of-plane bending stiffness for a shell model with the same high. As shown in the previous example, the proposed strain-gradient continuum formulation match the contributions of a gradient shell formulation, provided that the stiffness is chosen properly. Thus, if we resolve the thickness of sufficiently flat geometry with in the continuum model to obtain the same deformation as expected for the shell theory, a coincident bending behavior of the structure should result. Figure \ref{fig:fourpoint_results} shows the load deflection result for the investigated material settings. As expected, both results match in a good agreement, i.e.\ the tension/compression behavior of the continuum fiber model in this bending example can be described by the bending terms themselves. This is an important and well known result, as strain-gradient contributions emanate from a length-scale dependent microstructure and if this microstructure is already resolved by the first order continuum framework, the second-order contributions \textcolor{black}{must be} removed.

%% file: chapters/tension.tex
\begin{figure}[h]
\begin{center}
\footnotesize
\psfrag{a}[c][c]{\(\vartheta\)}
\psfrag{u1}[c][c]{\(u=0\)}
\psfrag{u2}[c][c]{\(u\)}
\includegraphics[width=0.85\textwidth]{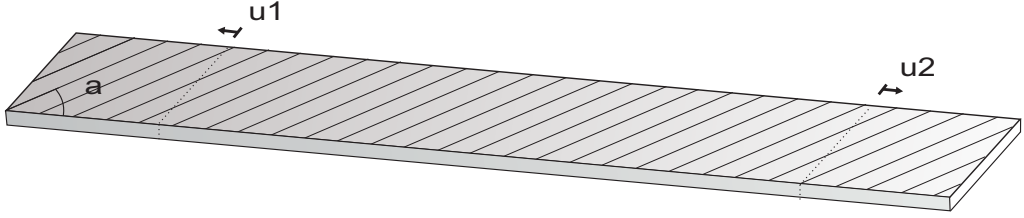}
\end{center}
\caption{\textbf{Tensile Test (unidirectional).} Problem setting. The lines illustrate the fiber structure.}
\label{fig:tensionUniRef}
\end{figure}

\begin{figure}[t]
\begin{center}
\footnotesize
\psfrag{a}[c][c]{\(\vartheta\)}
\psfrag{u1}[c][c]{\(u=0\)}
\psfrag{u2}[c][c]{\(u\)}
\includegraphics[width=0.85\textwidth]{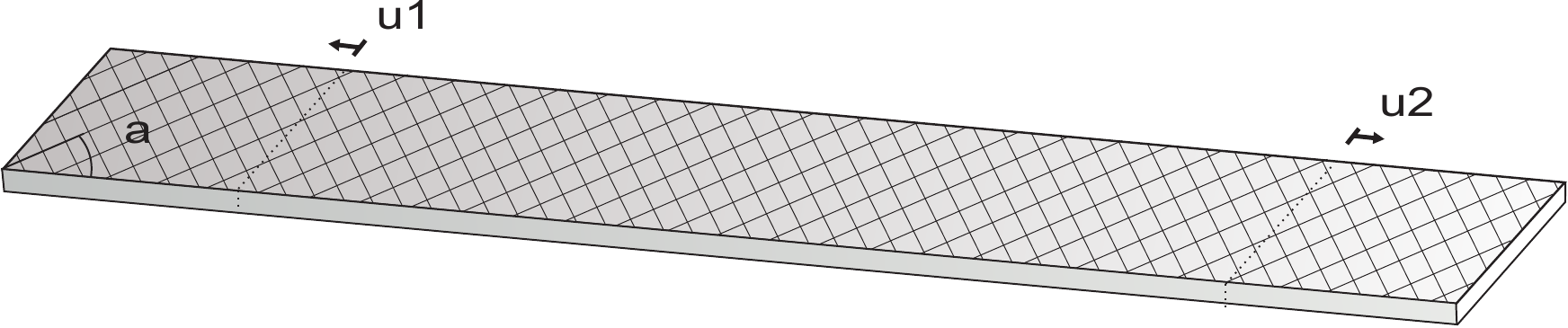}
\end{center}
\caption{\textbf{Tensile Test (bidirectional).} Problem setting. The lines illustrate the fiber structure.}
\label{fig:tensionBiRef}
\end{figure}

In this next example, we conduct a serious of tension tests to investigate the crack behavior of a prototypical roving glass composite material with different fiber configurations. Therefore, we consider a flat specimen of size \(L\times W\times H =125\,\mathrm{mm} \times 25\,\mathrm{mm} \times 2\,\mathrm{mm}\). Figure \ref{fig:tensionUniRef} and \ref{fig:tensionBiRef} show the geometry in the reference configuration along with the applied boundary conditions and the fiber configurations. The outer areas of length \(20\,\mathrm{mm}\) are subject to Dirichlet boundary conditions. To be specific, one flap is fixed and the other flap is moved by a displacement rate of \(0.5\,\mathrm{mm/s}\) within a quasi-static simulation setting neglecting inertia effects. The computational mesh consists of \(2432\) quadratic NURBS elements. The material setting of the composite is summarized in Table \ref{table:matData1}. We assume a quadratic cross section of the fibers with \(A_{\mathrm{fib}}=0.0025\,\mathrm{mm^2}\) and obtain a bending stiffness of \(c_{\perp}= c_{\#} = E_{\indi{fib}} A_{\mathrm{fib}}/12 = 16.46\,\mathrm{N}\).

\subsubsection{Unidirectional fiber reinforcement}
\begin{figure}[t]
\begin{center}
\psfrag{15}[c][c]{\ding{192}}
\psfrag{3}[c][c]{\ding{193}}
\psfrag{14}[c][c]{\ding{194}}
\psfrag{11}[c][c]{\ding{195}}
\psfrag{4}[c][c]{\ding{196}}
\psfrag{12}[c][c]{\ding{197}}
\psfrag{13}[c][c]{\ding{198}}
\psfrag{5}[c][c]{\ding{199}}
\psfrag{10}[c][c]{\ding{200}}
\psfrag{6}[c][c]{\ding{201}}
\psfrag{7}[c][c]{\ding{202}}
\psfrag{1}[c][c]{\ding{203}}
\psfrag{9}[c][c]{\ding{204}}
\psfrag{8}[c][c]{\ding{205}}
\includegraphics[width=0.98\textwidth]{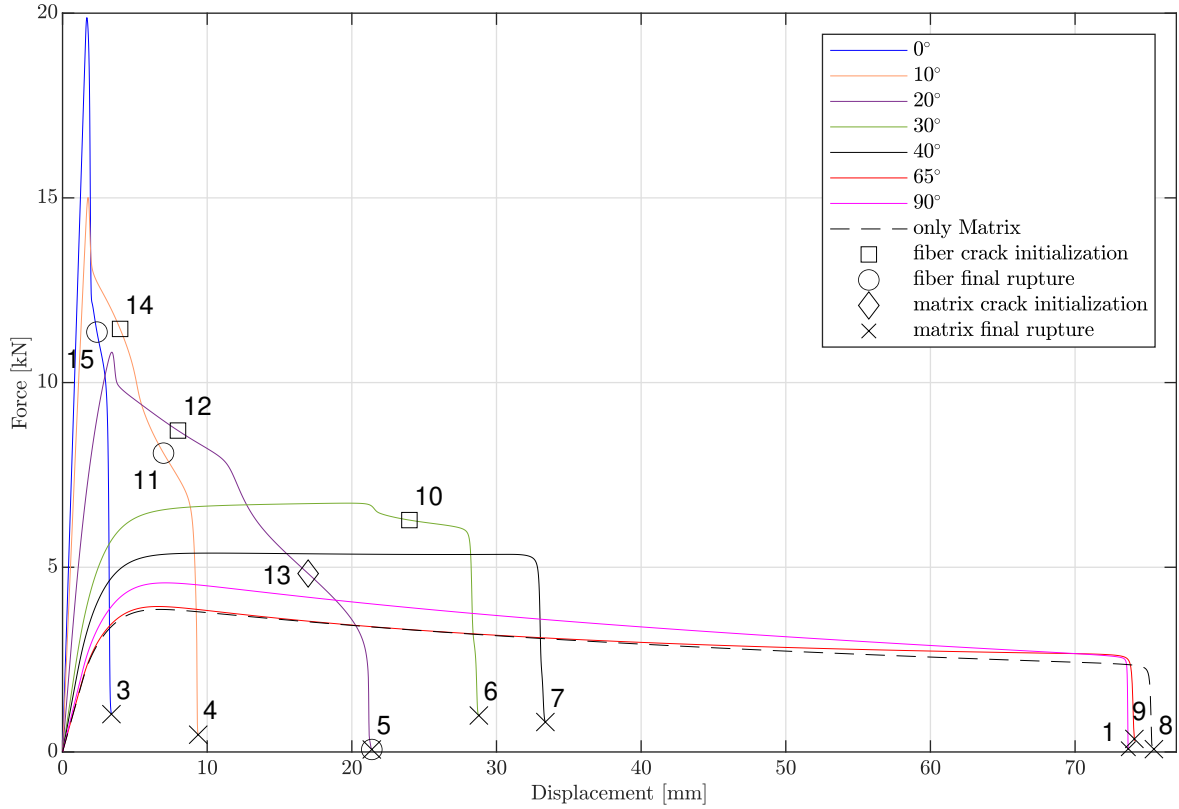}
\end{center}
\caption{\textbf{Tensile Test (unidirectional).} Load deflection results for unidirectional fiber reinforcements with different orientations.}
\label{fig:unitension}
\end{figure}

\begin{figure}
\begin{center}
\footnotesize
\psfrag{a1}[l][l]{0.7}
\psfrag{a0}[l][l]{0}
\psfrag{a}[l][l]{$\alpha$}
\psfrag{s1}[l][l]{1}
\psfrag{s0}[l][l]{0}
\psfrag{s}[l][l]{$\s$}
\psfrag{t}[l][l]{$\s_{\mathrm{L}}$}
\includegraphics[width=0.05\textwidth]{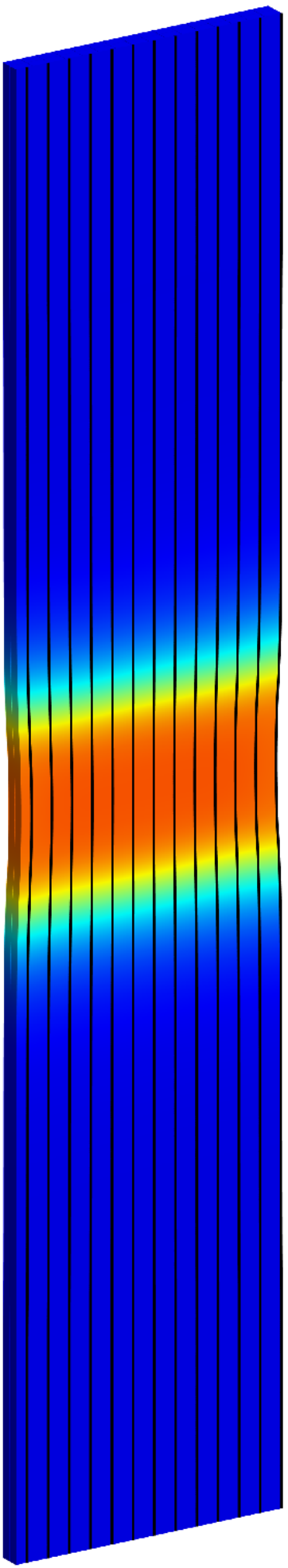}
\includegraphics[width=0.05\textwidth]{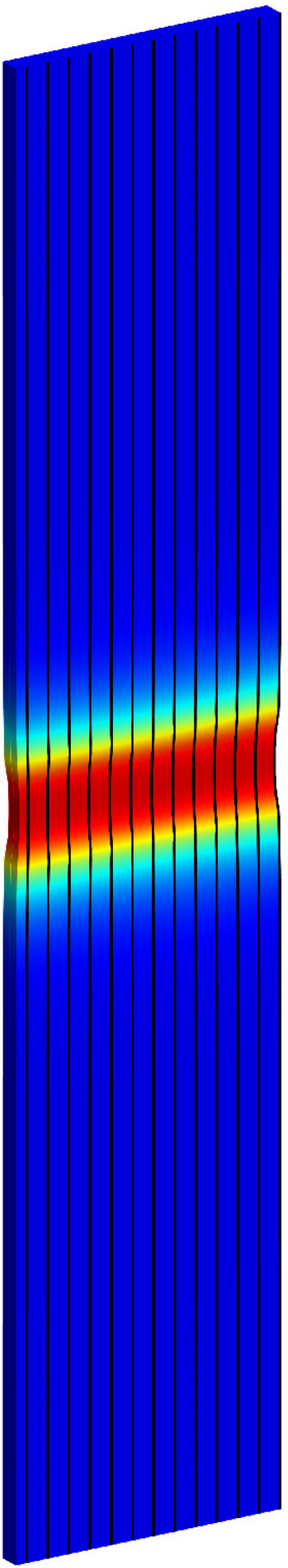}
\hspace{2mm}
\includegraphics[width=0.05\textwidth]{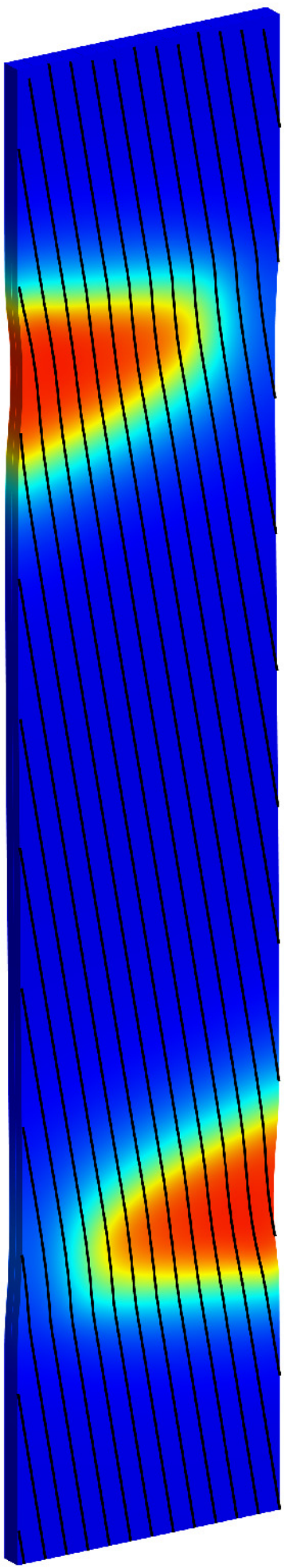}
\includegraphics[width=0.05\textwidth]{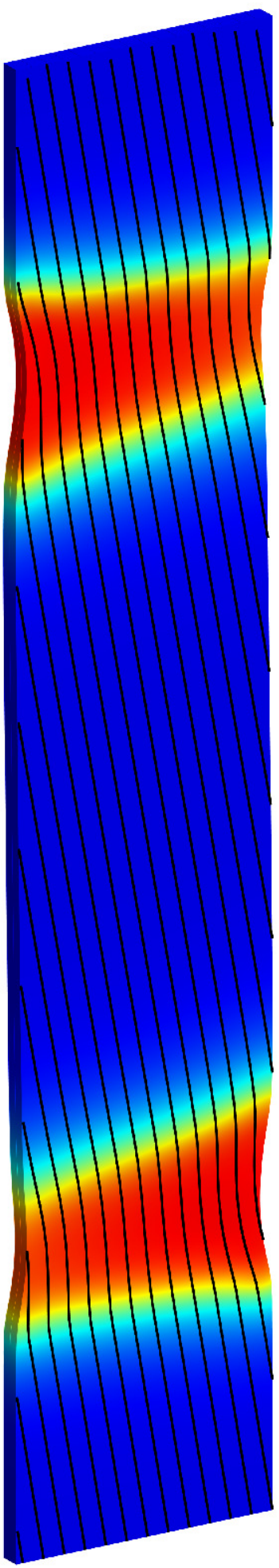}
\includegraphics[width=0.05\textwidth]{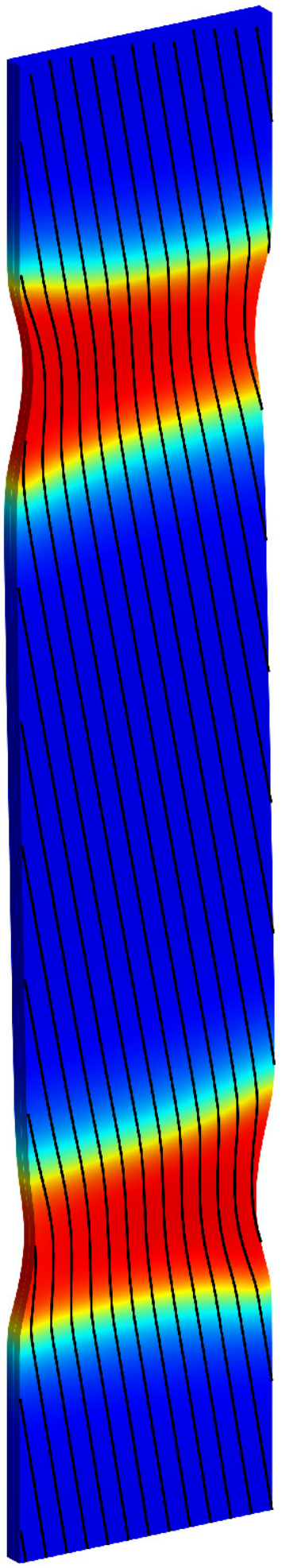}
\hspace{2mm}
\includegraphics[width=0.05\textwidth]{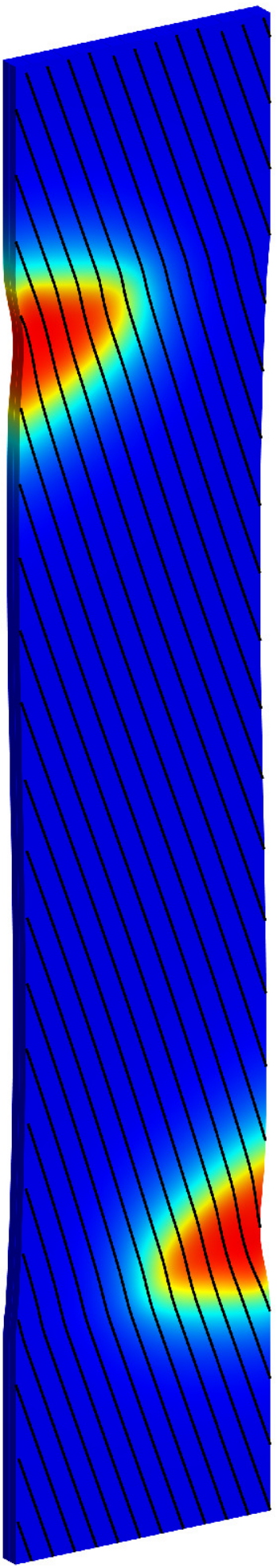}
\includegraphics[width=0.05\textwidth]{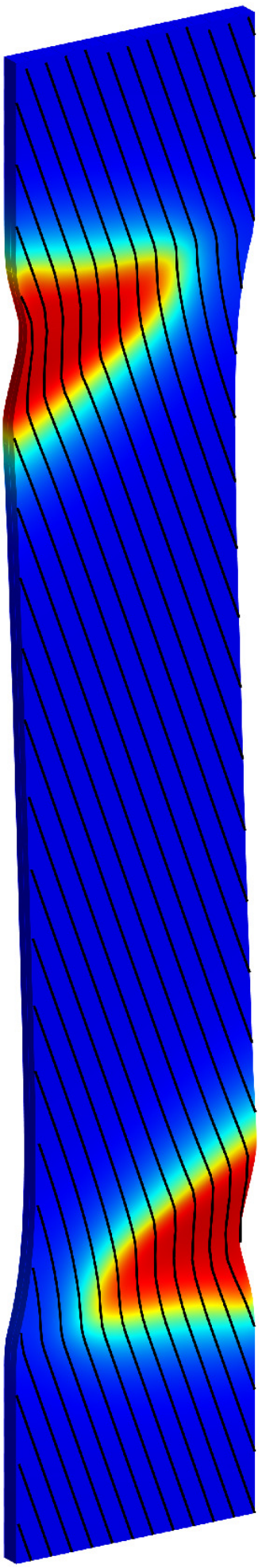}
\includegraphics[width=0.05\textwidth]{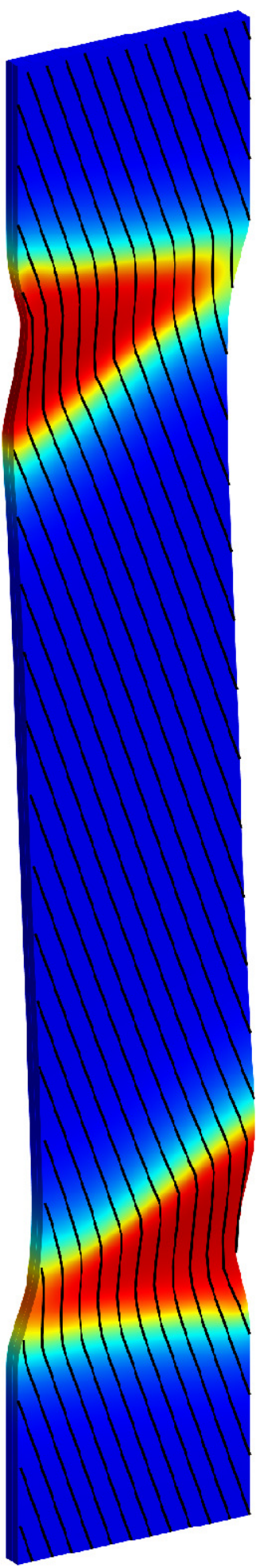}
\hspace{2mm}
\includegraphics[width=0.05\textwidth]{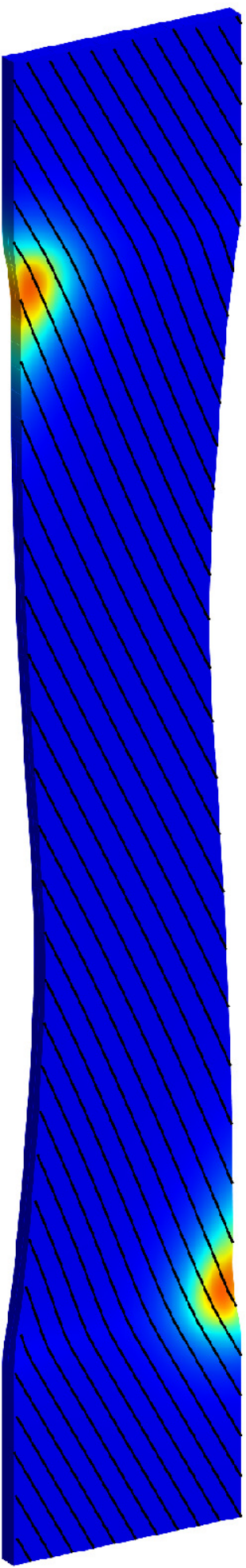}
\includegraphics[width=0.05\textwidth]{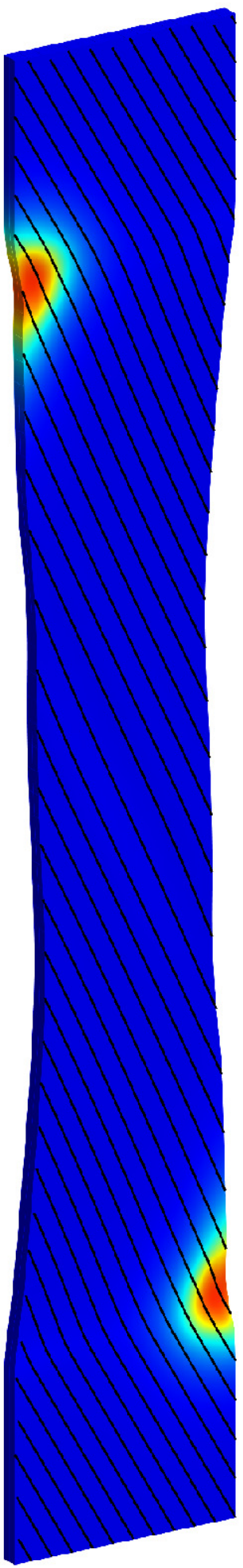}
\hspace{2mm}
\includegraphics[width=0.05\textwidth]{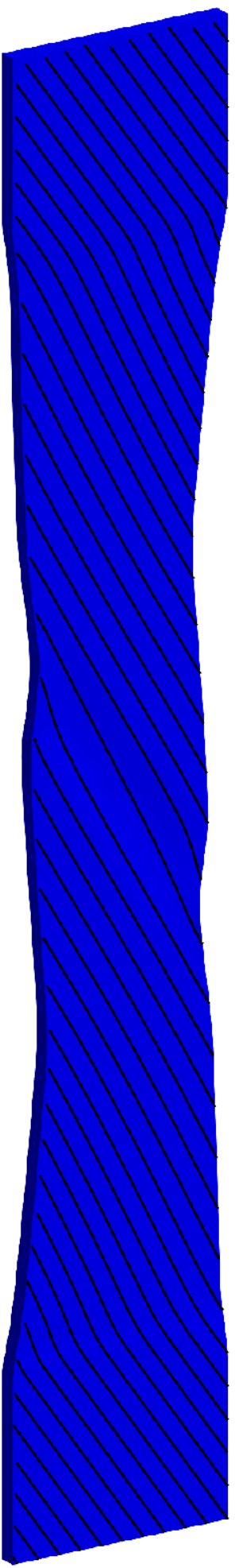}
\hspace{2mm}
\includegraphics[width=0.05\textwidth]{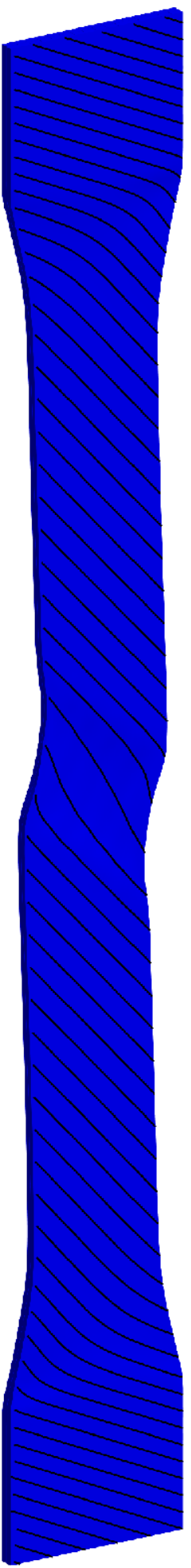}
\hspace{2mm}
\includegraphics[width=0.05\textwidth]{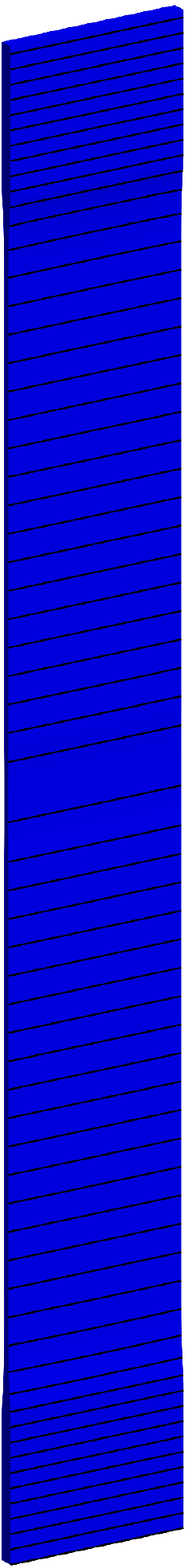}
\hspace{2mm}
\includegraphics[width=0.05\textwidth]{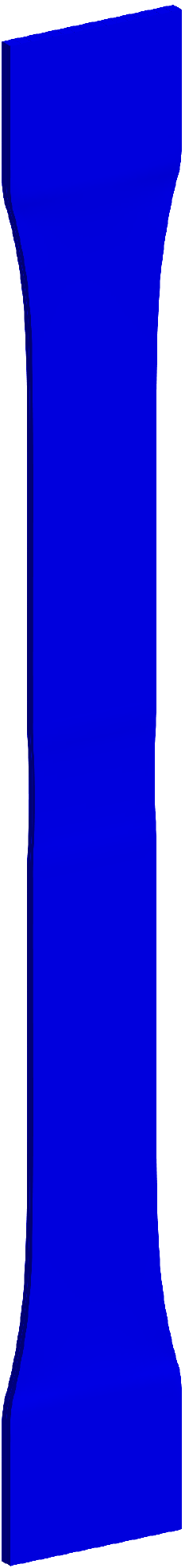}
\includegraphics[width=0.045\textwidth]{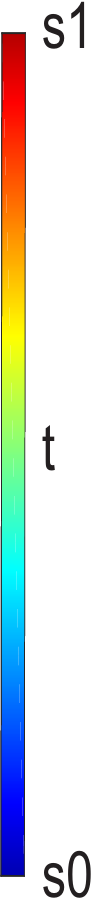}
\vspace{-1mm}
\includegraphics[width=0.05\textwidth]{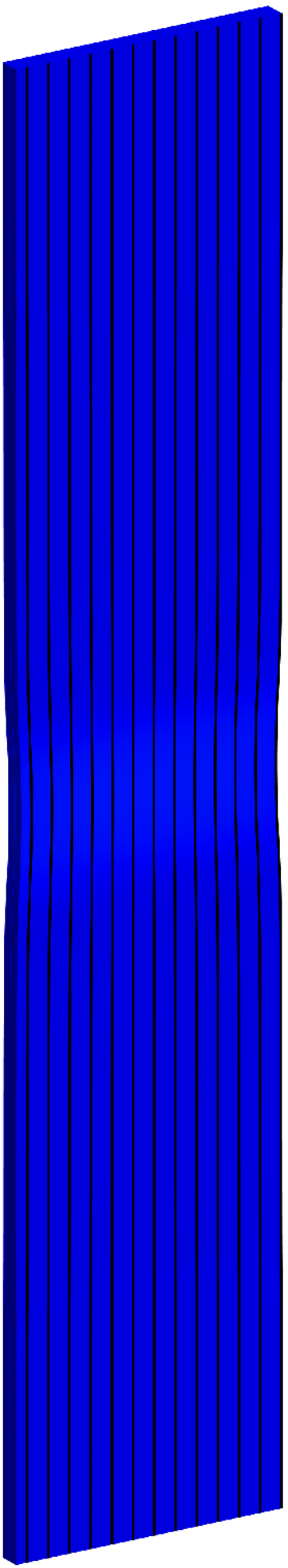}
\includegraphics[width=0.05\textwidth]{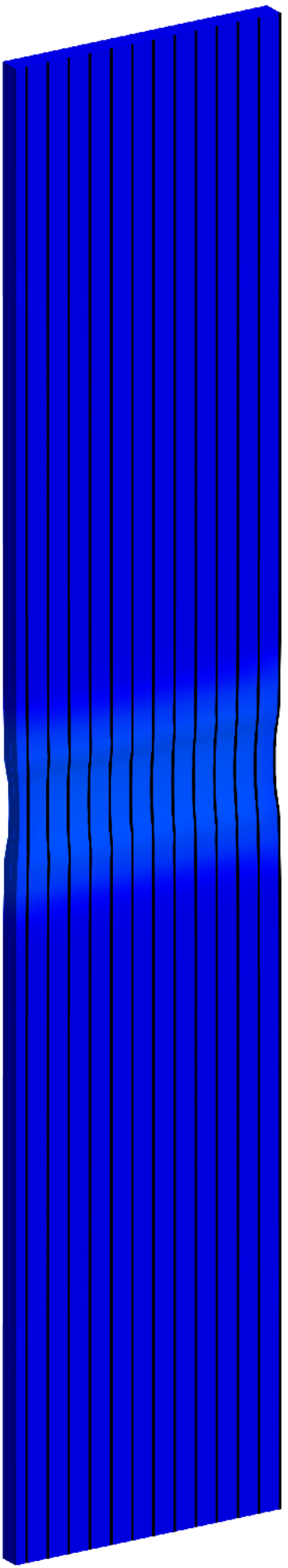}
\hspace{2mm}
\includegraphics[width=0.05\textwidth]{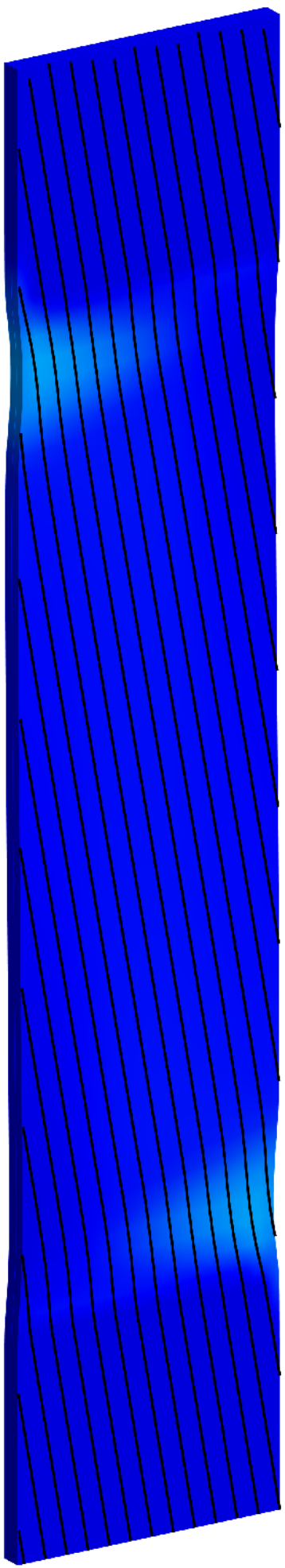}
\includegraphics[width=0.05\textwidth]{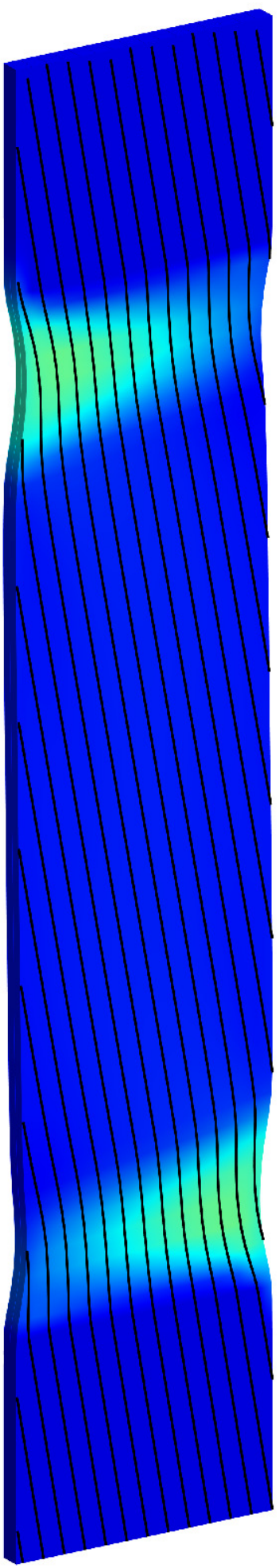}
\includegraphics[width=0.05\textwidth]{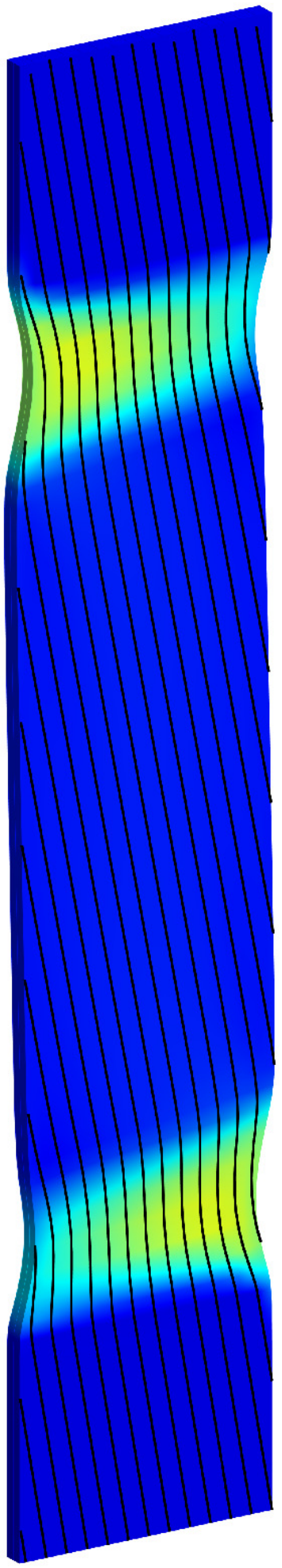}
\hspace{2mm}
\includegraphics[width=0.05\textwidth]{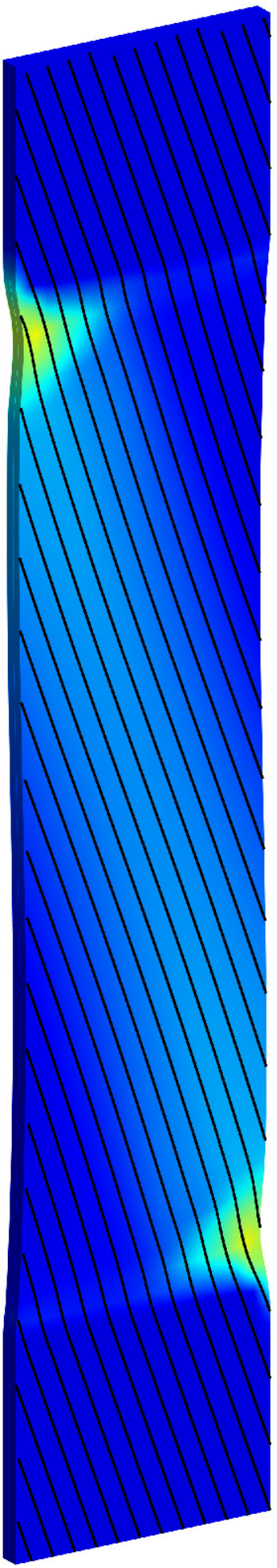}
\includegraphics[width=0.05\textwidth]{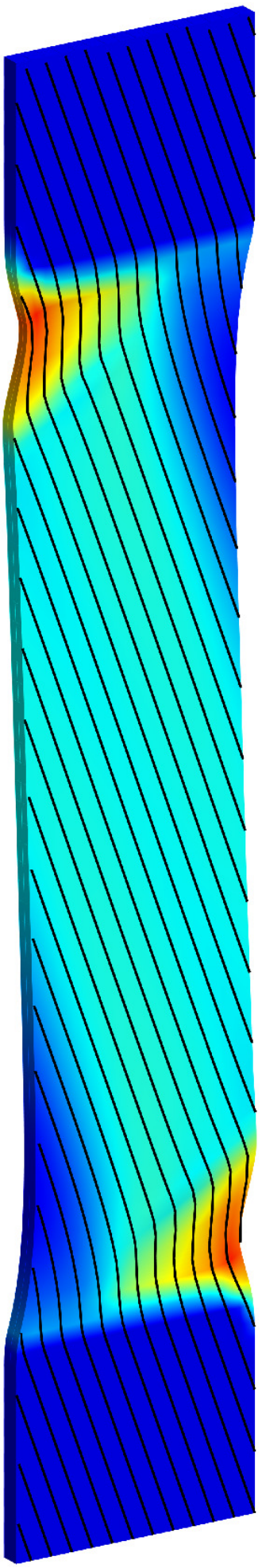}
\includegraphics[width=0.05\textwidth]{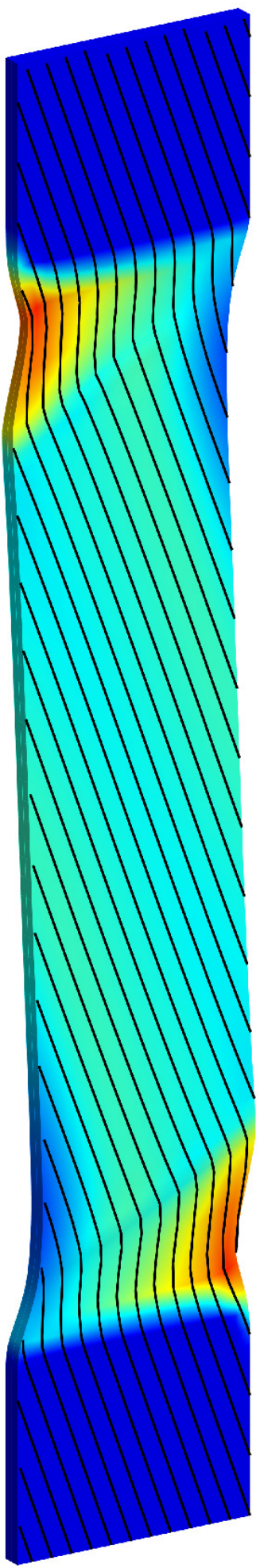}
\hspace{2mm}
\includegraphics[width=0.05\textwidth]{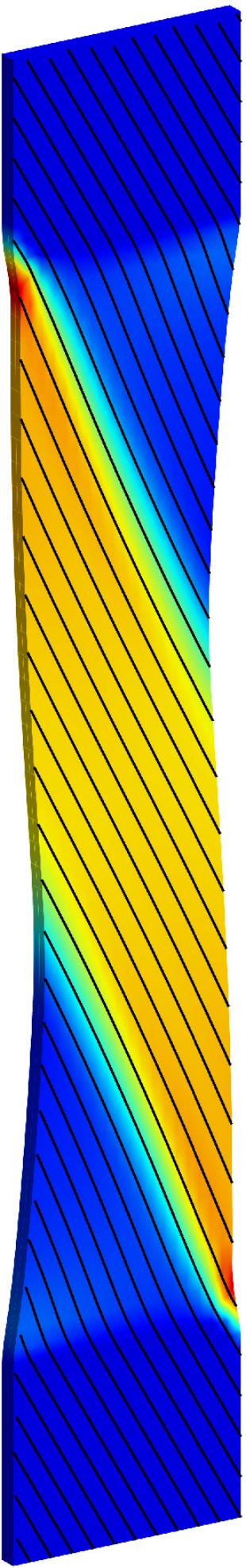}
\includegraphics[width=0.05\textwidth]{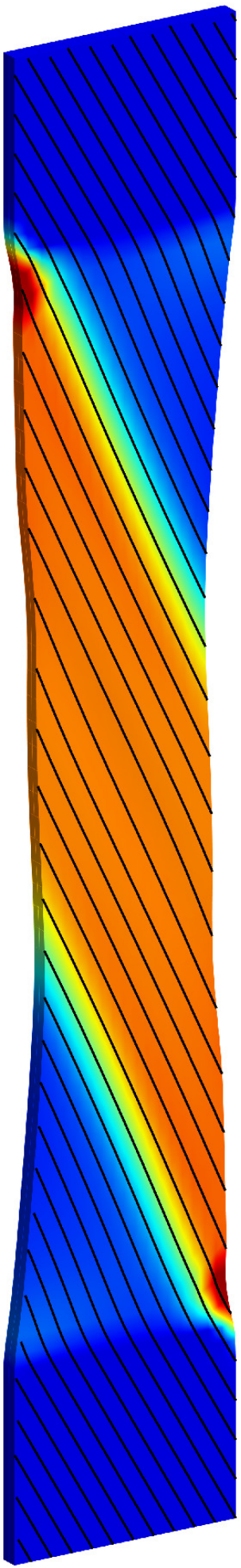}
\hspace{2mm}
\includegraphics[width=0.05\textwidth]{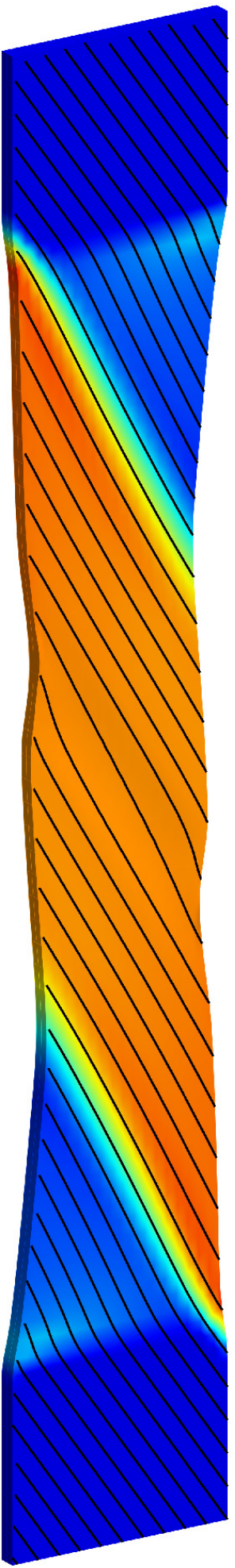}
\hspace{2mm}
\includegraphics[width=0.05\textwidth]{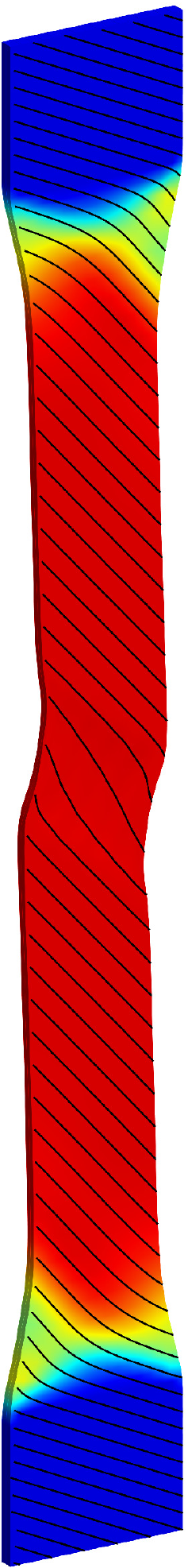}
\hspace{2mm}
\includegraphics[width=0.05\textwidth]{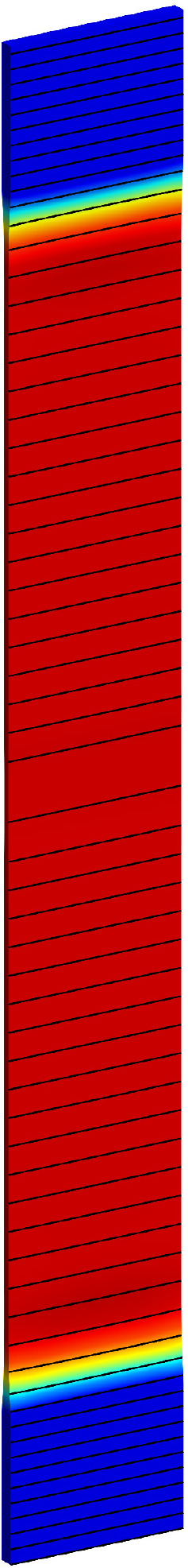}
\hspace{2mm}
\includegraphics[width=0.05\textwidth]{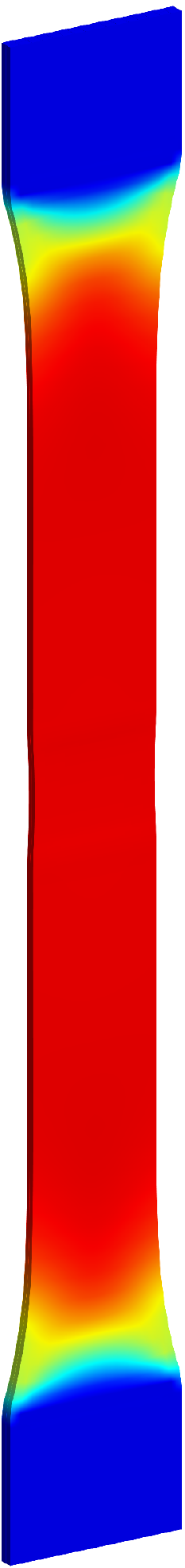}
\includegraphics[width=0.045\textwidth]{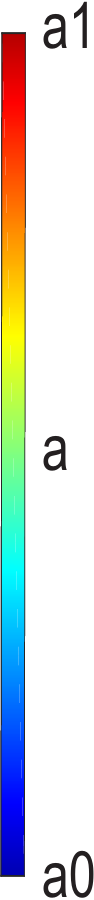}
\vspace{-1mm}
\includegraphics[width=0.05\textwidth]{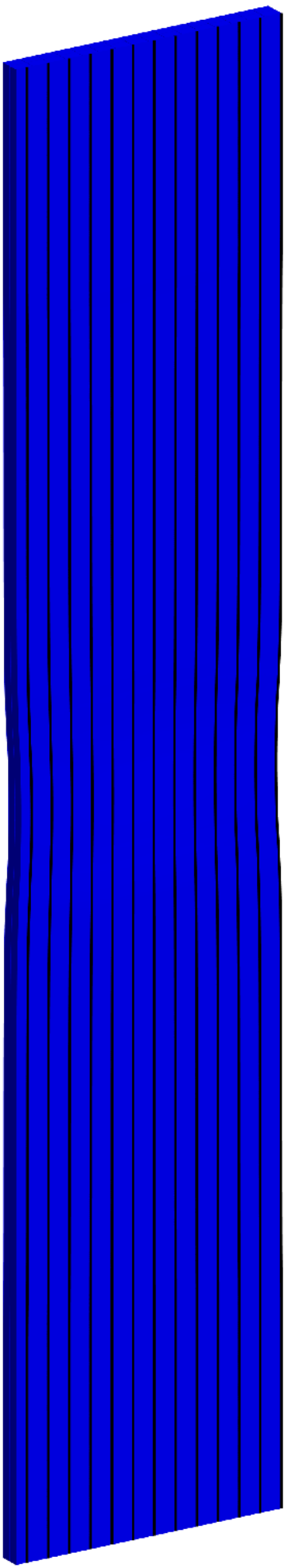}
\includegraphics[width=0.05\textwidth]{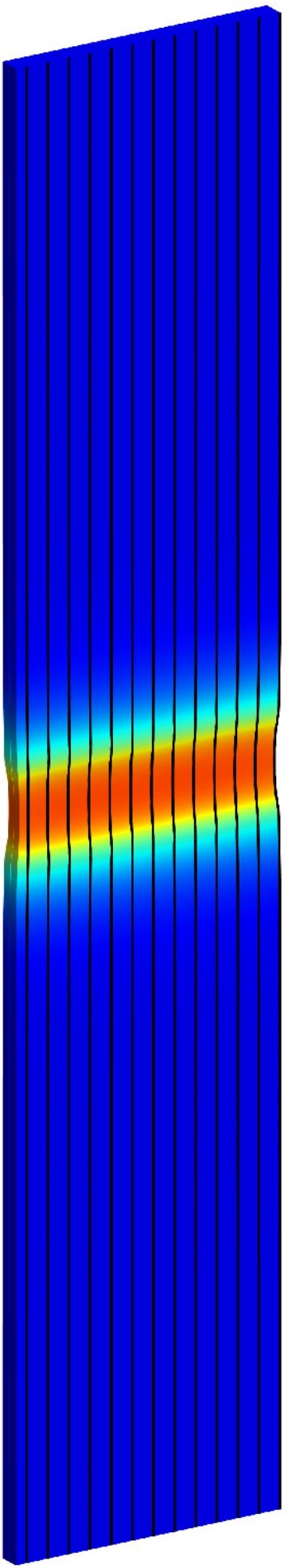}
\hspace{2mm}
\includegraphics[width=0.05\textwidth]{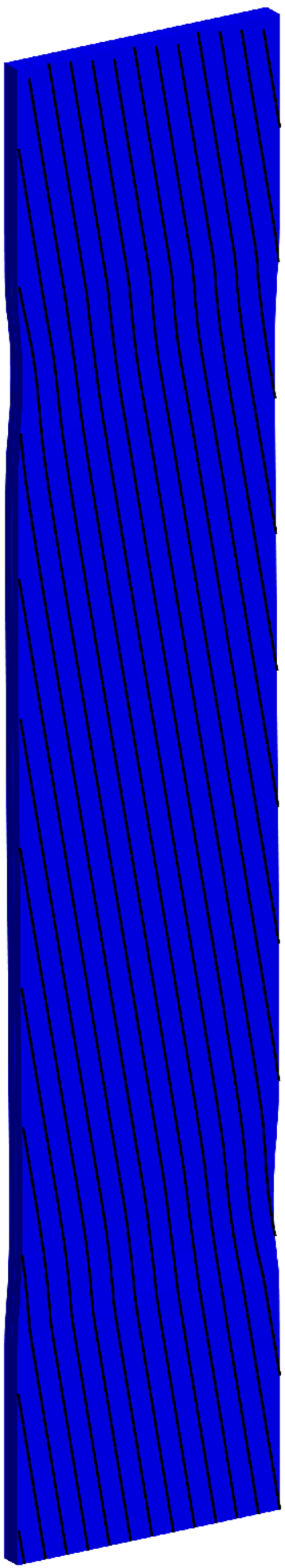}
\includegraphics[width=0.05\textwidth]{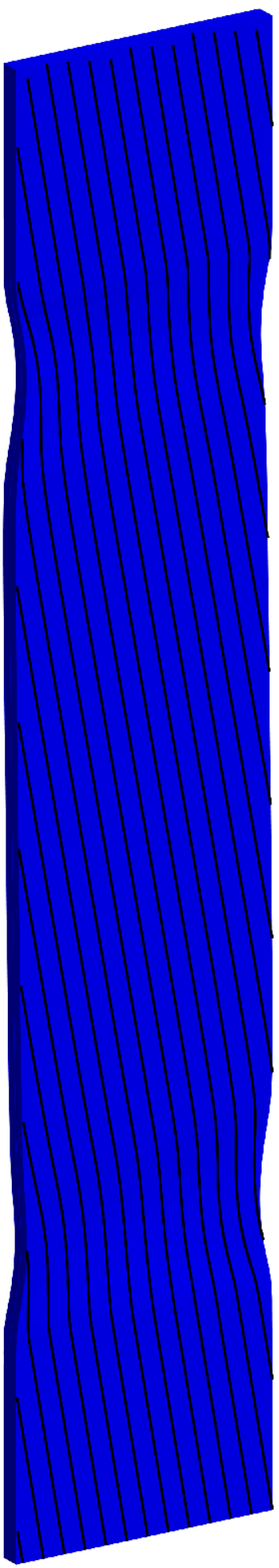}
\includegraphics[width=0.05\textwidth]{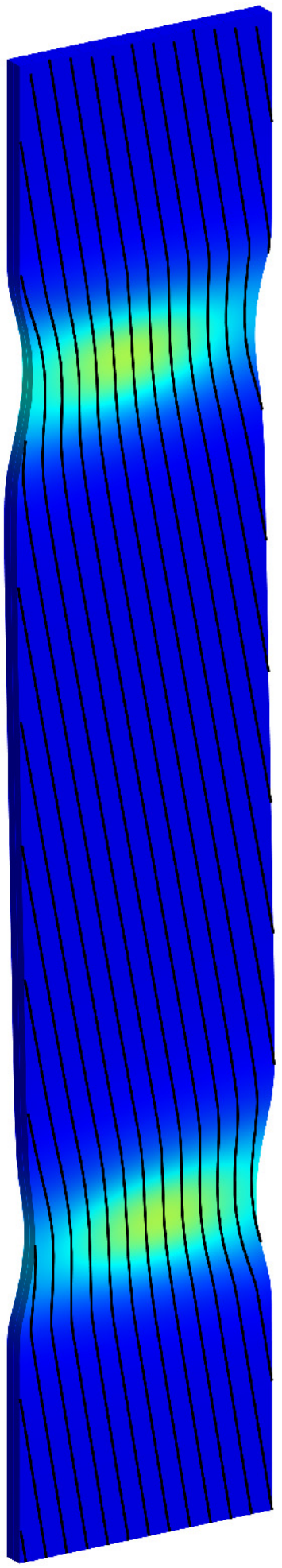}
\hspace{2mm}
\includegraphics[width=0.05\textwidth]{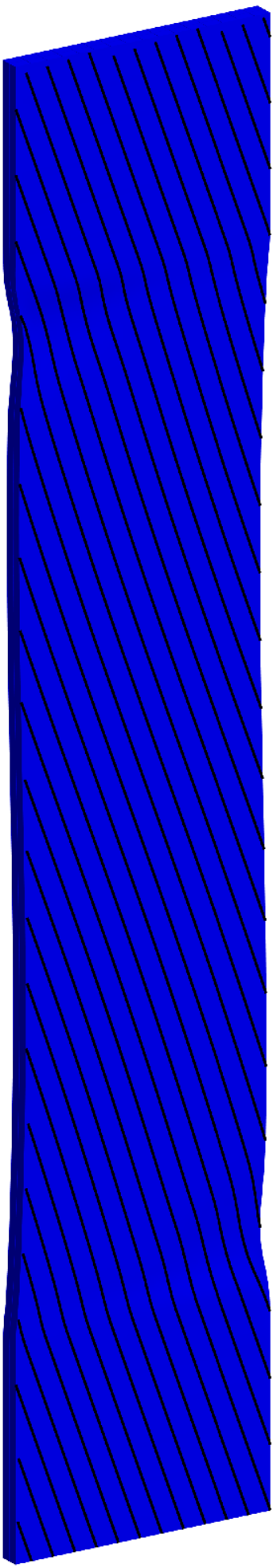}
\includegraphics[width=0.05\textwidth]{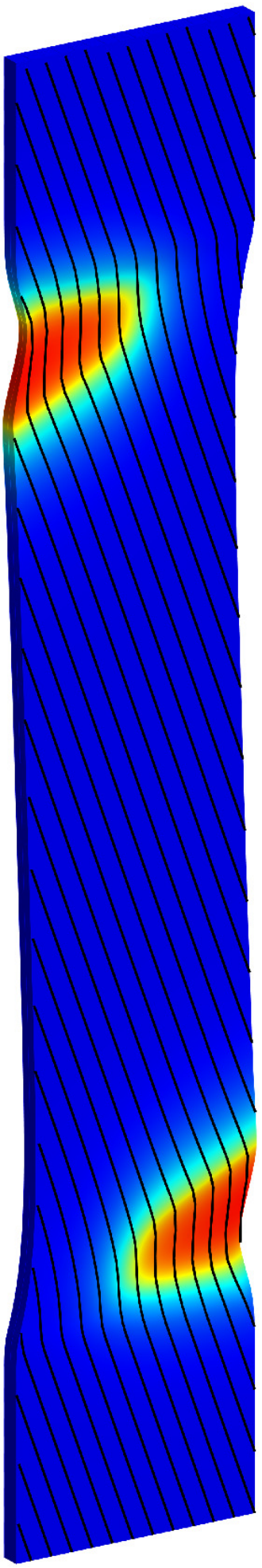}
\includegraphics[width=0.05\textwidth]{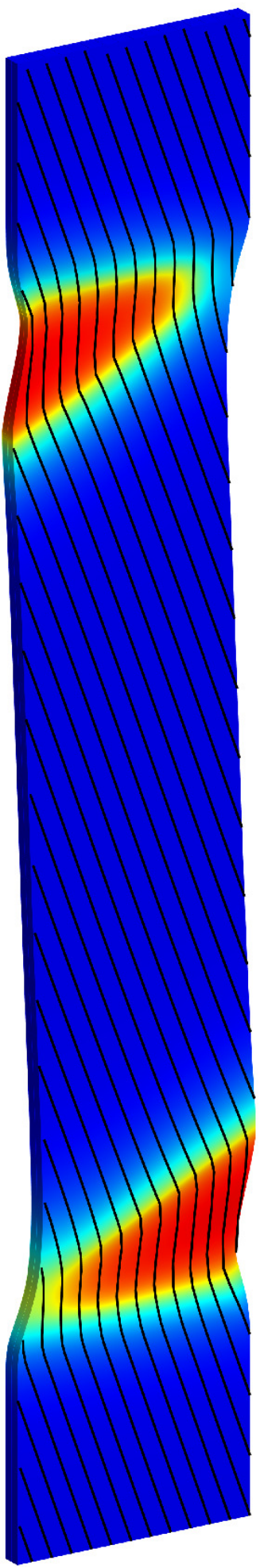}
\hspace{2mm}
\includegraphics[width=0.05\textwidth]{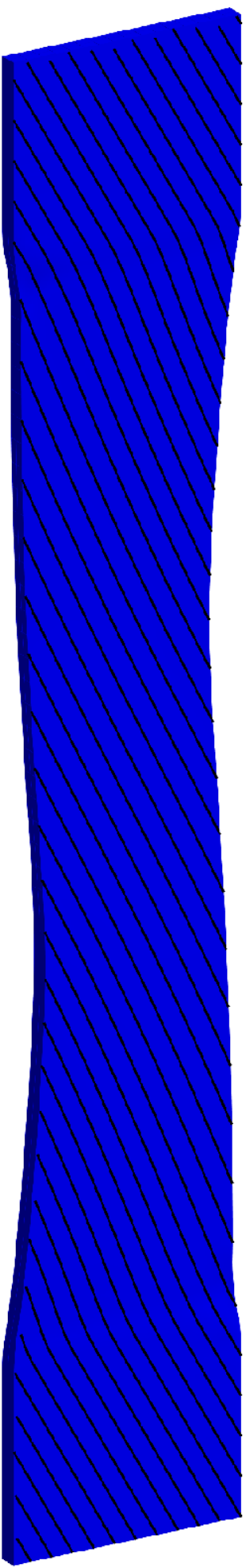}
\includegraphics[width=0.05\textwidth]{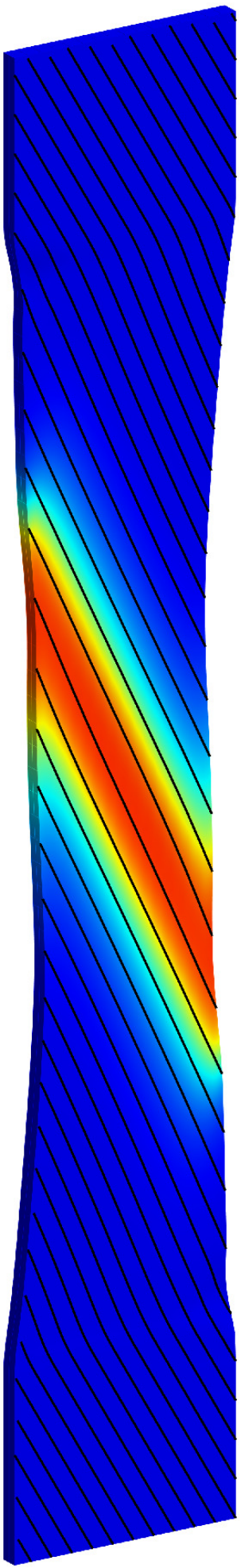}
\hspace{2mm}
\includegraphics[width=0.05\textwidth]{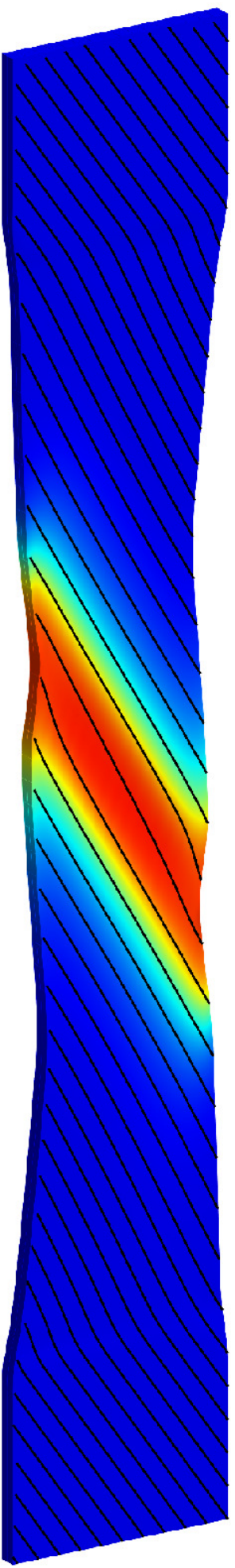}
\hspace{2mm}
\includegraphics[width=0.05\textwidth]{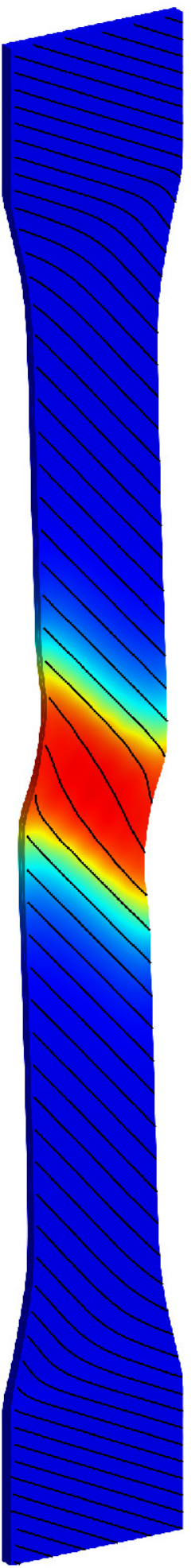}
\hspace{2mm}
\includegraphics[width=0.05\textwidth]{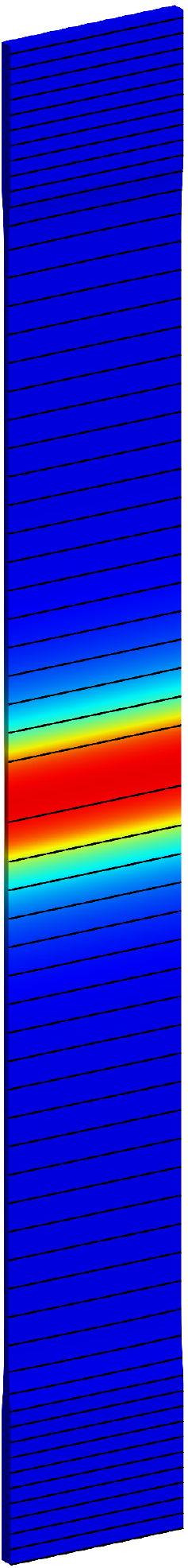}
\hspace{2mm}
\includegraphics[width=0.05\textwidth]{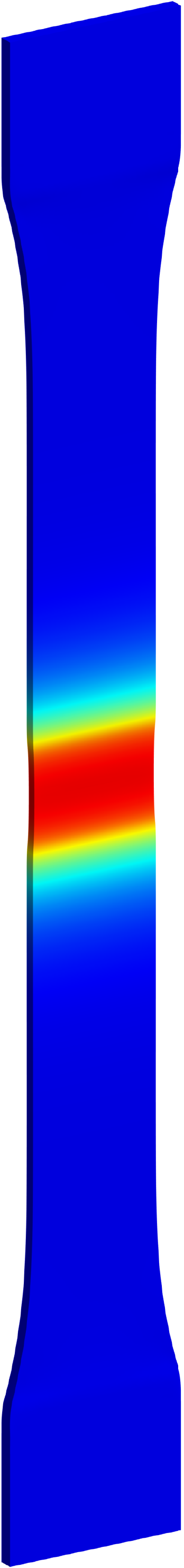}
\includegraphics[width=0.045\textwidth]{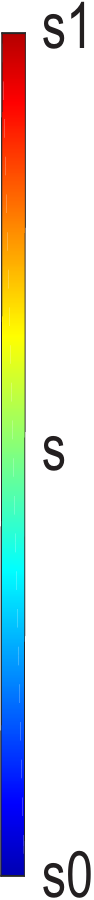}
\\
\hspace{-6.5mm}\ding{192}\hspace{6.5mm}\ding{193}\hspace{9.0mm}\ding{194}\hspace{6.5mm}\ding{195}\hspace{6.5mm}\ding{196}\hspace{9.5mm}\ding{197}\hspace{6.5mm}\ding{198}\hspace{6.5mm}\ding{199}\hspace{9.2mm}\ding{200}\hspace{6.5mm}\ding{201}\hspace{9.5mm}\ding{202}\hspace{9.5mm}\ding{203}\hspace{9.5mm}\ding{204}\hspace{9.5mm}\ding{205}
\vspace{-4.5mm}
\end{center}
\caption{\textbf{Tensile Test (unidirectional).} Results of the fiber crack phase-field (first row), the plastic strain field (second row) and crack phase-field of the matrix material (third row). The results are shown for the different deformation states and fiber configurations marked in Figure \ref{fig:unitension}.}
\label{fig:unitensionPfPs}
\end{figure}

We first analyze the behavior of a unidirectional reinforced composite material, with fiber orientations of \(\vartheta=[0^{\circ},10^{\circ},20^{\circ},30^{\circ},40^{\circ},65^{\circ},90^{\circ}]\), see Figure \ref{fig:tensionUniRef}. The load deflection results for isothermal simulations at \(\theta=293\,\mathrm{K}\) are shown in Figure \ref{fig:unitension}. Therein, crack initialization and final rupture of the fiber material are indicated by \(\square\) and \(\circ\), respectively. In addition, crack initialization and final rupture of the matrix material are indicated by \(\diamond\) and \(\cross\), respectively. In Figure \ref{fig:unitensionPfPs}, the crack phase field results of the fiber material and the matrix material are depicted along with results of the plastic strain field for the marked points. Note that black marker indicate states without fiber fracture, as the specimen is already fully broken.

For a fiber orientation of \(\vartheta=0^{\circ}\), the fibers account for most of the load transfer due to the different Young's modulus and fracture abruptly in the center of the specimen \ding{192}. Subsequently, the matrix material undergoes plastification and ductile fracture due to an abrupt load rearrangement \ding{193}. Note that the resulting high strain rates lead to a pronounced viscoplastic behavior within the matrix material which is controlled by the viscous regularization parameter.

Concerning the unidirectional \(10^{\circ}\) fiber configuration, the fibers start to crack near the clamping zones \ding{194} which is additionally driven by the bending contribution to the crack diving force. This process is slowed down due to the hardening behavior of the matrix material \ding{195}. At the state \ding{196}, the fibers are fully ruptured and the matrix material starts to fracture in the same region.

For a fiber orientation of \(\vartheta=20^{\circ}\), brittle fracture of the fibers starts again near the clamping zones \ding{197}. However, a more pronounced plastification and thus hardening of the matrix material occur \ding{198} such that the fiber and matrix material undergo final rupture nearly at the same deformation state \ding{199}.

Applying a fiber orientation of \(\vartheta=30^{\circ}\), a direct load transfer between both boundaries by the fibers is not possible since fibers which are clamped at the lower end do not reach the upper clamping zone. Hence, the load has to be transferred towards the matrix material leading to higher plastification \ding{200} and ductile fracture at the center of the specimen \ding{201}. Note that the fibers begin to fracture only in small areas near the clamping zones \ding{200}.

For fiber orientations of \(\vartheta=[40^{\circ},65^{\circ},90^{\circ}]\), the fiber material does not fracture due to small loads acting in fiber direction \ding{202}--\ding{205}. Instead, the matrix material undergoes suitable plastification and subsequently ductile fracture leading to failure orthogonal to the fiber orientations. Concerning the \(\vartheta=90^{\circ}\) fiber configuration, the fibers controls the necking which can be observed by comparing the deformation with the results obtained for pure matrix material \ding{206}. This can also be observed by a slightly higher stiffness within the load deflection results before crack initiation, whereas the results are nearly identical up to a displacement of \(u=40\,\mathrm{mm}\).

\begin{figure}[t]
\begin{center}
\psfrag{12}[c][c]{\ding{192}}
\psfrag{13}[c][c]{\ding{193}}
\psfrag{9}[c][c]{\ding{194}}
\psfrag{10}[c][c]{\ding{195}}
\psfrag{11}[c][c]{\ding{196}}
\psfrag{7}[c][c]{\ding{197}}
\psfrag{6}[c][c]{\ding{198}}
\psfrag{8}[c][c]{\ding{199}}
\psfrag{4}[c][c]{\ding{200}}
\psfrag{3}[c][c]{\ding{201}}
\psfrag{5}[c][c]{\ding{202}}
\psfrag{1}[c][c]{\ding{203}}
\psfrag{2}[c][c]{\ding{204}}
\includegraphics[width=0.99\textwidth]{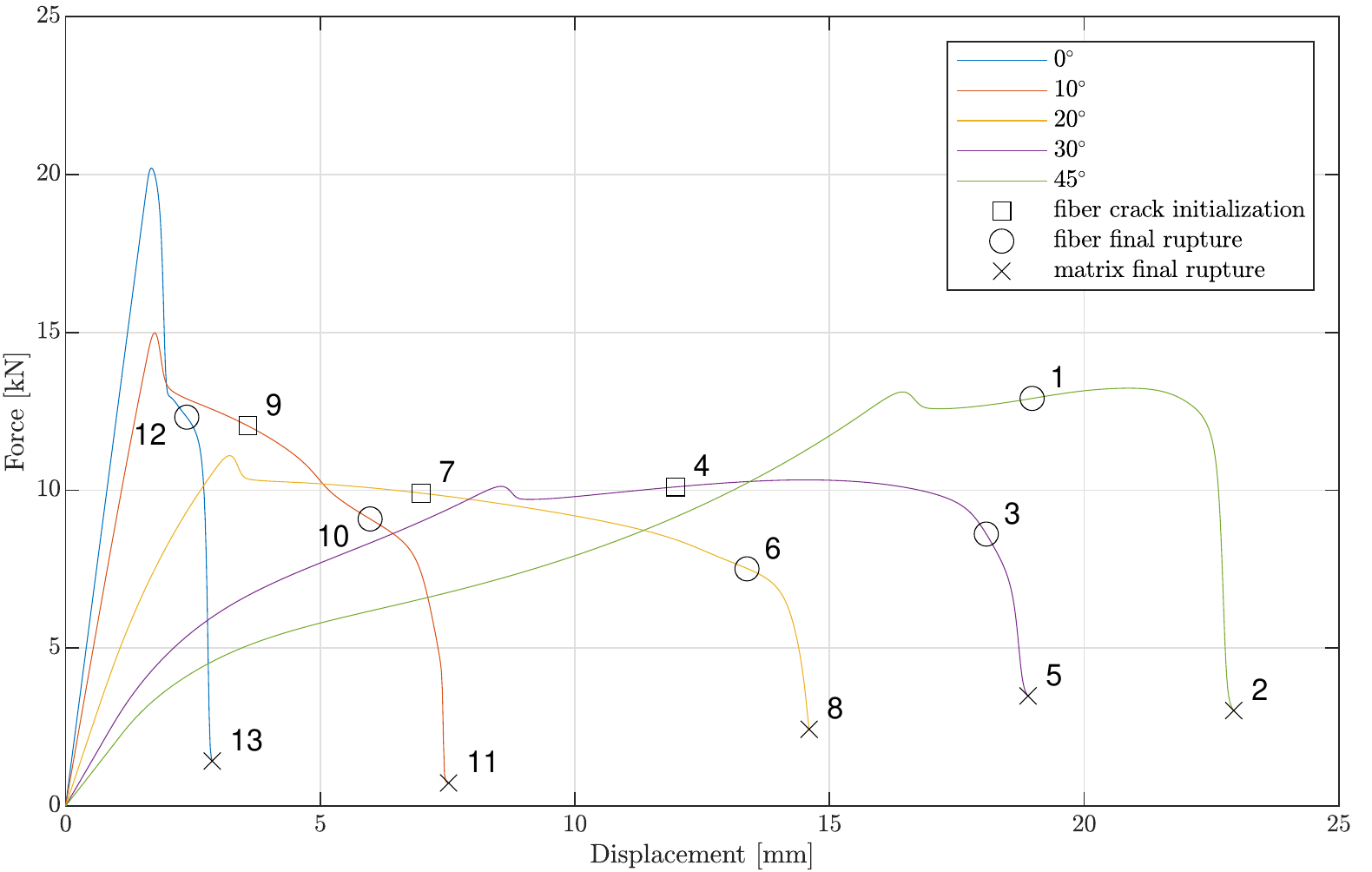}
\end{center}
\caption{\textbf{Tensile Test (bidirectional).} Load deflection results for bidirectional, orthotropic fiber reinforcements with different orientations.}
\label{fig:bitension}
\end{figure}

\subsubsection{Bidirectional fiber reinforcement}

\begin{figure}
\begin{center}
\footnotesize
\psfrag{a1}[l][l]{0.7}
\psfrag{a0}[l][l]{0}
\psfrag{a}[l][l]{$\alpha$}
\psfrag{s1}[l][l]{1}
\psfrag{s0}[l][l]{0}
\psfrag{s}[l][l]{$\s$}
\psfrag{t}[l][l]{$\s_{\mathrm{L}}$}
\includegraphics[width=0.05\textwidth]{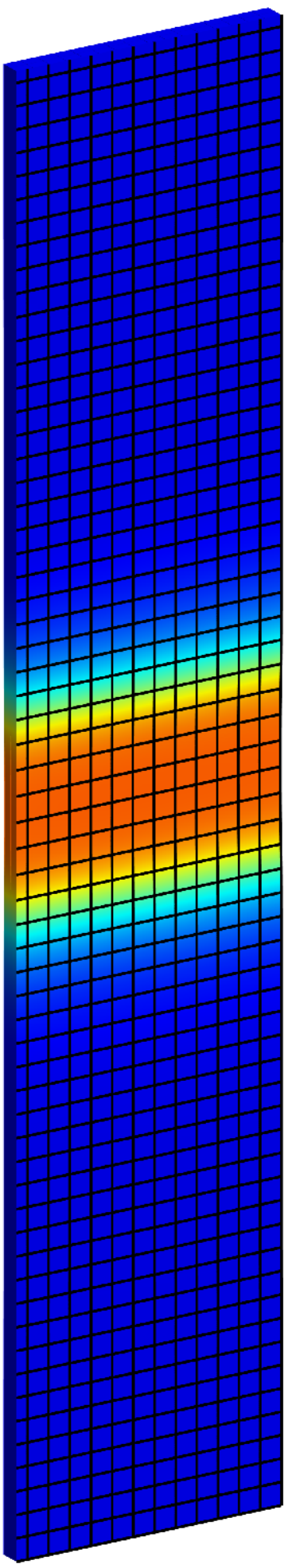}
\includegraphics[width=0.05\textwidth]{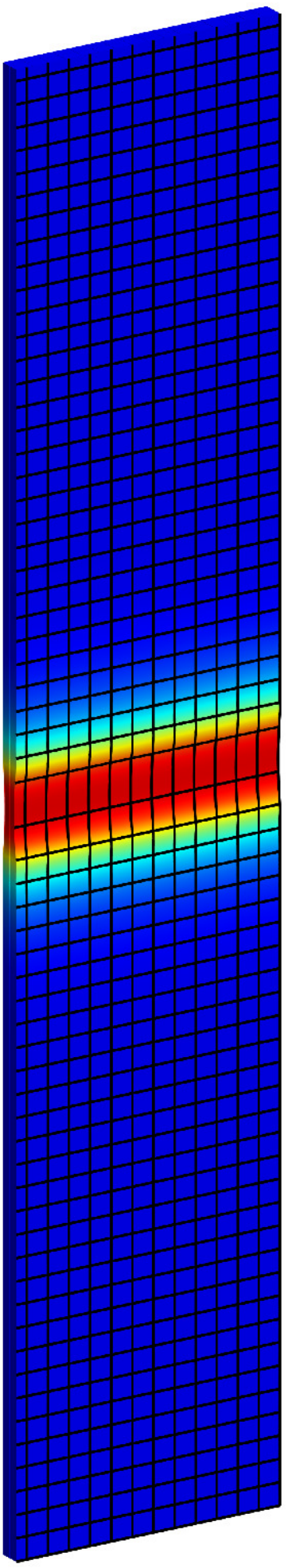}
\hspace{2mm}
\includegraphics[width=0.05\textwidth]{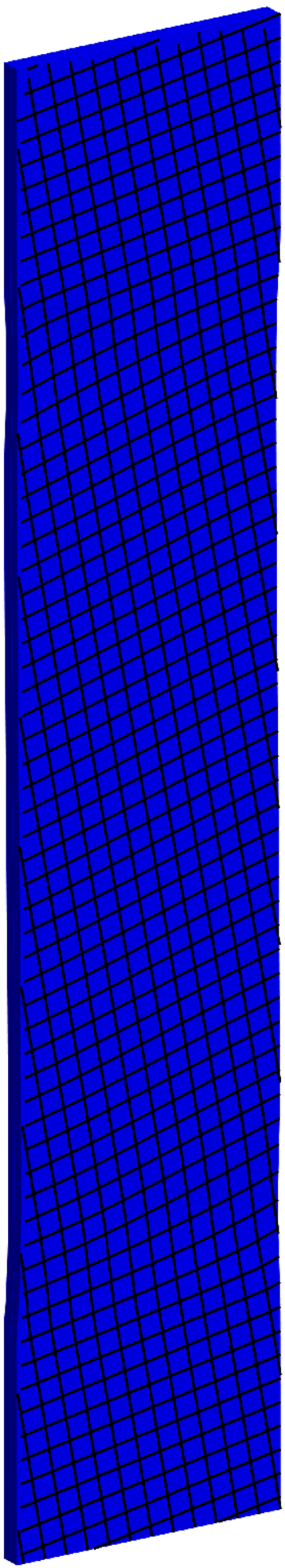}
\includegraphics[width=0.05\textwidth]{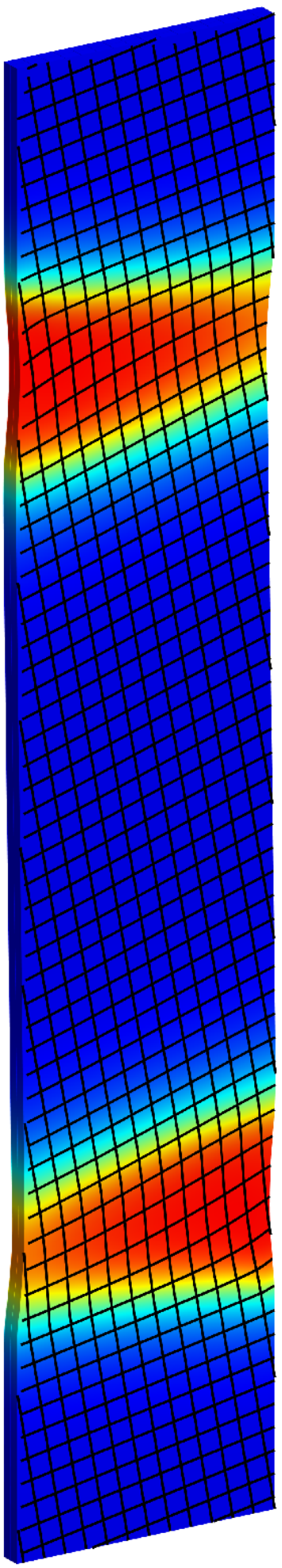}
\includegraphics[width=0.05\textwidth]{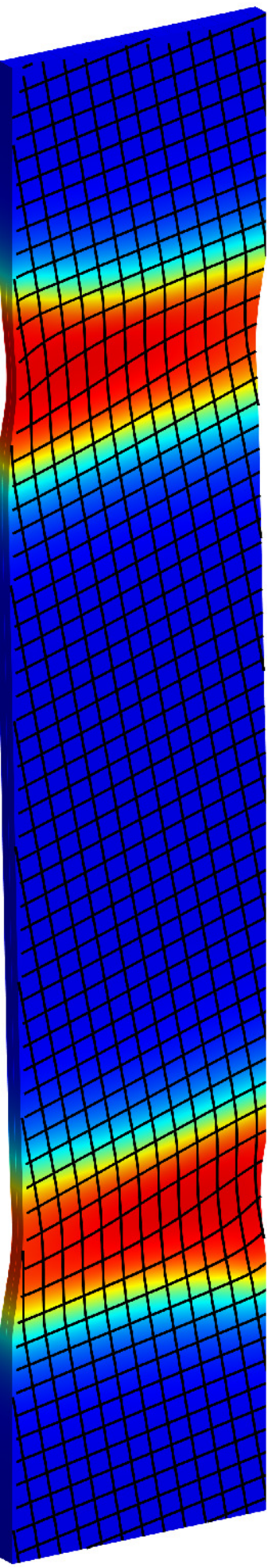}
\hspace{2mm}
\includegraphics[width=0.05\textwidth]{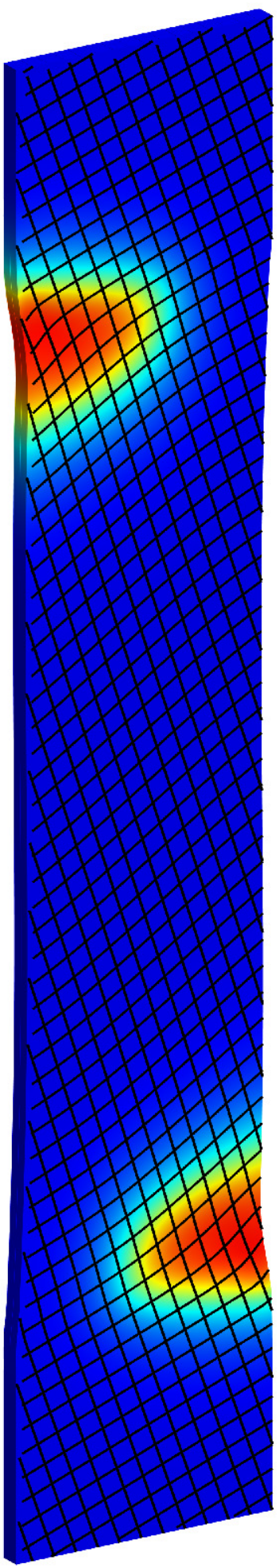}
\includegraphics[width=0.05\textwidth]{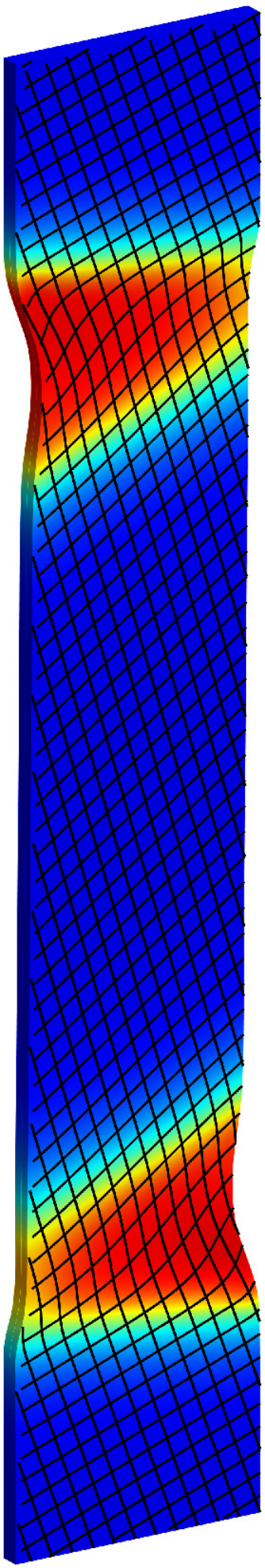}
\includegraphics[width=0.05\textwidth]{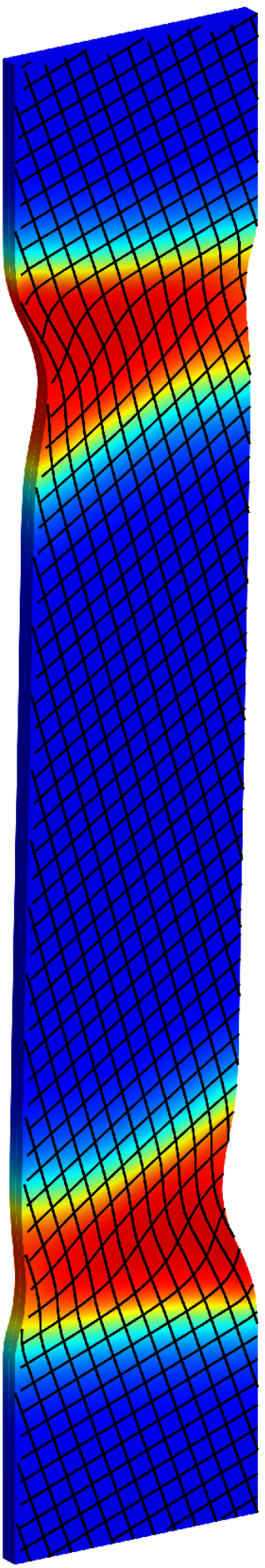}
\hspace{2mm}
\includegraphics[width=0.05\textwidth]{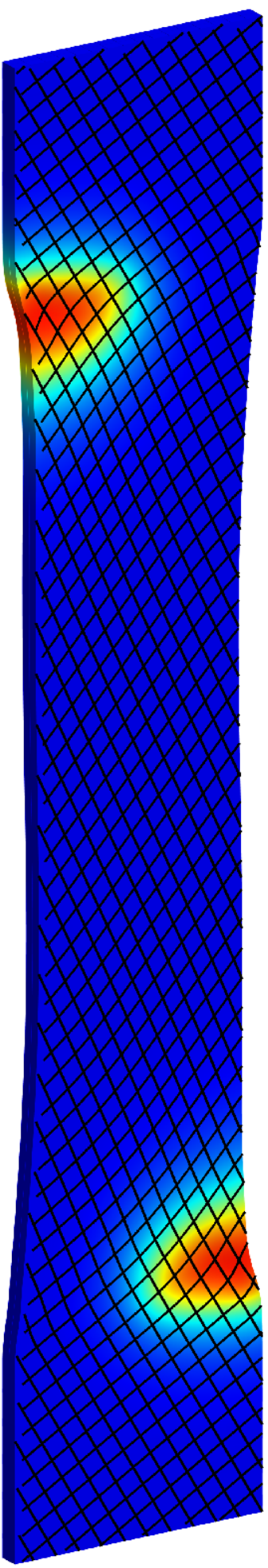}
\includegraphics[width=0.05\textwidth]{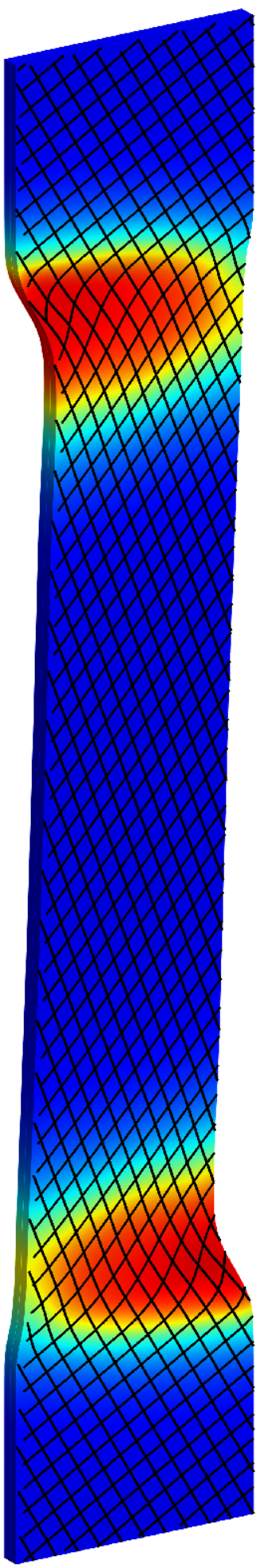}
\includegraphics[width=0.053\textwidth]{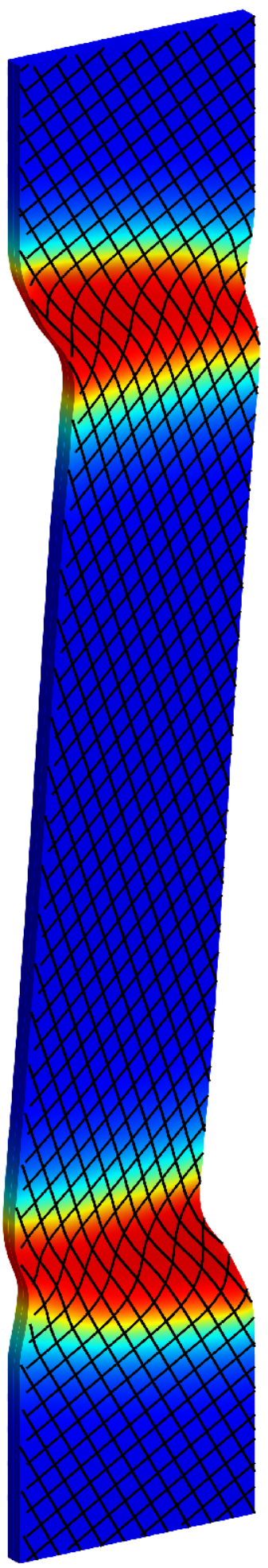}
\hspace{2mm}
\includegraphics[width=0.05\textwidth]{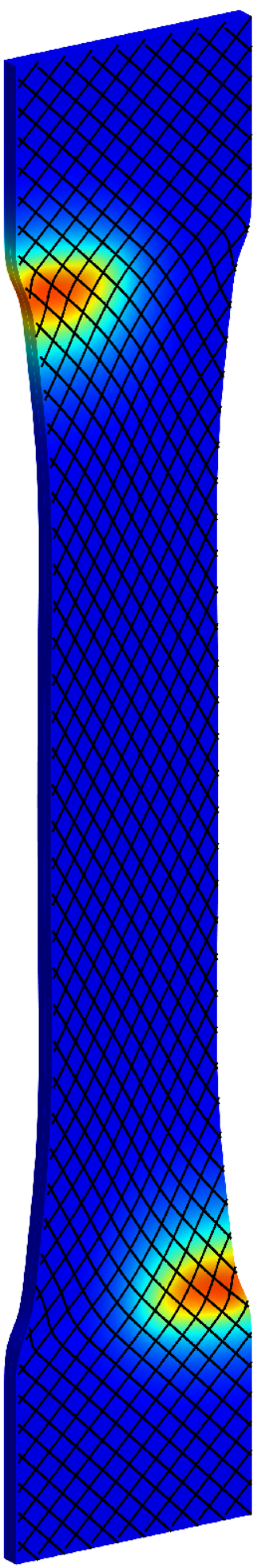}
\includegraphics[width=0.05\textwidth]{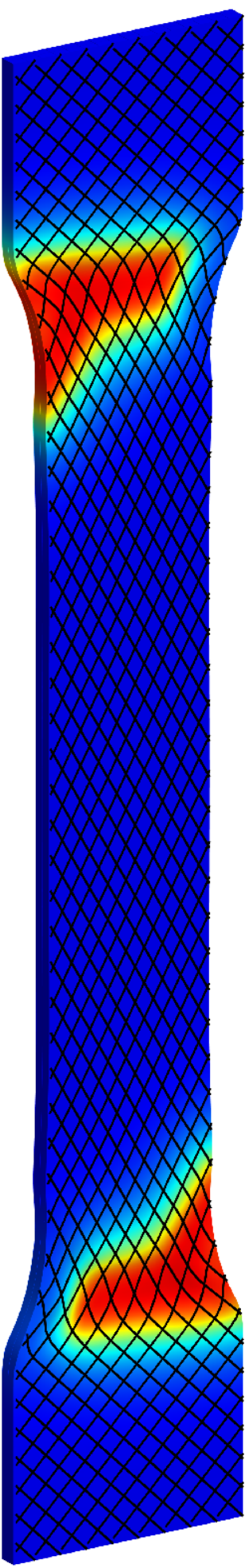}
\includegraphics[width=0.034\textwidth]{pictures/colorbarVphase2t}

\includegraphics[width=0.05\textwidth]{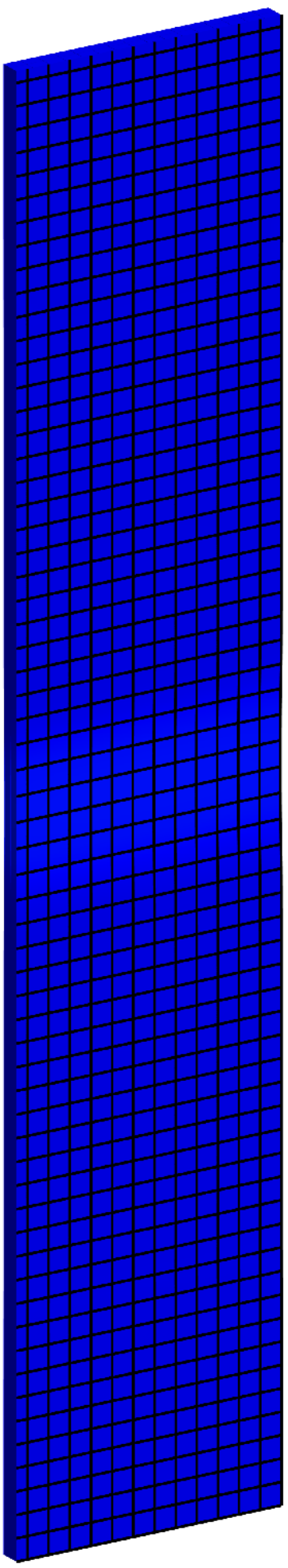}
\includegraphics[width=0.05\textwidth]{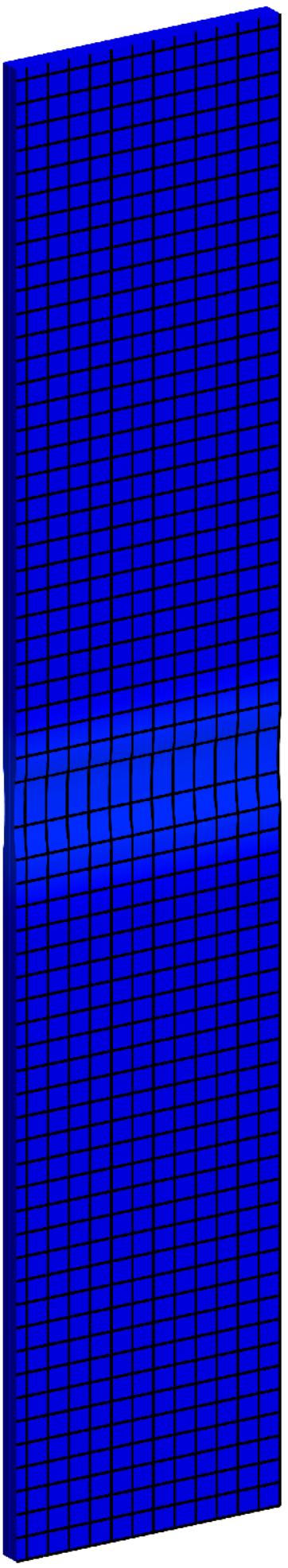}
\hspace{2mm}
\includegraphics[width=0.05\textwidth]{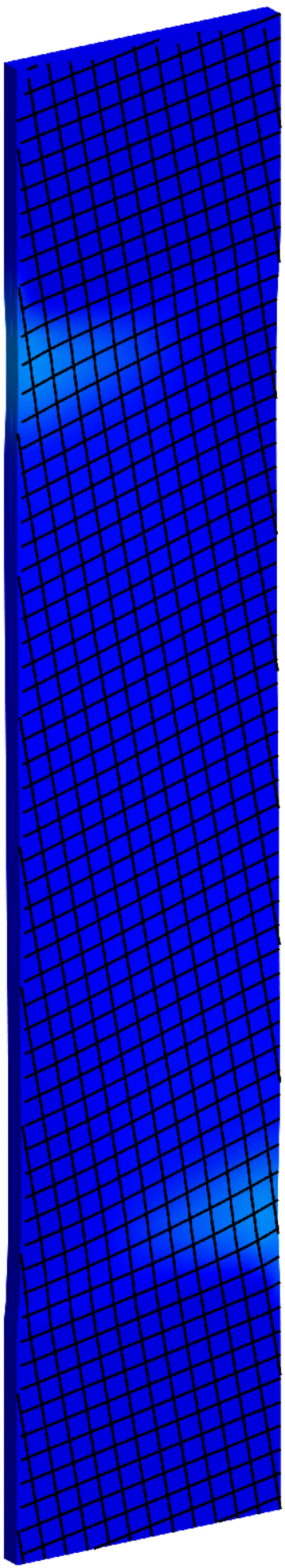}
\includegraphics[width=0.05\textwidth]{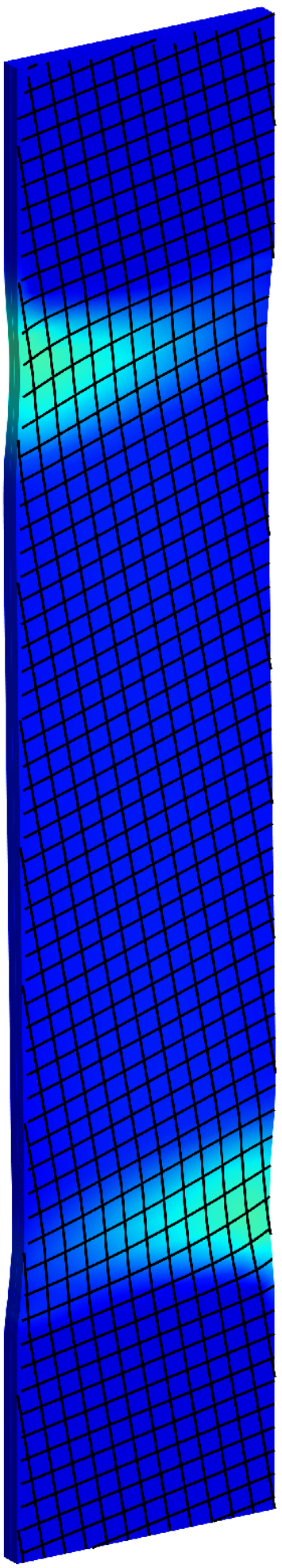}
\includegraphics[width=0.05\textwidth]{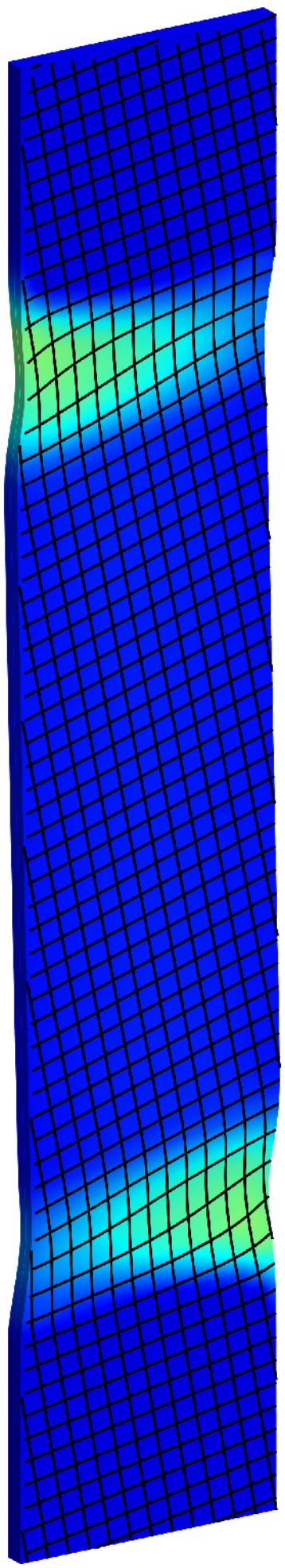}
\hspace{2mm}
\includegraphics[width=0.05\textwidth]{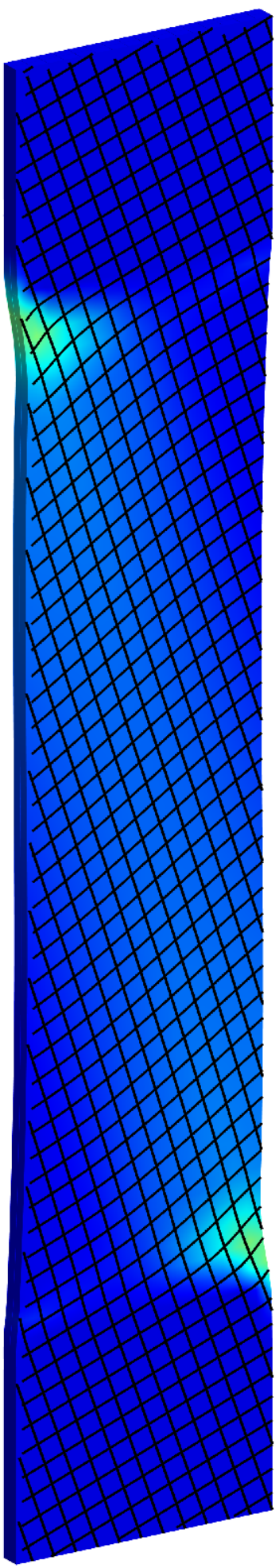}
\includegraphics[width=0.05\textwidth]{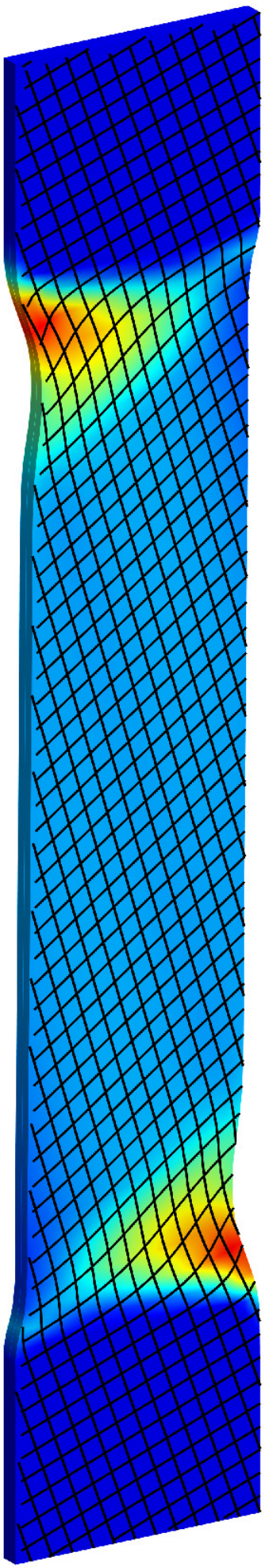}
\includegraphics[width=0.05\textwidth]{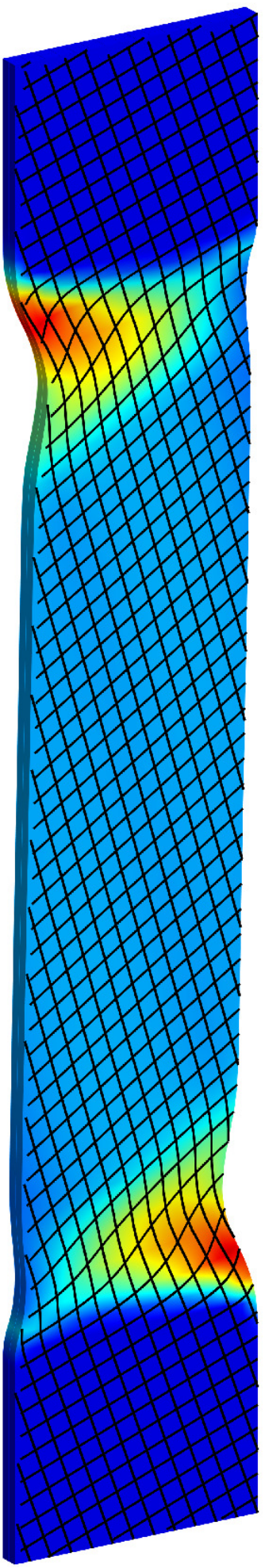}
\hspace{2mm}
\includegraphics[width=0.05\textwidth]{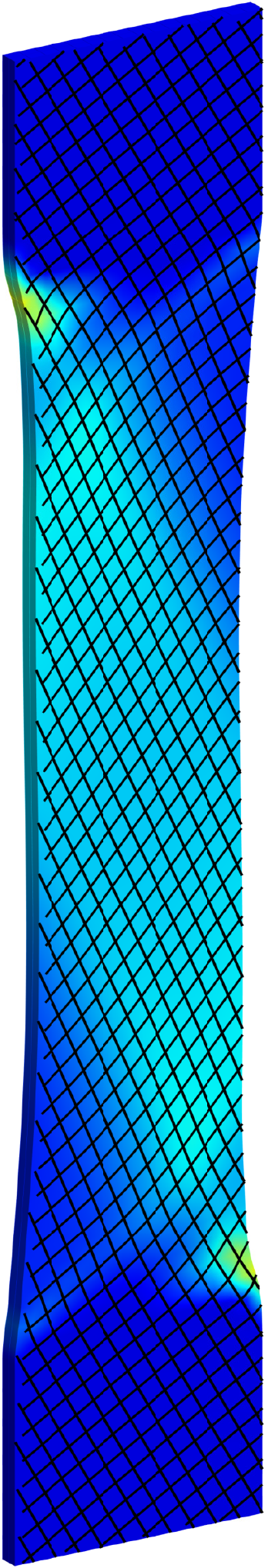}
\includegraphics[width=0.05\textwidth]{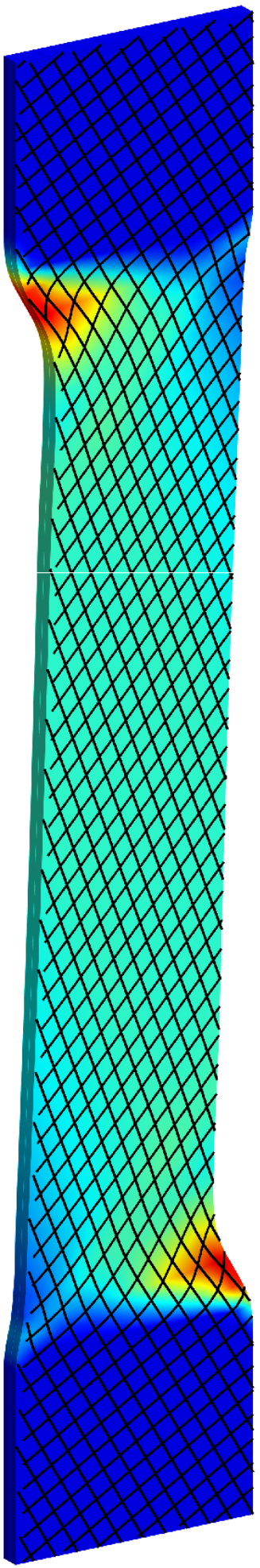}
\includegraphics[width=0.053\textwidth]{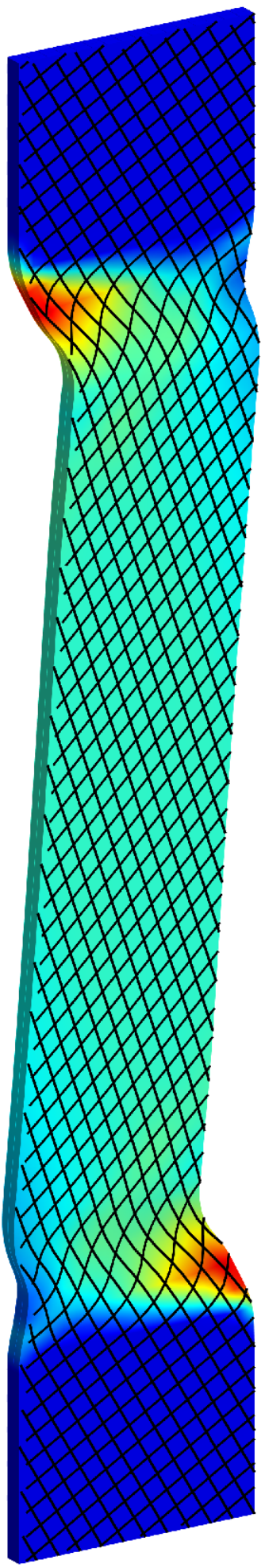}
\hspace{2mm}
\includegraphics[width=0.05\textwidth]{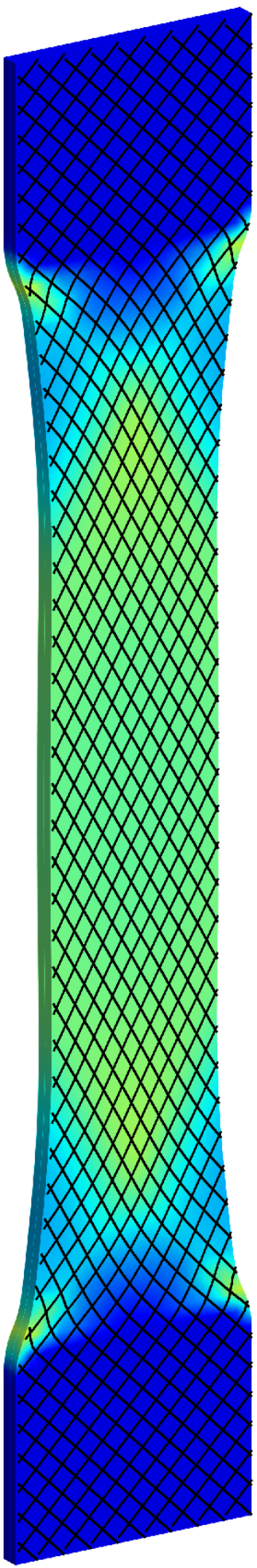}
\includegraphics[width=0.05\textwidth]{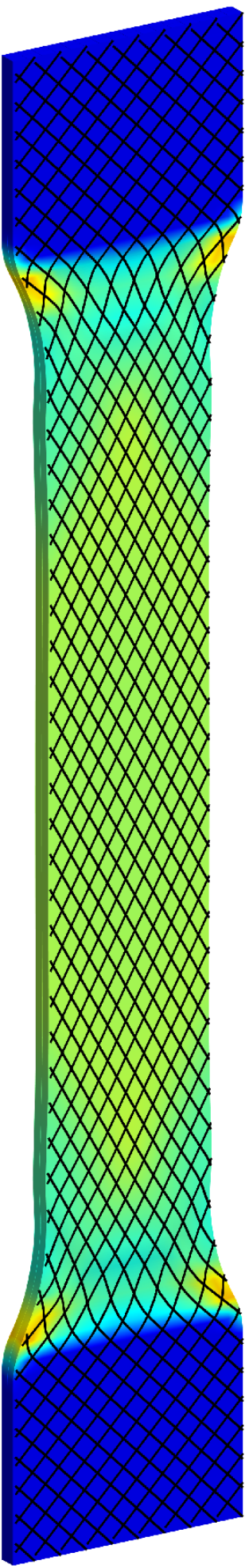}
\includegraphics[width=0.034\textwidth]{pictures/colorbarValpha2}

\includegraphics[width=0.05\textwidth]{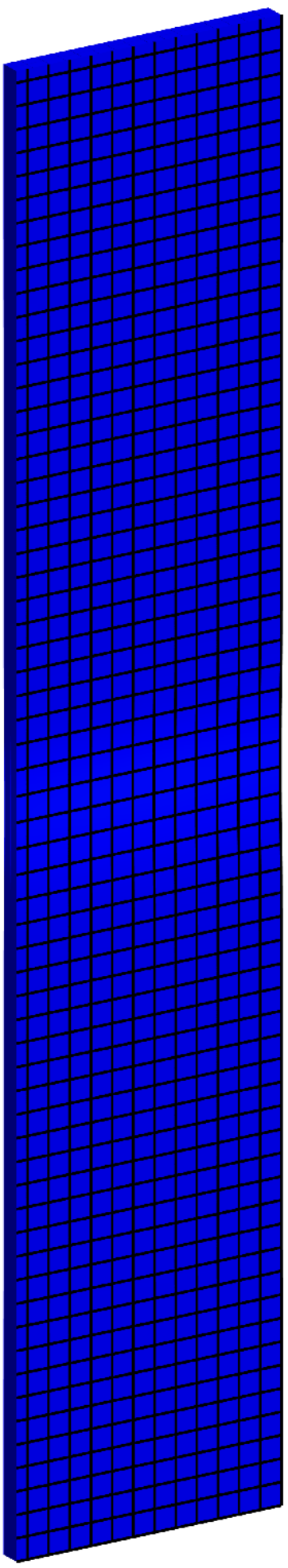}
\includegraphics[width=0.05\textwidth]{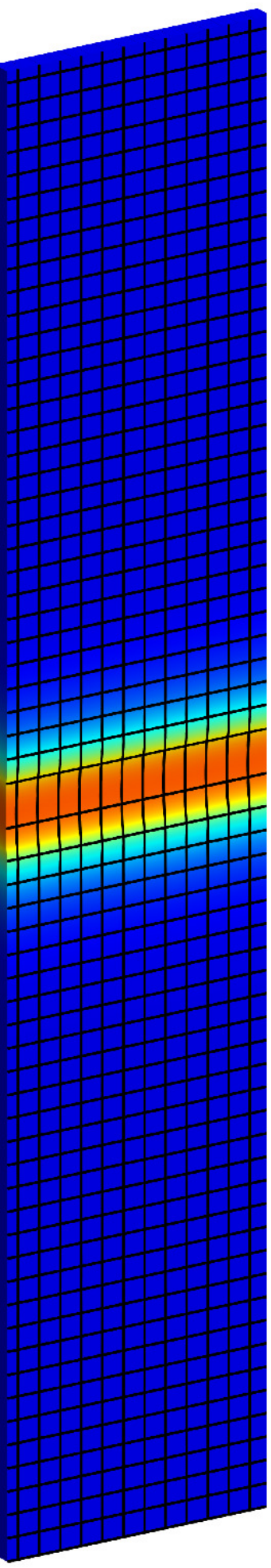}
\hspace{2mm}
\includegraphics[width=0.05\textwidth]{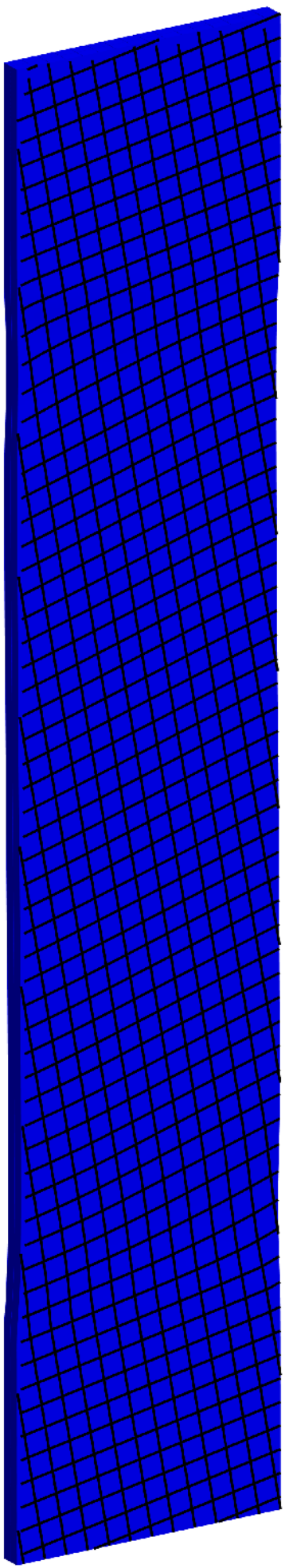}
\includegraphics[width=0.05\textwidth]{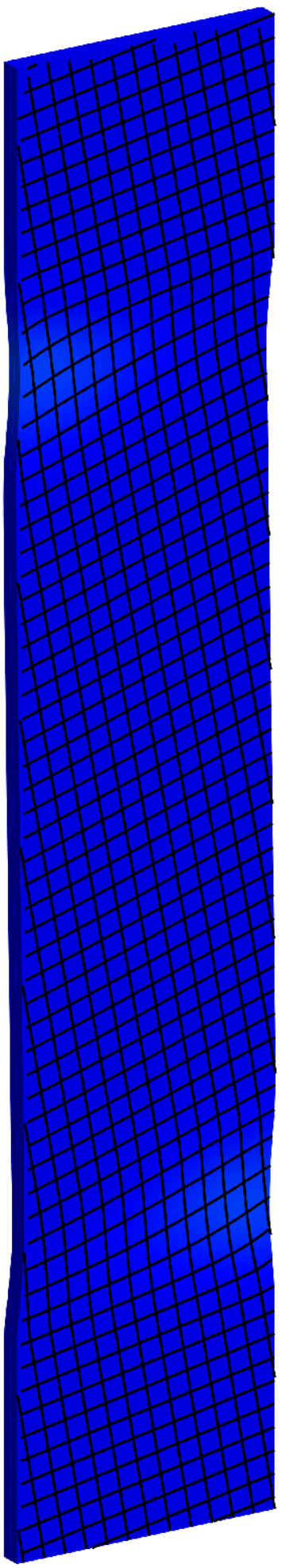}
\includegraphics[width=0.05\textwidth]{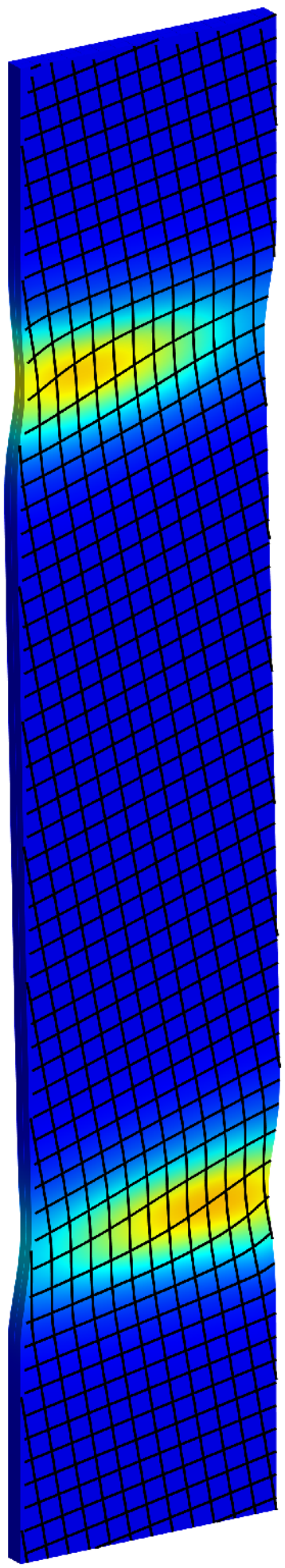}
\hspace{2mm}
\includegraphics[width=0.05\textwidth]{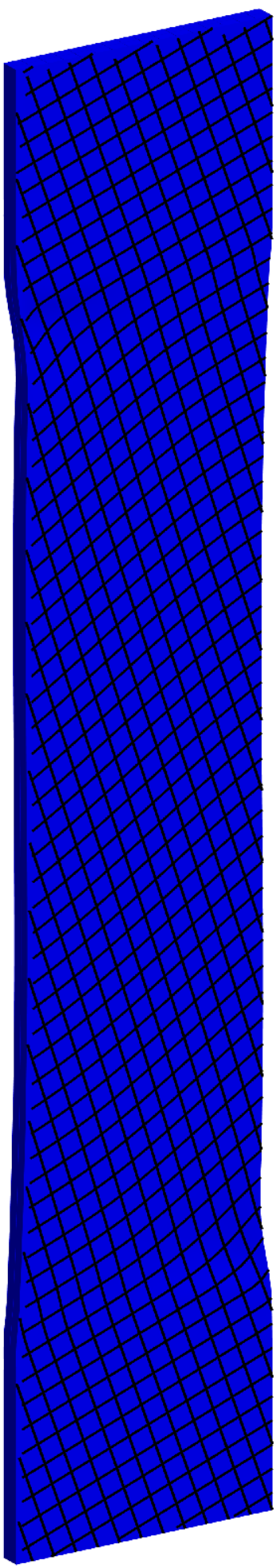}
\includegraphics[width=0.05\textwidth]{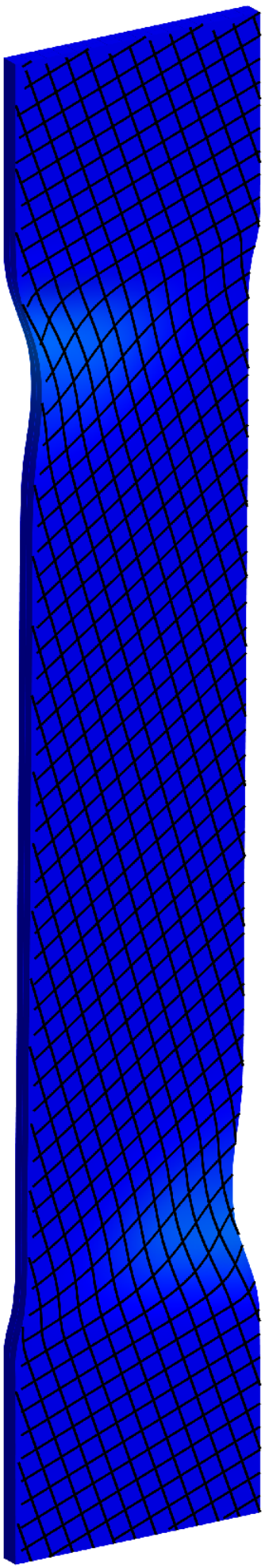}
\includegraphics[width=0.05\textwidth]{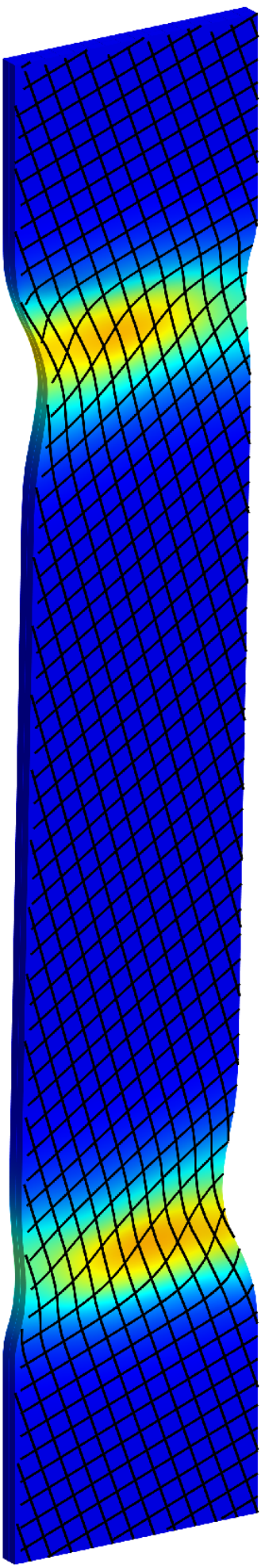}
\hspace{2mm}
\includegraphics[width=0.05\textwidth]{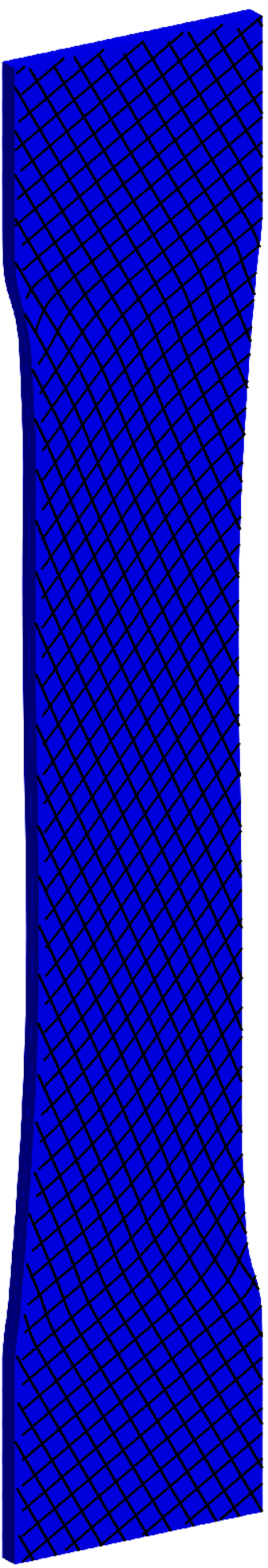}
\includegraphics[width=0.05\textwidth]{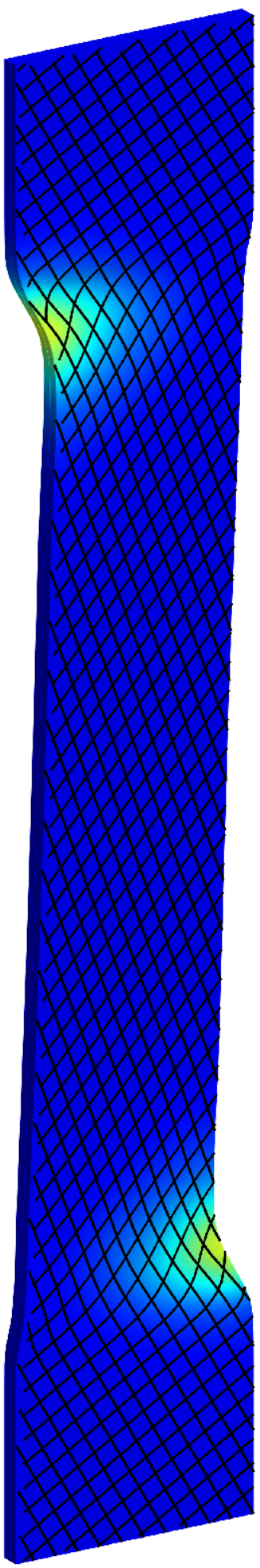}
\includegraphics[width=0.053\textwidth]{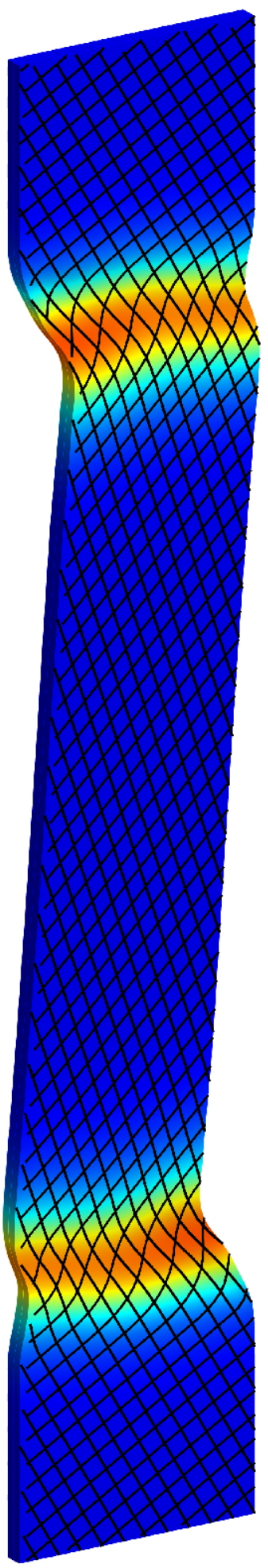}
\hspace{2mm}
\includegraphics[width=0.05\textwidth]{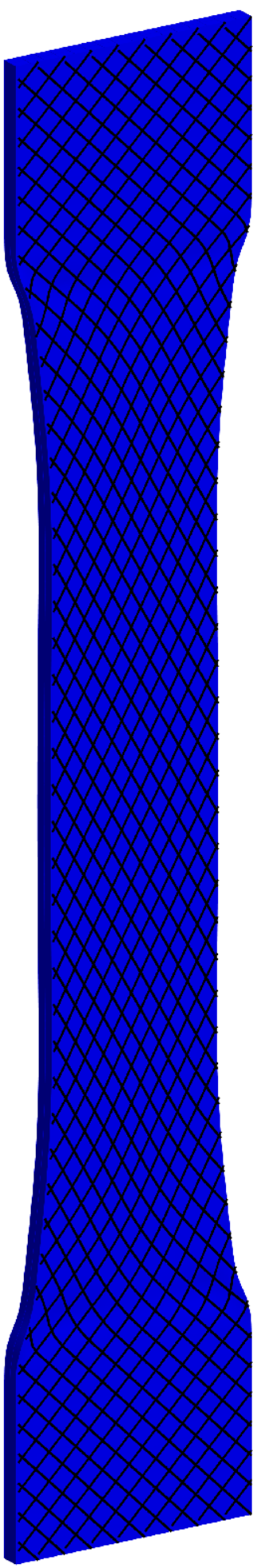}
\includegraphics[width=0.05\textwidth]{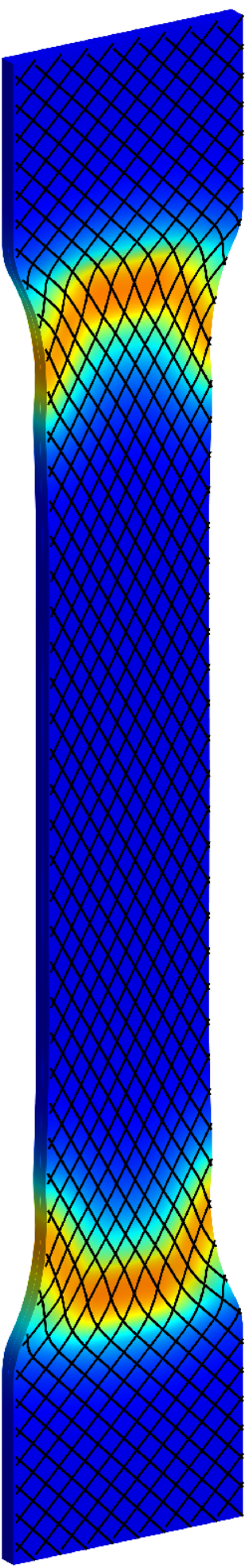}
\includegraphics[width=0.034\textwidth]{pictures/colorbarVphase2}
\\
\hspace{0.5mm}\ding{192}\hspace{6.4mm}\ding{193}\hspace{9.5mm}\ding{194}\hspace{6.4mm}\ding{195}\hspace{6.4mm}\ding{196}\hspace{9mm}\ding{197}\hspace{6.4mm}\ding{198}\hspace{6.4mm}\ding{199}\hspace{9mm}\ding{200}\hspace{6.4mm}\ding{201}\hspace{6.4mm}\ding{202}\hspace{10mm}\ding{203}\hspace{6.4mm}\ding{204}\hspace{5.5mm}
\vspace{-4.5mm}
\end{center}
\caption{\textbf{Tensile Test (bidirectional).} Results of the fiber crack phase-field (first row) as well as the plastic strain field (second row) and crack phase-field of the matrix material (third row). The results are shown for the different deformation states and fiber configurations marked in Figure \ref{fig:bitension}.}
\label{fig:bitensionPfPs}
\end{figure}

Next, we investigate the same tension test using a bidirectional reinforced material with fiber orientations of \(\vartheta=[0^{\circ},10^{\circ},20^{\circ},30^{\circ},45^{\circ}]\) as shown in Figure \ref{fig:tensionBiRef}. The load deflection results for isothermal simulations at \(\theta=293\,\mathrm{K}\) are depicted in Figure \ref{fig:bitension}. Again, crack initialization and final rupture of the fiber material are indicated by \(\square\) and \(\circ\), respectively. The final rupture of the matrix material is indicated by \(\cross\). Crack phase-field results of the fiber and matrix material as well as results of the plastic strain are depicted in Figure \ref{fig:bitensionPfPs}. Note that only phase-field results of the fiber aligned in \(\vartheta\)-direction is plotted.

For fiber orientations of \(\vartheta=[0^{\circ},10^{\circ},20^{\circ}]\), the bidirectional reinforced material shows a similar behavior compared to the corresponding unidirectional reinforced counterparts. The additional orthogonal fiber merely accounts for a higher necking resistance \ding{192}--\ding{199}. Moreover, the orthogonal fiber configuration restrict a relative movement between the respective fibers leading to a slightly stiffer material behavior. This has to be investigated in terms of experimental measurements which is out of the scope of present work.

Applying fiber orientations of \(\vartheta=[30^{\circ},45^{\circ}]\), this \textcolor{black}{stiffening} effect becomes more pronounced as can be observed in Figure \ref{fig:bitension}. As already discussed, the additional, orthogonal oriented fiber counteract the necking behavior due to the Poisson effect of the matrix material such that fractures within the matrix material emerges near the clamping zones and not in the center of the specimen \ding{200}--\ding{204}.

\subsubsection{Thermal investigation}

Eventually, we investigate the temperature dependency of the proposed model. Therefore, we reuse the tension test with a unidirectional fiber reinforcement as shown in Figure \ref{fig:tensionUniRef} and apply a fiber orientation of \(\vartheta=30^{\circ}\). 

\begin{figure}
\begin{center}
\psfrag{1}[c][c]{\ding{192}}
\psfrag{2}[c][c]{\ding{193}}
\psfrag{3}[c][c]{\ding{194}}
\includegraphics[width=0.98\textwidth]{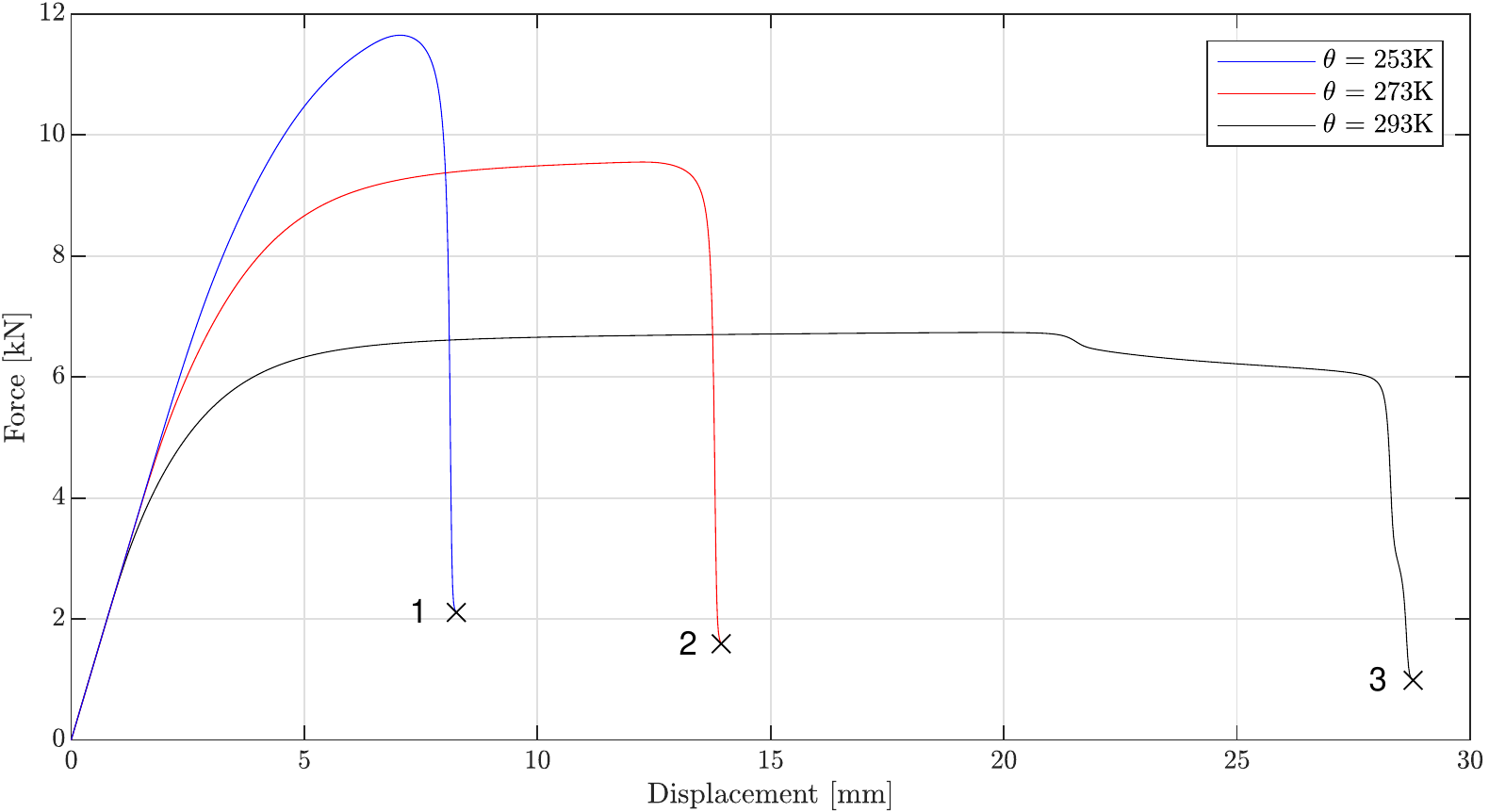}
\end{center}
\caption{\textbf{Thermal investigations.} Load deflection results for isothermal simulations at different temperatures of \(\theta= [253,\, 273,\, 293]\,\mathrm{K}\) and a fiber orientation of \(\vartheta=30^{\circ}\).}
\label{fig:unitensionIso}
\end{figure}

\begin{figure}
\begin{center}
\footnotesize
\psfrag{a1}[l][l]{0.7}
\psfrag{a0}[l][l]{0}
\psfrag{a}[l][l]{$\alpha$}
\psfrag{s1}[l][l]{1}
\psfrag{s0}[l][l]{0}
\psfrag{s}[l][l]{$\s$}
\psfrag{t}[l][l]{$\s_{\mathrm{L}}$}
\includegraphics[width=0.065\textwidth]{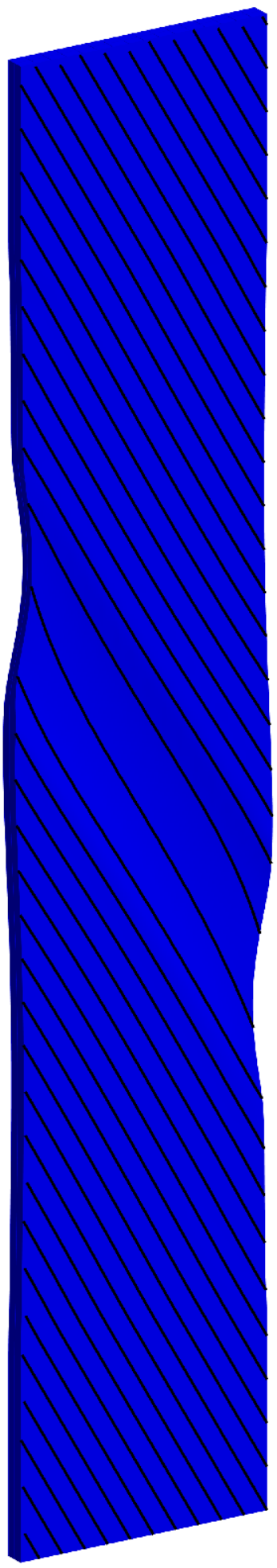}
\hspace{2mm}
\includegraphics[width=0.065\textwidth]{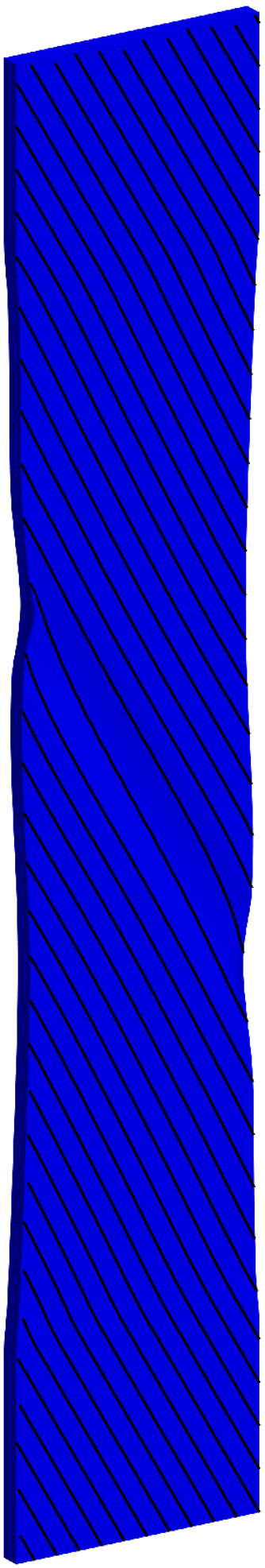}
\hspace{2mm}
\includegraphics[width=0.065\textwidth]{pictures/60PHF1440}
\includegraphics[width=0.047\textwidth]{pictures/colorbarVphase2t}
\hspace{4mm}
\includegraphics[width=0.065\textwidth]{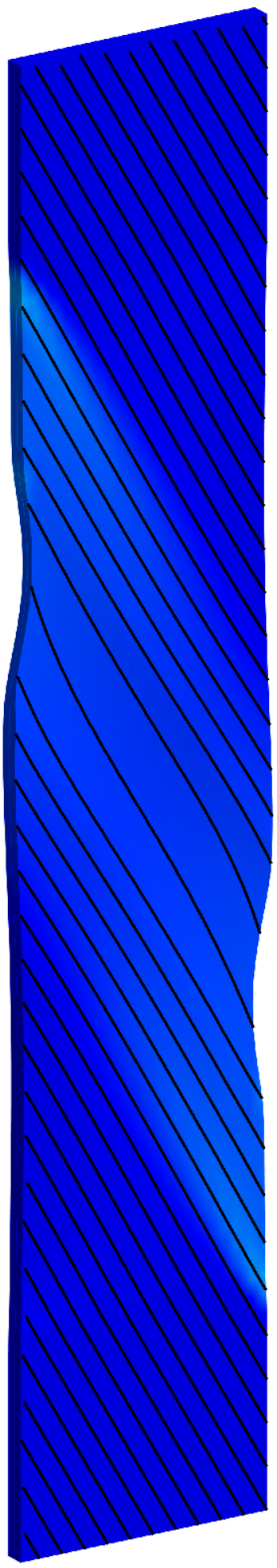}
\hspace{2mm}
\includegraphics[width=0.065\textwidth]{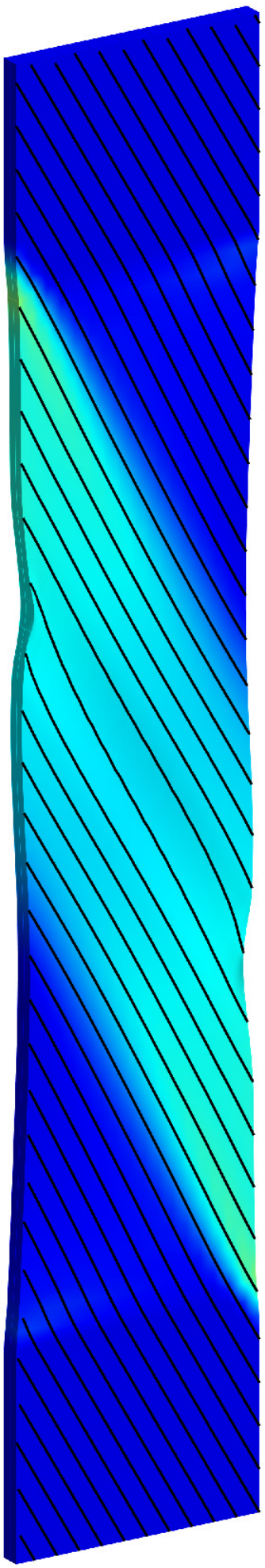}
\hspace{2mm}
\includegraphics[width=0.065\textwidth]{pictures/60PL1440}
\includegraphics[width=0.047\textwidth]{pictures/colorbarValpha2}
\hspace{4mm}
\includegraphics[width=0.065\textwidth]{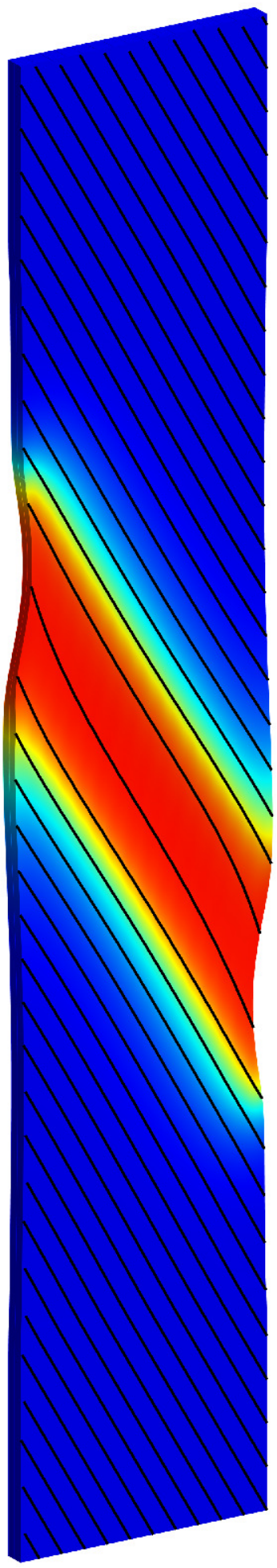}
\hspace{2mm}
\includegraphics[width=0.065\textwidth]{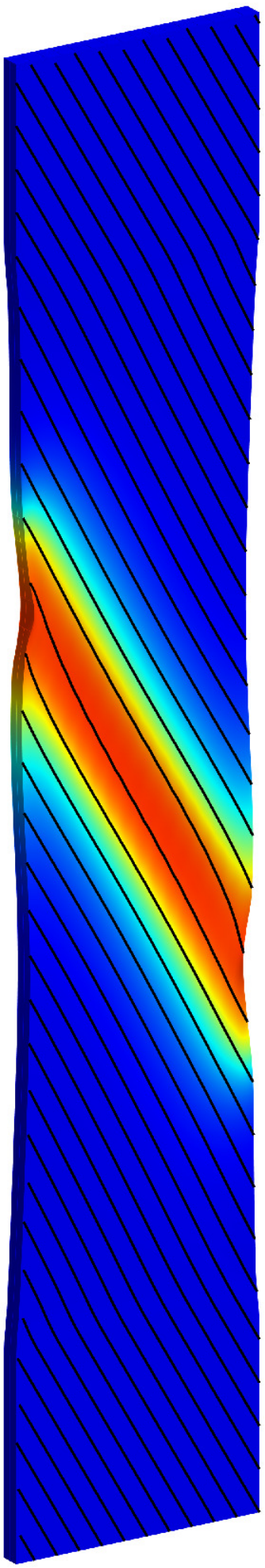}
\hspace{2mm}
\includegraphics[width=0.065\textwidth]{pictures/60PHM1440}
\includegraphics[width=0.047\textwidth]{pictures/colorbarVphase2}
\\
\hspace{0mm}\ding{192}\hspace{12mm}\ding{193}\hspace{12mm}\ding{194}\hspace{22mm}\ding{192}\hspace{12mm}\ding{193}\hspace{12mm}\ding{194}\hspace{22mm}\ding{192}\hspace{12mm}\ding{193}\hspace{12mm}\ding{194}\hspace{7mm}
\vspace{-2mm}
\end{center}
\caption{\textbf{Thermal investigations.} Results of isothermal simulations at different temperatures of \(\theta= [253,\, 273,\, 293]\,\mathrm{K}\) (each from left to right) and a fiber orientation of \(\vartheta=30^{\circ}\). Results are shown for the fiber crack phase-field (first block), the plastic strain field (second block) and crack phase-field of the matrix material (third block) at the last deformation state marked in Figure \ref{fig:unitensionIso}.}
\label{fig:unitensionIso2}
\end{figure}

Figure \ref{fig:unitensionIso} shows the load deflection result for isothermal simulations using temperatures of \(\theta= [253,\, 273,\, 293]\,\mathrm{K}\). The corresponding crack phase-field results of the fiber and matrix material as well as results of the plastic strain are depicted for the last deformation step in Figure \ref{fig:unitensionIso2}.
As already observed previously, for \(\theta=293\,\mathrm{K}\) the matrix material undergoes large plastic deformations followed by fiber fracture in small areas near the clamping zones and finally the matrix material undergoes ductile fracture at the center of the specimen. Lower temperatures increase the yield stress of the matrix material leading to a higher elastic energy and thus an earlier, less ductile fracture behavior of the matrix material. Note that for isothermal simulations with \(\theta=[ 253,\,273]\,\mathrm{K}\) the matrix material fails before any fiber cracks occur.

\begin{table}
\footnotesize
\centering
\begin{tabular}{@{}p{0.6\textwidth}@{}ll}
\hline
\textbf{Elastic parameters} & \\
Shear modulus \(\mu\) & \(1630\,\mathrm{MPa}\) \\
Shear exponent \(\alpha\) & \(2\) \\
Bulk modulus \(\kappa\) & \(6250\,\mathrm{MPa}\) \\
Bulk parameter \(\beta\) & \(-2\) \\
Matrix volume ratio  \(\zeta\) & \(0.53\) \\  
Tensile stiffness  \(a\) &   \(79 000\,\mathrm{MPa}\) \\
Shear stiffness  \(b\) &   \(0\,\mathrm{MPa}\) \\
Bending stiffness  \([c_{\perp},\,c_{\#}]\) &  \([16.46,\,16.46]\,\mathrm{N}\)\\
\hline
\textbf{Plastic parameters} & \\
Yield stress \([y_{0}(\theta_\mathrm{ref}),\,y_{1}(\theta_\mathrm{ref}),\,y_{2}(\theta_\mathrm{ref})]\) & \([22,\,56.8,\,30]\,\mathrm{MPa}\) \\
Saturation exponent \([\omega_{\mathrm{p1}},\,\omega_{\mathrm{p2}}]\) & \([1,\,115]\) \\
Thermal softening parameter \([\omega_{\mathrm{t0}},\,\omega_{\mathrm{t1}},\,\omega_{\mathrm{t2}}]\) & \([0.4,\,0.4,\,0.4]\,\mathrm{K}^{-1}\) \\
Viscoplastic parameter \(\eta_{\mathrm{p}}\) & \(5000\,\mathrm{MPa\cdot s}\) \\
Viscoplastic exponent \(\mathrm{n}_\mathrm{p}\) & \(1\) \\
Plastic length scale \(l_{\mathrm{p}}\) & \(3.1\,\mathrm{mm}\) \\
Initial void fraction  \(f_{\mathrm{0}}\) & \(0.01\) \\
Gurson fitting parameter \([q_{\mathrm{1}},\,q_{\mathrm{2}}]\) & \([3,\,0.8]\) \\
\hline
\textbf{Phase-field fracture parameters} & \\
Brittle critical fracture energy \([g_{\mathrm{c,e}},\,g_{\mathrm{c_L}},\,g_{\mathrm{c_M}}]\) & \([500,\,500,\,500]\,\mathrm{kJ/m^{2}}\) \\
Ductile critical fracture energy \(g_{\mathrm{c,p}}\) & \(50\,\mathrm{kJ/m^{2}}\) \\
Saturation exponent \(\omega_{\mathrm{f}}\) & \(3\) \\
Fracture viscosity \([\eta_{\mathrm{f}},\,\eta_{\mathrm{f_L}},\,\eta_{\mathrm{f_M}}]\) & \([1,\,1,\,1]\cdot10^{-7}\,\mathrm{MPa\cdot s}\) \\
Fracture length scale \([l_{\mathrm{f}},\,l_{\mathrm{f_L}},\,l_{\mathrm{f_M}}]\) & \([3.1,\,3.1,\,3.1]\,\mathrm{mm}\) \\
Degradation parameter \([a_{\mathrm{g}},\,a_{\mathrm{g_L}},\,a_{\mathrm{g_M}}]\) & \([0.001,\,0.001,\,0.001]\) \\
\hline
\textbf{Thermal parameters} &  \\
Specific heat capacity  \([c_{\mathrm{mat}},\,c_{\mathrm{fib}}]\) &  \( [1860,\,2080] \,\mathrm{kJ}/(\mathrm{m^3}\cdot\mathrm{K})\)  \\ 
Thermal expansion coefficient \([\epsilon,\,\upsilon]\) & \([106,\,5]\,\cdot10^{-6}\,\mathrm{K}^{-1}\) \\
Thermal expansion parameter \(\gamma\) & \(1\) \\
Conductivity \(K\) & \(0.25\,\mathrm{W}/(\mathrm{m}\cdot\mathrm{K})\) \\
Convection \(K_{\mathrm{conv}}\) & \(0\,\mathrm{W}/(\mathrm{m}\cdot\mathrm{K})\) \\ 
Reference temperature \(\theta_\mathrm{ref}\) & \(293\,\mathrm{K}\) \\  
Fracture \& plastic dissipation factor \([\nu_{\mathrm{p_{mat}}},\,\nu_{\mathrm{f_{mat}}},\,\nu_{\mathrm{f_{fib}}}]\) & \([0.9,\,0.9,\,0.9]\) \\
\hline
\end{tabular}
\caption{Material setting of the fiber reinforced composite (PA 6/Roving glass).}
\label{table:matData1}
\end{table}

%% file: chapters/conclusion.tex
\section{Conclusions}\label{sec:conclusions}
The non-linear framework presented in this work allows for a comprehensive investigation of damage and fracture in fiber reinforced polymers. The combination of a second-gradient theory, a novel hybrid phase-field model and a temperature dependent GTN-type plasticity model provides a numerical framework which is able to describe different failure mechanisms in detail. This approach allows for improvements in the design of such composite materials since we are able to predict fiber and matrix failure and their sequence dependent on the fiber orientation. Moreover, due to the fully-coupled, thermomechanical approach we can optimize the fiber orientation for specific loads and thermal states. Several numerical tests conducted within this work have demonstrate the capability of the proposed framework to investigate such a complex behavior including the growth of microvoids, plasification and necking, crack initiation and propagation within the composite material and its components, respectively.

\section*{Acknowledgements}
Support for the present research was provided by the Deutsche Forschungsgemeinschaft (DFG) under grant HE5943/8-1 and DI2306/1-1. The authors C. Hesch and M. Dittmann gratefully acknowledge this support.

This is a preprint of an article published in Computational Mechanics. The final authenticated version is available online at 
\href{https://doi.org/10.1007/s00466-021-02018-0}{doi.org/10.1007/s00466-021-02018-0}.